\documentclass[onecolumn,amsmath,amssymb,aps]{revtex4-1}
\usepackage[latin9]{inputenc}
\usepackage{color}
\usepackage{amsmath}
\usepackage{graphicx}
\usepackage{esint}

\makeatletter

\newcommand{\lyxmathsym}[1]{\ifmmode\begingroup\def\b@ld{bold}
  \text{\ifx\math@version\b@ld\bfseries\fi#1}\endgroup\else#1\fi}

\providecommand{\tabularnewline}{\\}

\usepackage{dcolumn}
\usepackage{bm}

\makeatother

\begin{document}

\preprint{APS/123-QED}

\title{On the Theory of Light Propagation in Crystalline Dielectrics }

\author{Marius Dommermuth}

\author{Nils Schopohl}

\email[Corresponding author: ]{nils.schopohl@uni-tuebingen.de }

\affiliation{Institut für Theoretische Physik and CQ Center for Collective Quantum
Phenomena and their Applications in LISA$^{+}$,\\
 Eberhard-Karls-Universität Tübingen, Auf der Morgenstelle 14, D-72076
Tübingen, Germany}

\date{\today}
\begin{abstract}
A synoptic view on the long-established theory of light propagation
in crystalline dielectrics is presented, where charges, tightly bound
to atoms (molecules, ions) interact with the microscopic local electromagnetic
field. Applying the Helmholtz-Hodge decomposition to the current density
in Maxwell's equations, two decoupled sets of equations result determining
separately the divergence-free (transversal) and curl-free (longitudinal)
parts of the electromagnetic field, thus facilitating the restatement
of Maxwell's equations as equivalent field-integral equations. Employing
a suitably chosen basis system of Bloch functions we present for dielectric
crystals an exact solution to the inhomogenous field-integral equations
determining the \emph{local} electromagnetic field that polarizes
individual atoms or ionic subunits in reaction to an external electromagnetic
wave. From the solvability condition of the associated homogenous
integral equation then the propagating modes and the photonic bandstructure
$\omega_{n}\left(\mathbf{q}\right)$ for various crystalline symmetries
$\Lambda$ are found solving a \emph{small} sized $3M\times3M$ matrix
eigenvalue problem, with {\normalsize{}$M$} denoting the number of
polarizable atoms (ions) in the unit cell. Identifying the \emph{macroscopic}
electric field inside the sample with the spatially low-pass filtered
\emph{microscopic} local electric field, the dielectric $3\times3$-tensor
$\varepsilon_{\Lambda}\left(\mathbf{q},\omega\right)$ of crystal
optics emerges, relating the accordingly low-pass filtered microscopic
polarization field to the macroscopic electric field, solely with
the individual microscopic polarizabilities $\alpha\left(\mathbf{R},\omega\right)$
of atoms (molecules, ions) at a site $\mathbf{R}$ and the crystalline
symmetry as input into the theory. Decomposing the microscopic local
electric field into longitudinal and transversal parts, an effective
wave equation determining the radiative part of the macroscopic field
in terms of the transverse dielectric tensor $\varepsilon_{ab}^{\left(T\right)}\left(\mathbf{q},\omega\right)$
is deduced from the exact solution to the field-integral equations.
The Taylor expansion $\varepsilon_{ab}^{\left(T\right)}\left(\mathbf{q},\omega\right)=\varepsilon_{ab}^{\left(T\right)}\left(\omega\right)+i\sum_{c}\gamma_{abc}\left(\omega\right)q_{c}+\sum_{c,d}\alpha_{abcd}\left(\omega\right)q_{c}q_{d}+...$
around $\mathbf{q}=\mathbf{0}$ provides then insight into various
optical phenomena connected to retardation and non-locality of the
dielectric tensor, in full agreement with the phenomenological reasoning
of Agranovich and Ginzburg in ``Crystal Optics with Spatial Dispersion,
and Excitons'' (Springer Berlin Heidelberg, 1984): the eigenvalues
of the tensor $\varepsilon_{ab}^{\left(T\right)}\left(\omega\right)$
describing chromatic dispersion of the index of refraction and birefringence,
the first order term $\gamma_{abc}\left(\omega\right)$ specifying
rotary power (natural optical activity), the second order term $\alpha_{abcd}\left(\omega\right)$
shaping the effects of a spatial-dispersion-induced birefringence,
a critical parameter for the design of lenses made from $CaF_{2}$
and $BaF_{2}$ for optical lithography systems in the ultraviolet.
In the \emph{static} limit an exact expression for $\varepsilon_{\varLambda}$
is deduced, that conforms with general thermodynamic stability criteria
and reduces for cubic symmetry to the Clausius-Mossotti relation.
Considering various dielectric crystals comprising atoms with known
polarizabilities from the literature, in all cases the calculated
indices of refraction, the rotary power and the spatial-dispersion-induced
birefringence coincide well with the experimental data, thus illustrating
the utility of the theory. For ionic crystals, exemplarily for $CsI$
and $RbCl$, a satisfactory agreement between theory and the measured
chromatic dispersion of the index of refraction is shown over a wide
frequency interval, ranging from ultraviolet to far infra-red, accomplishing
this with an appreciably smaller number of adjustable parameters compared
to the well known Sellmeier fit. 
\end{abstract}

\pacs{Valid PACS appear here}

\maketitle

\section{\label{sec:Intro}Introduction}

When optical signals traverse a transparent dielectric, for example
a fused quartz (silica) prism, the light travels at different speed
depending on frequency $f=\frac{\omega}{2\pi}$ , so that the shape
of a wave packet, say composed of mixed frequencies $\left|f-f_{c}\right|\leq\frac{\varDelta f_{c}}{2}$
around a carrier frequency $f_{c}$ in a frequency interval of width
$\varDelta f_{c}$, tends to spread out. This is the well known chromatic
dispersion effect resulting from the frequency dependence of the refractive
index $n=n\left(\omega\right)$. Microscopic considerations based
on first principles reveal, that the frequency dependence of the refractive
index $n\left(\omega\right)$ is directly connected to the\emph{ retarded}
response of the polarizable constituents of matter, the latter distinguishing
themselves as atoms, molecules or ions. The chromatic dispersion effect
is further supplemented by the effects of crystalline anisotropy and
also by the effects of spatial dispersion brought about by the \emph{non
local} dependence of this response, that is a charge at point $\mathbf{r}$
recollects the action exerted on it at another position $\mathbf{r}'$
\citep{L.V.Keldysh2012,Maksimov2012}.

Both fundamental features of the electromagnetic response, retardation
and non locality, can be described jointly by a dielectric (tensor)
function $\varepsilon_{\Lambda}\left(\mathbf{q},\omega\right)$ depending
on (circular) frequency $\omega$ and on wave vector $\mathbf{q}$.
While the dependence of the dielectric function on $\omega$ explains
the chromatic dispersion effect, and its anisotropy for $\mathbf{q}=\mathbf{0}$
describes birefringence, the dependence on wave vector $\mathbf{q}$
is directly connected to phenomena like natural optical activity and
spatial dispersion induced birefringence. Furthermore, under the influence
of a \emph{static} magnetic induction field $\mathbf{B}^{\left(0\right)}$,
respectively a \emph{static} electric field $\mathbf{E}^{\left(0\right)}$,
the additional dependence of the dielectric function on those static
fields gives rise to many other magneto-optic and electro-optic phenomena,
for instance the Faraday effect, the Pockels effect and also the Kerr
effect \citep{Agranovich1984}.

When electromagnetic waves propagate inside a dielectric material,
the microscopic local electric field convoyed by those waves exerts
a small supplemental force onto charge carriers inside the atoms (ions,
molecules) comprising that material, pulling apart inside each atom
the positively charged nucleus and the negatively charged bound electrons
by a (small) shift in \emph{opposite} direction. Of course the position
of the total mass of a charge neutral atom, considering valence electrons
and nucleus together, remains then unchanged. Since the atomic nucleus
is certainly much heavier than the bound electrons, the resulting
shift of the barycenter of the bound electrons by far surpasses the
associated tiny shift of the position of the atomic nucleus. This
is the effect of \emph{induced electronic polarization.}

A fully microscopic theory of the optical properties of a material
certainly requires to consider the positive charged atomic nuclei
and the electrons, taking into account the laws of quantum statistical
physics and low energy quantum electrodynamics, for example \citep{L.V.Keldysh2012,H.Bilz1984}.
In the ensuing discussion we shall accept a phenomenological (semiclassical)
picture of the matter light interaction based on the fundamental fact,
that matter is stable, i.e. the energy of an atom located at a site
$\mathbf{R}$ is, in the absence of external fields, a minimum against
any (small) displacement of its bound electrons. Accordingly, in reaction
to the presence of a (weak) local time dependent electric field $\mathbf{E}\left(\mathbf{R},t\right)$
the electrons bound to an atom redistribute themselves, so that considered
from outside, an atom comprising a number $Z$ of electrons, acquires
an \emph{induced} electric dipole moment. This is the basic idea of
the phenomenological\emph{ classical }model of atomic polarizability
due to Lorentz, who described the induced dipole moment of electrons,
tied to a heavy (immobile) nucleus by a harmonic spring, solving a
harmonic oscillator problem with that local electric field acting
as a drive. Incidentally, the Lorentz model predicts a frequency dependent
Fourier amplitude

\footnote{We conform to the convention, that the Fourier transformation of a
function $F(t)$ of time $t$ and its Fourier inverse $\tilde{F}\left(\omega\right)$
as a function of (circular) frequency $\omega$ are defined by
\begin{eqnarray*}
F\left(t\right) & = & \int_{-\infty}^{\infty}\frac{d\omega}{2\pi}e^{-i\omega t}\tilde{F}\left(\omega\right)\\
\tilde{F}\left(\omega\right) & = & \int_{-\infty}^{\infty}dte^{i\omega t}F\left(t\right)\:,
\end{eqnarray*}
whereas the Fourier transformation of a function $f(\mathbf{r})$
of position $\mathbf{r}$ and its Fourier inverse $\bar{f}(\mathbf{q})$
as a function of wave vector $\mathbf{q}$ are defined by: 
\begin{eqnarray*}
f\left(\mathbf{r}\right) & = & \int\frac{d^{3}q}{\left(2\pi\right)^{3}}e^{i\mathbf{q}\cdot\mathbf{r}}\bar{f}\left(\mathbf{q}\right)\\
\bar{f}\left(\mathbf{q}\right) & = & \int d^{3}re^{-i\mathbf{q}\cdot\mathbf{r}}f\left(\mathbf{r}\right)
\end{eqnarray*}
} of the induced electronic dipole moment of an atom at site $\mathbf{R}$
that coincides with a full quantum mechanical calculation, see supplemental
material \citep{Supplementary}: 
\begin{eqnarray}
\tilde{d}_{a}\left(\mathbf{R},\omega\right) & = & \sum_{a'}\alpha_{a,a'}\left(\mathbf{R},\omega\right)\tilde{E}_{a'}\left(\mathbf{R},\omega\right)\label{eq: induced dipole moment of one atom}\\
a,a' & \in & \left\{ x,y,z\right\} \nonumber 
\end{eqnarray}
Here $\alpha_{a,a'}\left(\mathbf{R},\omega\right)$ denotes the atom-individual
electric polarizability given by 
\begin{equation}
\alpha_{a,a'}\left(\mathbf{R},\omega\right)=\frac{\left|e\right|^{2}}{m}\sum_{\nu\neq0}\frac{f_{a,a'}\left(\nu;\mathbf{R}\right)}{\omega_{\nu}^{2}\left(\mathbf{R}\right)-\left(\omega+\frac{i}{2\tau_{\nu}\left(\mathbf{R}\right)}\right)^{2}}\:,\label{eq: atom polarizability}
\end{equation}
 the summation index $\nu$ running here over all eigenstates $\left|\Phi_{\nu}\left(\mathbf{R}\right)\right\rangle $
of the multi-electron configration of that atom except the groundstate
$\left|\Phi_{0}\left(\mathbf{R}\right)\right\rangle $. Indeed, this
expression looks like it was derived from an ensemble of (classical)
harmonic oscillators with resonance frequencies $\omega_{\nu}\left(\mathbf{R}\right)$,
each oscillator being driven by the local electric field with a factor
of proportionality $f_{a,a'}\left(\nu;\mathbf{R}\right)>0$, the so
called oscillator strength. However, the physical meaning of the oscillator
strength is only revealed by quantum mechanics. With $\hat{d}_{a}=-\left|e\right|\sum_{n=1}^{Z}\hat{r}_{a}^{\left(n\right)}$
denoting the dipole operator of the count $Z$ of electrons tied to
an atom at position $\mathbf{R}$, then according to quantum mechanics
\begin{eqnarray}
f_{a,a'}\left(\nu;\mathbf{R}\right) & = & \frac{1}{\left|e\right|^{2}}\frac{2m}{\hbar}\omega_{\nu}\left(\mathbf{R}\right)\left\langle \Phi_{0}\left(\mathbf{R}\right)|\hat{d}_{a}|\Phi_{\nu}\left(\mathbf{R}\right)\right\rangle \left\langle \Phi_{\nu}\left(\mathbf{R}\right)|\hat{d}_{a'}|\Phi_{0}\left(\mathbf{R}\right)\right\rangle \:,
\end{eqnarray}
explaining why the oscillator strength is a measure for how much a
bound electron contributes to the electric polarizability of an atom,
say under a transition from the multi-electron ground state $\left|\Phi_{0}\left(\mathbf{R}\right)\right\rangle $
of the atom Hamiltonian with eigenvalue $\hbar\Omega_{0}\left(\mathbf{R}\right)$
to an excited multi-electron eigenstate $\left|\Phi_{\nu}\left(\mathbf{R}\right)\right\rangle $
with eigenvalue $\hbar\Omega_{\nu}\left(\mathbf{R}\right)$. The differences
$\omega_{\nu}\left(\mathbf{R}\right)\equiv\Omega_{\nu}\left(\mathbf{R}\right)-\Omega_{0}\left(\mathbf{R}\right)$
in (\ref{eq: atom polarizability}) are the optical transition frequencies,
the life-time parameter $\tau_{\nu}\left(\mathbf{R}\right)>0$ describes
spontaneous emission as reasoned by quantum electrodynamics and thus
being always present if the atom was excited to an eigenstate state
$\nu\neq0$. Expanded details how the result (\ref{eq: atom polarizability})
can be derived within first order time dependent perturbation theory
in response to a weak time dependent electric field, including a discussion
of the $f$-sum rule, see supplemental material \citep{Supplementary}. 

De facto, the spectrum of atoms with $Z>1$ cannot be calculated ab
initio with sufficient precision, the exact taking into account of
electronic correlations being (alas) an unsolved problem. In what
follows we therefore conceive the optical transition frequencies $\omega_{\nu}\left(\mathbf{R}^{\left(j\right)}\right)$,
the life-time parameter $\tau_{\nu}\left(\mathbf{R}^{\left(j\right)}\right)$
and the oscillator strengths $f_{a,a'}\left(\nu,\mathbf{R}^{\left(j\right)}\right)$
of each atom species positioned at a site $\mathbf{R}^{\left(j\right)}=\mathbf{R}+\boldsymbol{\eta}^{\left(j\right)}$,
with $\mathbf{R}\in\Lambda$ a lattice vector and $\boldsymbol{\eta}^{\left(j\right)}$
indicating a position of an atom (ion, molecule) inside a unit cell
$C_{\Lambda}$ of the crystal $\Lambda$, as fitting parameters, so
that the optical properties as calculated from the dielectric tensor
$\varepsilon_{\Lambda}\left(\mathbf{q},\omega\right)$ of the crystal
coincide with experiment. How this objective can be accomplished,
and in particular how $\varepsilon_{\Lambda}\left(\mathbf{q},\omega\right)$
depends on the individual atom polarizabilities $\alpha_{a,a'}\left(\boldsymbol{\eta}^{\left(j\right)},\omega\right)$,
and thus via (\ref{eq: atom polarizability}) on the atom individual
multi-electron spectrum, we shall elaborate in what follows. 

A time dependent external electromagnetic signal $\mathbf{E}_{ext}\left(\mathbf{r},t\right)$
incident upon a material probe polarizes the atoms inside, and for
weak amplitude of the incident field the induced polarization at the
atom sites $\mathbf{R}$ will then be proportional to that amplitude.
However the microscopic local electric field $\mathbf{E}\left(\mathbf{r},t\right)$
polarizing an atom (ion, molecule) positioned at site $\mathbf{R}$
is not known a priory, because all atoms in the sample will get polarized
and hence all act back with a (retarded) reaction in response to the
primary external field $\mathbf{E}_{ext}\left(\mathbf{r},t\right)$.
Then everywhere inside (and outside) of the probe there holds 
\begin{equation}
\mathbf{E}\left(\mathbf{r},t\right)=\mathbf{E}_{ext}\left(\mathbf{r},t\right)+\mathbf{E}_{ind}\left(\mathbf{r},t\right),\label{eq:total electric field I}
\end{equation}
the secondary induced electric field $\mathbf{E}_{ind}\left(\mathbf{r},t\right)$
being a superposition of the individually radiated and retarded electric
fields emitted by all the atoms inside the sample, that have been
polarized in turn by that field $\mathbf{E}\left(\mathbf{r},t\right)$.
With a suitable model of the polarizability of individual atoms thus
an integral equation determining the microscopic local electric field
$\mathbf{E}\left(\mathbf{r},t\right)$ emerges.

So far, everything said is well known from the original (early) literature
\citep{Ewald1916,Ewald1916a,Ewald1917,Ewald1938,Laue1931} and from
highly cited textbooks on crystal optics, for example \citep{Born1999,Fluegge2013,Agranovich1984,Laue1960}.
For a concise summary of the pioneering works on crystal optics of
Ewald and v. Laue (and later authors) see \citep{Authier2012}. Nevertheless,
we believe the approach we present in what follows truely discerns
from traditional presentations of the subject. For example, nowhere
do we make use of the Ewald-Oseen extinction theorem to recover the
correct index of refraction $n$ of a material, it being customary
practice in the so called rigorous theory of dispersion \citep{Born1999,Fluegge2013}
to consider the wave incident from free space to be propagated with
vacuum light velocity $c$ and the signal induced in the sample to
be propagated with velocity $c/n$, the Lorentz-Lorenz formula connecting
the index of refraction $n$ with the polarizability of individual
atoms thus emerging as a solvability condition for the field-integral
equation stated (implicitely) in (\ref{eq:total electric field I}).

\subsection*{Outline}

In Sec. II we first establish on the basis of the fundamental Maxwell
equations an exact\emph{ integral equation} for the microscopic local
electric field $\mathbf{E}\left(\mathbf{r},t\right)$ in (\ref{eq:total electric field I})
with an explicit formula for the integral kernel that derives directly
from the atom polarizability (\ref{eq: atom polarizability}). Posing
boundary conditions for the components of the electromagnetic field
at the boundary of a material probe is then redundant. Moreover, the
frequency dependence of longitudinal and transversal parts of the
electromagnetic field are treated consistently, thus making everywhere
in our calculations the correct static limit $\omega\rightarrow0$
accessible.

In Sec. III we solve the field-integral equation for crystalline dielectric
materials exactly making use of a set of non-standard Bloch functions,
not constructed from plane waves but designed from eigenfunctions
of the position operator, representing a \emph{complete} orthonormal
system of eigenfunctions of the translation operator $T_{\mathbf{R}}$
under a shift by a lattice vector $\mathbf{R}\in\Lambda$. Accordingly,
instead of expanding the kernel of the field-integral equation in
the well known basis of plane waves $e^{i\left(\mathbf{q}+\mathbf{G}\right)\mathbf{r}}$,
thus requiring to handle for each wave vector $\mathbf{q}$ in the
Brillouin zone $C_{\Lambda^{-1}}$ of a lattice $\Lambda$ then (infinite)
matrices labelled by reciprocal lattice vectors $\mathbf{G},\mathbf{G}'\in\Lambda^{-1}$,
our choice of eigenfunctions of the translation operator $T_{\mathbf{R}}$
sidesteps the inversion (and truncation) of such large matrices, thus
easing notably the determination of the photonic bandstructure of
a crystal. Also we show, if the incident electromagnetic wave was
purely transversal, yet the \emph{microscopic} local electric field
features both, a transversal and a longitudinal component, see Fig.(\ref{fig:local_field_projections}),
the strength of the longitudinal component being strongly dependent
on the density of polarizable atoms (ions, molecules) in the crystal.

Thereafter we present, exemplifying our calculation method, results
for the photonic bandstructure of diamond ($M=2$), that have been
calculated previously with other (computationally more time-consuming)
methods. Our findings for the photonic bandstructures for various
monoatomic Bravais lattices ($M=1$) we present and discuss in the
supplemental material \citep{Supplementary}. In comparison to well
known (phenomenological) work on photonic bandstructures \citep{Leung1990,Zhang1990},
within which the (macroscopic) Maxwell's equations in a superlattice
are solved assuming a spatially repetetive varying index of refraction,
it turns out that in our approach based on the field-integral equations
the need to eliminate unphysical ``longitudinal'' modes doesn't
arise. 

In section \ref{sec:Macroscopic-Electric-Field-and-dielectric-function}
we introduce the notion of a \emph{macroscopic} electric field $\mathcal{\boldsymbol{E}}\left(\mathbf{r},t\right)$
inside a material, conceiving it with regard to spatial variations
as a low pass filtered signal, with the solution to the field-integral
equation (\ref{eq:local field integral equation}) as input. Relating
the macroscopic polarization to that macroscopic electric field, thereafter
the macroscopic dielectric $3\times3-$tensor $\varepsilon_{\Lambda}\left(\mathbf{q},\omega\right)$
of a crystalline dielectric material emerges, with chromatic and spatial
dispersion fully taken into account. The exact expression for $\varepsilon_{\Lambda}\left(\mathbf{q},\omega\right)$
thus obtained is solely dependent on the symmetry of the lattice $\varLambda$
under consideration and on the polarizability $\alpha\left(\mathbf{R}^{\left(j\right)},\omega\right)$
of individual atoms (ions, molecules) positioned at their equilibrium
sites $\mathbf{R}^{\left(j\right)}$ in the crystal. We also confirm,
exemplarily for the ionic crystal $CsI$, that our formula for the
dielectric tensor with regard to its frequency dependence indeed obeys
to the Lyddane-Sachs-Teller relation. With a view to the key role
of locality claimed by \emph{macroscopic electrodynamics} \citep{Liu2009}
we caution the reader not to discard the $\mathbf{q}$-dependence
of the dielectric tensor $\varepsilon_{\Lambda}\left(\mathbf{q},\omega\right)$.
As emphasized by Ginzburg and Agranovich \citep{Agranovich1984},
a rich variety of optical effects, including rotary power and spatial
dispersion induced birefringence, manifestly proof the importance
of the non local nature of the dielectric response.

We next derive directly from the microscopic field-integral equations
a set of \emph{differential equations} that determine the spatial
variation of the\emph{ transversal }and\emph{ longitudinal }parts
of the \emph{macroscopic} electric field $\mathcal{\boldsymbol{E}}\left(\mathbf{r},t\right)$,
without prior knowledge of the microscopic local field $\mathbf{E}\left(\mathbf{r},t\right)$. 

If the external field was purely transversal, this set of coupled
differential equations reduces to a wave equation determining the
radiative part of the macroscopic field. In this way the parts of
the dielectric tensor $\varepsilon_{\Lambda}\left(\mathbf{q},\omega\right)$
are indentified that determine the \emph{transversal} dielectric tensor
$\varepsilon^{\left(T\right)}\left(\mathbf{q},\omega\right)$ comprising
the optical properties of a dielectric crystal. Assuming spatial dispersion
is a weak effect, which assumption applies for many transparent media,
the Taylor expansion of $\varepsilon_{ab}^{\left(T\right)}\left(\mathbf{q},\omega\right)$
around $\mathbf{q}=\mathbf{0}$ then provides (yet in an implicit
manner) access to various optical phenomena featuring the propagation
of light in dielectric crystals \citep{Agranovich1984}, for instance
chromatic dispersion and birefringence, rotary power (natural optical
activity) and also the (weak) effects of a spatial-dispersion-induced
birefringence, the latter being a critical problem for the design
of lens elements made from crystalline materials like $CaF_{2}$ and
$BaF_{2}$ widely used in optical lithograpy systems in the ultraviolet
\citep{Burnett2002,Serebryakov2003}. Further we summarize in section
\ref{sec:chromatic-dispersion,-optical activity and spatial dispersion induced birefringence},
see Table \ref{results}, Fig.\ref{Sellmeier} and also Fig.\ref{fig:CaF2_and_BaF2},
to what large extend our theory of the dielectric tensor for crystalline
dielectrics agrees with measurements over a wide range of optical
frequencies for a series of well known crystalline materials, including
for example Bi$_{12}$TiO$_{20}$ and also Bi$_{12}$SiO$_{20}$,
both crystals featuring a large number of basis atoms ($M=66$) in
the unit cell, thus demonstrating the utility of our approach. 

If on the other hand the external field was purely longitudinal, the
differential equations derived from the field-integral equations reduce
to a Poisson type equation for a scalar potential function determining
the macroscopic electric field, thus identifying the parts of the
dielectric tensor $\varepsilon_{\Lambda}\left(\mathbf{q},\omega\right)$
featuring electric-field \emph{screening, }like in electrostatics.
Also we deduce an exact analytic formula for the \emph{static} dielectric
tensor $\varepsilon_{\Lambda}$, that conforms with general thermodynamic
stability criteria \citep{Kirzhnitz2012} and applies for all $14$
monoatomic Bravais lattices ($M=1$). For the special case of cubic
symmetry the long-known Clausius-Mossotti relation is recovered.

\section{\label{sec:local field}The Field-Integral Equations}

Consider a \emph{fixed} inertial frame, with a polarizable dielectric
material at rest occupying a volume $\Omega$ in that frame. Without
loss of generality let the microscopic charge density inside $\Omega$
be given the representation
\begin{equation}
\rho\left(\mathbf{r},t\right)=\boldsymbol{-\nabla}\cdot\mathbf{P}\left(\mathbf{r},t\right),\label{eq: representation of charge density}
\end{equation}
with $\mathbf{P}\left(\mathbf{r},t\right)$ denoting the vector field
of electric polarization \citep{L.V.Keldysh2012}. We find it then
convenient to decompose the associated microscopic current density
\begin{equation}
\mathbf{j}\left(\mathbf{r},t\right)=\partial_{t}\mathbf{P}\left(\mathbf{r},t\right)\label{eq: representation of current density}
\end{equation}
 flowing inside $\Omega$ into \emph{longitudinal} and \emph{transversal}
parts. To serve this purpose we introduce integral kernels 
\begin{eqnarray}
\Pi_{aa'}^{\left(L\right)}\left(\mathbf{r}-\mathbf{r}'\right) & = & \lim_{\kappa\rightarrow\infty}\int\frac{d^{3}q}{\left(2\pi\right)^{3}}e^{i\mathbf{q}\cdot\mathbf{\left(\mathbf{r}-\mathbf{r}'\right)}}\bar{\Pi}_{aa'}^{\left(L\right)}\left(\mathbf{q}\right)\frac{1}{1+\frac{\left|\mathbf{q}\right|^{2}}{\kappa^{2}}}\label{eq: projection operators I}\\
\Pi_{aa'}^{\left(T\right)}\left(\mathbf{r}-\mathbf{r}'\right) & = & \lim_{\kappa\rightarrow\infty}\int\frac{d^{3}q}{\left(2\pi\right)^{3}}e^{i\mathbf{q}\cdot\mathbf{\left(\mathbf{r}-\mathbf{r}'\right)}}\bar{\Pi}_{aa'}^{\left(T\right)}\left(\mathbf{q}\right)\frac{1}{1+\frac{\left|\mathbf{q}\right|^{2}}{\kappa^{2}}}\nonumber 
\end{eqnarray}
with labels $a,a',a''\in\left\{ x,y,z\right\} $ specifying Cartesian
components, and 
\begin{eqnarray}
\bar{\Pi}_{aa'}^{\left(L\right)}\left(\mathbf{q}\right) & = & \frac{q_{a}q_{a'}}{\left|\mathbf{q}\right|^{2}}\;,\label{eq:Fourier transform projection operators}\\
\bar{\Pi}_{aa'}^{\left(T\right)}\left(\mathbf{q}\right) & = & \delta_{a,a'}-\frac{q_{a}q_{a'}}{\left|\mathbf{q}\right|^{2}}\;.\nonumber 
\end{eqnarray}
Denoting the convolution of two kernels $A_{aa'}\left(\mathbf{r},\mathbf{r}'\right)$
and $B_{aa'}\left(\mathbf{r},\mathbf{r}'\right)$ with the symbol
\begin{equation}
\left[A\circ B\right]_{aa'}\left(\mathbf{r},\mathbf{r}'\right)=\int_{\Omega_{P}}d^{3}r''\sum_{a''}A_{aa''}\left(\mathbf{r},\mathbf{r}''\right)B_{a''a'}\left(\mathbf{r}'',\mathbf{r}'\right),\label{eq:kompact notation convolution}
\end{equation}
and writing for the kernel representing unity 
\begin{equation}
\left[I\right]_{aa'}\left(\mathbf{r},\mathbf{r}'\right)=\delta_{aa'}\delta^{\left(3\right)}\left(\mathbf{r}-\mathbf{r}'\right),\label{eq: unity kernel}
\end{equation}
 then the validity of all the relations distinctive for projection
operators are readily confirmed:
\begin{eqnarray}
\Pi^{\left(L\right)}+\Pi^{\left(T\right)} & = & I\label{eq:general relations projection operator}\\
\Pi^{\left(L\right)}\circ\Pi^{\left(L\right)} & = & \Pi^{\left(L\right)}\nonumber \\
\Pi^{\left(T\right)}\circ\Pi^{\left(T\right)} & = & \Pi^{\left(T\right)}\nonumber \\
\Pi^{\left(T\right)}\circ\Pi^{\left(L\right)} & =\:0\:= & \Pi^{\left(L\right)}\circ\Pi^{\left(T\right)}\nonumber 
\end{eqnarray}
 In position space, for $\left|\mathbf{\mathbf{r}}\right|\neq0$,
the kernel of the longitudinal and the transverse projection operator
both correspond to a dipole field, see \citep{ClaudeCohen-Tannoudji1989}
for a concise derivation: 
\begin{eqnarray}
\Pi_{aa'}^{\left(T\right)}\left(\mathbf{r}\right) & = & \frac{2}{3}\delta_{a,a'}\delta^{\left(3\right)}\left(\mathbf{\mathbf{r}}\right)+\Theta\left(\left|\mathbf{\mathbf{r}}\right|-0^{+}\right)\nabla_{a}\nabla_{a'}\left(\frac{1}{4\pi\left|\mathbf{r}\right|}\right)\label{eq:projection operators II}\\
\Pi_{aa'}^{\left(L\right)}\left(\mathbf{r}\right) & = & \frac{1}{3}\delta_{a,a'}\delta^{\left(3\right)}\left(\mathbf{\mathbf{r}}\right)-\Theta\left(\left|\mathbf{\mathbf{r}}\right|-0^{+}\right)\nabla_{a}\nabla_{a'}\left(\frac{1}{4\pi\left|\mathbf{r}\right|}\right)\nonumber 
\end{eqnarray}
 It follows from what has been said that the longitudinal part (L)
, respectively the transversal part (T) of the current distribution
$j_{a}\left(\mathbf{r},t\right)$ conforms with the convolution integrals
\begin{equation}
j_{a}^{(L,T)}(\mathbf{r},t)=\int d^{3}r'\sum_{a'}\Pi_{aa'}^{\left(L,T\right)}\left(\mathbf{r}-\mathbf{r}'\right)j_{a'}(\mathbf{r}',t).\label{eq:longitudinal-transverse part of current density}
\end{equation}
It should be underlined, this link between the original vector field
$\mathbf{j}(\mathbf{r},t)$ and the associated longitudinal (transversal)
part $\mathbf{j}^{\left(L,T\right)}(\mathbf{r},t)$ is \emph{non local}.
There holds by construction
\begin{eqnarray}
\mathbf{j}(\mathbf{r},t) & = & \mathbf{j}^{(L)}(\mathbf{r},t)+\mathbf{j}^{(T)}(\mathbf{r},t)\label{eq: Helmholtz decomposition current density}\\
\boldsymbol{\nabla}\mathbf{\cdot j}^{(T)}(\mathbf{r},t) & = & 0\nonumber \\
\mathbf{\boldsymbol{\nabla}}\wedge\mathbf{j}^{(L)}(\mathbf{r},t) & = & \mathbf{0}.\nonumber 
\end{eqnarray}
According to the Helmholtz-Hodge theorem such a decomposition of a
vector field $\mathbf{j}\left(\mathbf{r},t\right)$ is unique. A recent
thourough discussion and compilation of the literature on the subject
can be found in \citep{Bhatia2013}. 

Because Maxwell's equations are linear, particular monochromatic solutions
\begin{eqnarray}
\mathbf{E}^{(L,T)}\left(\mathbf{r},t\right) & = & \mathbf{\tilde{E}}^{(L,T)}\left(\mathbf{r},\omega\right)e^{-i\omega t}\label{eq: time harmonic fields}\\
\mathbf{B}\left(\mathbf{r},t\right) & = & \mathbf{\tilde{B}}\left(\mathbf{r},\omega\right)e^{-i\omega t},\nonumber 
\end{eqnarray}
with Fourier amplitudes $\mathbf{\tilde{E}}^{(L,T)}\left(\mathbf{r},\omega\right)$
and $\mathbf{\tilde{B}}\left(\mathbf{r},\omega\right)$, can be superposed
to construct any wanted time dependence. Applying next the projection
operators $\Pi_{aa'}^{\left(L,T\right)}$ to the in time Fourier transformed
Maxwell equations, there emerge two groups of decoupled equations
in the space-frequency domain for the respective Fourier amplitudes.
The first group relates to the components $\tilde{E}_{a}^{\left(L\right)}(\mathbf{r},\omega)$
of the Fourier amplitudes of the longitudinal electric field, 
\begin{eqnarray}
\boldsymbol{\nabla}\mathbf{\cdot\tilde{E}}^{(L)}(\mathbf{r},\omega) & = & \frac{1}{\varepsilon_{0}}\tilde{\varrho}\left(\mathbf{r},\omega\right)\label{eq: longitudinal Maxwell equations}\\
-i\omega\varepsilon_{0}\tilde{E}_{a}^{\left(L\right)}(\mathbf{r},\omega)+\tilde{j}_{a}^{(L)}(\mathbf{r},\omega) & = & 0\;,\nonumber 
\end{eqnarray}
the second group involves the Cartesian components of the Fourier
amplitudes of the transversal electromagnetic field, which obey to
the following six inhomogenous \emph{scalar} Helmholtz equations,
the respective source terms being provided by the Fourier amplitudes
of the transversal current density, with $c=\frac{1}{\sqrt{\varepsilon_{0}\mu_{0}}}$
the speed of light in free space: 
\begin{eqnarray}
\left(-\nabla^{2}-\frac{\omega^{2}}{c^{2}}\right)\tilde{E}_{a}^{(T)}\left(\mathbf{r},\omega\right) & = & \mu_{0}i\omega\tilde{j}_{a}^{(T)}\left(\mathbf{r},\omega\right)\label{eq: transversal Maxwell equations}\\
\left(-\nabla^{2}-\frac{\omega^{2}}{c^{2}}\right)\tilde{B}_{a}\left(\mathbf{r},\omega\right) & = & \mu_{0}\left[\mathbf{\boldsymbol{\nabla}}\wedge\tilde{\mathbf{j}}(\mathbf{r},\omega)\right]_{a}\nonumber 
\end{eqnarray}

\begin{figure}
\includegraphics[scale=0.78]{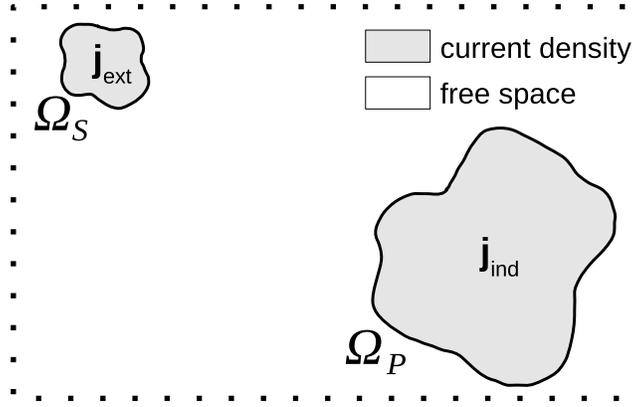} \caption{\label{fig:SourceRegions}Schematic illustration of a source domain
$\Omega_{\text{S}}$ and a probe volume $\Omega_{\text{P}}$ indicating
non vanishing current densities inside. }
\end{figure}

We are interested to solve (\ref{eq: longitudinal Maxwell equations})
and (\ref{eq: transversal Maxwell equations}) considering now a geometry
consisting of two \emph{disjoint} material bodies at rest, see Fig.\ref{fig:SourceRegions},
i.e. $\Omega=\Omega_{S}\cup\Omega_{P}$ and $\Omega_{S}\cap\Omega_{P}=\emptyset$.

Let us refer to the body $\Omega_{S}$ as the source, and to the body
$\Omega_{P}$ as the probe. Accordingly, we split the current density
$\tilde{j}_{a}\left(\mathbf{r},\omega\right)$ into an \emph{externally}
controlled partial current $\tilde{j}_{ext,a}\left(\mathbf{r},\omega\right)$
flowing solely inside $\Omega_{S}$, and into an \emph{induced} partial
current $\tilde{j}_{ind,a}\left(\mathbf{r},\omega\right)$ flowing
solely inside the probe $\Omega_{P}$:
\begin{eqnarray}
\tilde{j}_{a}\left(\mathbf{r},\omega\right) & = & \begin{cases}
\tilde{j}_{ext,a}\left(\mathbf{r},\omega\right) & \textrm{for}\:\mathbf{r}\in\Omega_{S}\\
\tilde{j}_{ind,a}\left(\mathbf{r},\omega\right) & \textrm{for}\:\mathbf{r}\in\Omega_{P}\\
0 & \textrm{for}\:\mathbf{r}\notin\Omega_{S}\cup\Omega_{P}
\end{cases}\label{eq:decomposition of current into external+induced part}
\end{eqnarray}
 If the source $\Omega_{S}$ was a cavity producing a laser beam,
and if the surface of the probe $\Omega_{P}$ would be reflecting
(parts) of the radiation incident back into that cavity, certainly
there would exist a backaction from the probe to the source, leading
then to the existence of an induced partial current flowing also inside
the source region $\Omega_{S}$. We preclude here and in the following
any such backaction effects. Provided transfer of charge between $\Omega_{S}$
and $\Omega_{P}$ is prohibited, there holds charge conservation separately
(!) inside $\Omega_{S}$ and inside $\Omega_{P}$. Consequently, inside
the domain $\Omega_{S}$ there holds
\begin{equation}
i\omega\tilde{\rho}_{ext}(\mathbf{r},\omega)=\boldsymbol{\nabla}\cdot\mathbf{\tilde{j}}_{ext}(\mathbf{r},\omega).\label{eq: continuity equation}
\end{equation}
Of course $\tilde{\rho}_{ext}(\mathbf{r},\omega)\equiv0$ and $\tilde{j}_{ext,a}\left(\mathbf{r},\omega\right)\equiv0$
for $\mathbf{r}\notin\Omega_{S}$.

It follows from (\ref{eq: longitudinal Maxwell equations}) that the
longitudinal electric field $\tilde{E}_{a}^{\left(L\right)}(\mathbf{r},\omega)$
is already determined by the longitudinal part of the current distribution
(\ref{eq:decomposition of current into external+induced part}), a
seemingly simple result. Indeed from (\ref{eq: longitudinal Maxwell equations})
and (\ref{eq:decomposition of current into external+induced part})
then 
\begin{equation}
\tilde{E}_{a}^{(L)}\left(\mathbf{r},\omega\right)=\tilde{E}_{ext,a}^{(L)}\left(\mathbf{r},\omega\right)+\frac{1}{\epsilon_{0}}\frac{1}{i\omega}\tilde{j}_{ind,a}^{\left(L\right)}\left(\mathbf{r},\omega\right),\label{eq:longitudinal electric field}
\end{equation}
where the \emph{external} longitudinal field
\begin{equation}
\tilde{E}_{ext,a}^{(L)}\left(\mathbf{r},\omega\right)=\frac{1}{\epsilon_{0}}\frac{1}{i\omega}\tilde{j}_{ext,a}^{\left(L\right)}\left(\mathbf{r},\omega\right)\equiv-\frac{\partial}{\partial r_{a}}\tilde{\phi}_{ext}(\mathbf{r},\omega)\label{eq:external longitudinal electric field}
\end{equation}
is derived from a scalar potential
\begin{equation}
\tilde{\phi}_{ext}(\mathbf{r},\omega)=\int_{\Omega_{S}}d^{3}r'\frac{\tilde{\rho}_{ext}(\mathbf{r}',\omega)}{4\pi\varepsilon_{0}\left|\mathbf{r}-\mathbf{r}'\right|},\label{eq:external scalr potential}
\end{equation}
like in electrostatics. Even though $\tilde{j}_{ext,a}(\mathbf{r},\omega)$
was restricted to be non vanishing solely inside the source domain
$\Omega_{S}$, its longitudinally projected part $\tilde{j}_{ext,a}^{\left(L\right)}\left(\mathbf{r},\omega\right)$
also exists outside of $\Omega_{S}$, the non locality of that projection
thus becoming manifest. 

The electromagnetic fields radiated by the transversal current density
$\tilde{j}_{a}^{(T)}\left(\mathbf{r},\omega\right)$ in (\ref{eq: transversal Maxwell equations})
can be readily determined introducing the \emph{retarded} Green's
function
\begin{eqnarray}
\tilde{g}\left(\mathbf{r}-\mathbf{r}',\omega\right) & = & \frac{\exp\left(i\frac{\omega}{c}\left|\mathbf{r}-\mathbf{r}'\right|\right)}{4\pi\left|\mathbf{r}-\mathbf{r}'\right|}\label{eq: Helmholtz propagator}\\
\mathtt{Im}\left(\omega\right) & \rightarrow & 0^{+},\nonumber 
\end{eqnarray}
with $\tilde{g}\left(\mathbf{r}-\mathbf{r}',\omega\right)$ the solution
of the three-dimensional inhomogeneous \emph{scalar} Helmholtz equation
in free space $\mathbb{R}^{3}$ with a point source (of strength unity)
at position $\mathbf{r}'$: 
\begin{equation}
\left(-\nabla^{2}-\frac{\omega^{2}}{c^{2}}\right)\tilde{g}\left(\mathbf{r}-\mathbf{r}',\omega\right)=\delta^{\left(3\right)}\left(\mathbf{r}-\mathbf{r}'\right)\label{eq: Helmholtz propagator II}
\end{equation}
As the distance $\left|\mathbf{r}-\mathbf{r}'\right|$ to the source
position at $\mathbf{r}'$ increases then $\tilde{g}\left(\mathbf{r}-\mathbf{r}',\omega\right)$
behaves like an \emph{outgoing} spherical wave. 

Based on Green's identity applied to the domain $\Omega=\Omega_{S}\cup\Omega_{P}$,
and again applied to the complementary domain $\mathbb{R}^{3}\setminus\Omega$,
there follow directly from (\ref{eq: transversal Maxwell equations})
and (\ref{eq: Helmholtz propagator II}) for all points $\mathbf{r}\notin\Omega_{S}$
the integral representations
\begin{equation}
\tilde{E}_{a}^{(T)}\left(\mathbf{r},\omega\right)=\tilde{E}_{ext,a}^{(T)}\left(\mathbf{r},\omega\right)+i\omega\mu_{0}\int d^{3}r'\tilde{g}\left(\mathbf{r}-\mathbf{r}',\omega\right)\tilde{j}_{ind,a}^{(T)}\left(\mathbf{r}',\omega\right)\label{eq:Integral representation transversal electric field}
\end{equation}
and
\begin{equation}
\tilde{B}_{a}\left(\mathbf{r},\omega\right)=\tilde{B}_{ext,a}\left(\mathbf{r},\omega\right)+\mu_{0}\int d^{3}r'\tilde{g}\left(\mathbf{r}-\mathbf{r}',\omega\right)\left[\mathbf{\boldsymbol{\nabla}'}\wedge\tilde{\mathbf{j}}_{ind}(\mathbf{r}',\omega)\right]_{a}.\label{eq:Integral equation magnetic induction field}
\end{equation}
For details of the derivation of (\ref{eq:Integral representation transversal electric field})
and (\ref{eq:Integral equation magnetic induction field}), see supplemental
material \citep{Supplementary}.

According to (\ref{eq: representation of current density}) the Fourier
amplitude of the \emph{induced} current density flowing inside the
probe volume $\Omega_{P}$ is directly proportional to the Fourier
amplitude of the microscopic electric polarization: 
\begin{equation}
\tilde{j}_{ind,a}\left(\mathbf{r},\omega\right)=-i\omega\tilde{P}_{a}\left(\mathbf{r},\omega\right)\label{eq: connection induced current - electric polarizabilty}
\end{equation}
Combining now the respective longitudinal and transverse parts by
adding the integral representations (\ref{eq:longitudinal electric field})
and (\ref{eq:Integral representation transversal electric field})
there follows 
\begin{equation}
\tilde{E}_{a}\left(\mathbf{r},\omega\right)=\tilde{E}_{ext,a}\left(\mathbf{r},\omega\right)+\frac{1}{\epsilon_{0}}\int d^{3}r'\sum_{a'}\mathcal{\tilde{G}}_{aa'}(\mathbf{r}-\mathbf{r}',\omega)\tilde{P}_{a'}\left(\mathbf{r}',\omega\right),\label{eq:local field in terms of polarization}
\end{equation}
with
\begin{eqnarray}
\mathcal{\tilde{G}}_{aa'}(\mathbf{r}-\mathbf{r}',\omega) & = & \frac{\omega^{2}}{c^{2}}\int d^{3}s\:\tilde{g}\left(\mathbf{r}-\mathbf{s},\omega\right)\Pi_{aa'}^{\left(T\right)}\left(\mathbf{s}-\mathbf{r}'\right)-\Pi_{aa'}^{\left(L\right)}\left(\mathbf{r}-\mathbf{r}'\right)\label{eq:3x3 matrix propagator}\\
 & = & \lim_{\kappa\rightarrow\infty}\int\frac{d^{3}q}{\left(2\pi\right)^{3}}\frac{e^{i\mathbf{q}\cdot\left(\mathbf{r}-\mathbf{r}'\right)}}{1+\frac{\left|\mathbf{q}\right|^{2}}{\kappa^{2}}}\left[\frac{\omega^{2}}{c^{2}}\frac{\delta_{a,a'}-\frac{q_{a}q_{a'}}{\left|\mathbf{q}\right|^{2}}}{\left|\mathbf{q}\right|^{2}-\frac{\omega^{2}}{c^{2}}-i0^{+}}-\frac{q_{a}q_{a'}}{\left|\mathbf{q}\right|^{2}}\right]\nonumber 
\end{eqnarray}
denoting the electromagnetic kernel. The Fourier transformed kernel
$\mathcal{\tilde{G}}_{aa'}(\mathbf{r},\omega)$ in the wave vector
domain we denote as 
\begin{eqnarray}
\mathcal{\bar{G}}_{aa'}(\mathbf{q},\omega) & = & \int d^{3}re^{-i\mathbf{q}\cdot\mathbf{r}}\mathcal{\tilde{G}}_{aa'}(\mathbf{r},\omega)\label{eq: Fourier transform of 3x3 matrix propagator}\\
 & = & \frac{\omega^{2}}{c^{2}}\frac{\delta_{a,a'}-\frac{q_{a}q_{a'}}{\left|\mathbf{q}\right|^{2}}}{\left|\mathbf{q}\right|^{2}-\frac{\omega^{2}}{c^{2}}-i0^{+}}-\frac{q_{a}q_{a'}}{\left|\mathbf{q}\right|^{2}}\nonumber \\
 & = & \frac{\frac{\omega^{2}}{c^{2}}\delta_{a,a'}-q_{a}q_{a'}}{\left|\mathbf{q}\right|^{2}-\frac{\omega^{2}}{c^{2}}-i0^{+}}\:.\nonumber 
\end{eqnarray}

Assuming a small amplitude of the perturbing external field $\tilde{E}_{ext,a}\left(\mathbf{r},\omega\right)$
there results inside the probe at a position $\mathbf{r}\in\Omega_{P}$
the (total) microscopic polarization 
\begin{equation}
\tilde{P}_{a}\left(\mathbf{r},\omega\right)=\varepsilon_{0}\int_{\Omega_{P}}d^{3}r'\sum_{a'}\tilde{\chi}_{ext,aa'}\left(\mathbf{r},\mathbf{r}',\omega\right)\tilde{E}_{ext,a'}\left(\mathbf{r}',\omega\right)\:.\label{eq:polarization vs external field}
\end{equation}
Within a fully microscopic approach the response kernel $\tilde{\chi}_{ext,aa'}\left(\mathbf{r},\mathbf{r}',\omega\right)$
is calculated from Kubo's formula in reaction to the presence of the
\emph{external} field $\tilde{E}_{ext,a'}\left(\mathbf{r}',\omega\right).$
However, what we are really interested in here, is not the response
kernel $\tilde{\chi}_{ext,aa'}\left(\mathbf{r},\mathbf{r}',\omega\right)$
connecting the polarization $\tilde{P}_{a}\left(\mathbf{r},\omega\right)$
inside the probe with the \emph{external} field $\tilde{E}_{ext,a'}\left(\mathbf{r}',\omega\right)$,
but the dielectric susceptibility kernel $\tilde{\chi}_{aa'}\left(\mathbf{r},\mathbf{r}',\omega\right)$
connecting $\tilde{P}_{a}\left(\mathbf{r},\omega\right)$ with the
microscopic\emph{ local} electric field $\tilde{E}_{a'}\left(\mathbf{r}',\omega\right)$
that acts on each atom (ion, molecule): 
\begin{equation}
\tilde{P}_{a}\left(\mathbf{r},\omega\right)=\varepsilon_{0}\int_{\Omega_{P}}d^{3}r'\sum_{a'}\tilde{\chi}_{aa'}\left(\mathbf{r},\mathbf{r}',\omega\right)\tilde{E}_{a'}\left(\mathbf{r}',\omega\right)\label{eq:polarization vs local field}
\end{equation}
As emphasized by Keldysh \citep{L.V.Keldysh2012}, there is no need
to consider dielectric and magnetic susceptibilities separately, the
latter being already incorporated in the \emph{non locality} of the
dielectric kernel. 

The fundamental field-integral equation determining the microscopic
\emph{local} electric field is then given by an inhomogenous integral
equation comprising the dielectric kernel $\tilde{\chi}$: 
\begin{equation}
\tilde{E}_{a}\left(\mathbf{r},\omega\right)=\tilde{E}_{ext,a}\left(\mathbf{r},\omega\right)+\int_{\Omega_{P}}d^{3}r'\sum_{a'}\left[\mathcal{\tilde{G}}\circ\tilde{\chi}\right]_{aa'}\left(\mathbf{r},\mathbf{r}',\omega\right)\tilde{E}_{a'}\left(\mathbf{r}',\omega\right)\label{eq:local field integral equation}
\end{equation}
The link between the dielectric susceptibility kernel $\tilde{\chi}$
and the response kernel $\tilde{\chi}_{ext}$ is readily identified
\citep{L.V.Keldysh2012}, combining (\ref{eq:polarization vs external field})
with (\ref{eq:polarization vs local field}): 
\begin{equation}
\tilde{\chi}=\tilde{\chi}_{ext}\circ\left[I+\mathcal{\tilde{G}}\circ\tilde{\chi}_{ext}\right]^{-1}=\left[I+\tilde{\chi}_{ext}\circ\mathcal{\tilde{G}}\right]^{-1}\circ\tilde{\chi}_{ext}\label{eq: connection dielectric kernel to response kernel}
\end{equation}
For a more detailed explanation, see supplemental material \citep{Supplementary}.
So, if $\tilde{\chi}_{ext}$ was known, say from a full microscopic
calculation with Kubo's formula, then $\tilde{\chi}$ follows from
(\ref{eq: connection dielectric kernel to response kernel}). In case
the probe volume $\Omega_{P}$ was of finite size, solving the integral
equation (\ref{eq: connection dielectric kernel to response kernel})
for the microscopic dielectric susceptibility kernel $\tilde{\chi}$
poses a formidable (numerical) problem, the non-locality radius of
$\tilde{\chi}_{ext}$ being substantially enhanced up to the macroscopic
scale by the presence of a boundary \citep{L.V.Keldysh2012}. In the
next section III we circumvent this problem and introduce a phenomenological
model for $\tilde{\chi}$ that proves a posteriori to be appropriate
to describe in detail the propagation of light in dielectric crystals.

\section{Microscopic Local Electric Field in Crystalline Dielectrics\label{sec:Microscopic-Local-Electric-Field-in crystalline-dielectrics}}

Crystalline order assumes each equilibrium position of atoms (ions,
molecules) inside a material corresponds to a site vector $\mathbf{R}^{\left(j\right)}=\mathbf{R}+\boldsymbol{\eta}^{\left(j\right)}$,
where $\mathbf{R}$ denotes a lattice vector in a Bravais lattice
$\Lambda$, and $\boldsymbol{\eta}^{\left(j\right)}$ with $j\in\left\{ 1,2,...,M\right\} $
is a set of position vectors indicating equilibrium positions of the
atoms (molecules, ions) inside the unit cell $C_{\Lambda}$. The entire
probe volume $\Omega_{P}$ can be thought of to be filled translating
a number $\left|\Lambda_{P}\right|$ of Wigner-Seitz unit cells $C_{\Lambda}$
by lattice vectors $\mathbf{R}\in\Lambda$. This subset of all such
Bravais lattice vectors $\mathbf{R}\in\Lambda$ we denote as $\Lambda_{P}\subset\Lambda$.
Under the action of the external electric field $\tilde{E}_{ext,a'}\left(\mathbf{r}',\omega\right)$
each atom (ion, molecule) gets polarized proportional to the \emph{local
}electric field $\tilde{E}_{a}\left(\mathbf{R}^{\left(j\right)},\omega\right)$
acting at position $\mathbf{R}^{\left(j\right)}$ of that atom. Then
a simple \emph{phenomenological} model for the microscopic dielectric
susceptibility kernel in crystalline insulators emerges assembling
first all individual atom contributions located at positions $\boldsymbol{\eta}^{\left(j\right)}$
inside the unit cell positioned around the origin $\mathbf{R}=\mathbf{0}$,
and then sum over all such cells filling the probe volume $\Omega_{P}$
of the crystal lattice $\Lambda$ under consideration: 
\begin{eqnarray}
\tilde{\chi}_{aa'}\left(\mathbf{r},\mathbf{r}',\omega\right) & = & \frac{1}{\varepsilon_{0}}\sum_{\mathbf{R}\in\Lambda_{P}}\:\sum_{1\leq j,j'\leq M}\alpha_{aa'}\left(\boldsymbol{\eta}^{\left(j\right)},\boldsymbol{\eta}^{\left(j'\right)},\omega\right)\delta^{\left(3\right)}\left(\mathbf{r}-\mathbf{R}-\mathbf{\boldsymbol{\eta}}^{\left(j\right)}\right)\delta^{\left(3\right)}\left(\mathbf{r}'-\mathbf{R}-\boldsymbol{\eta}^{\left(j'\right)}\right)\label{eq: phenomenological dielectric kernel}\\
\nonumber \\
\alpha_{aa'}\left(\boldsymbol{\eta}^{\left(j\right)},\boldsymbol{\eta}^{\left(j'\right)},\omega\right) & = & \delta_{j,j'}\alpha_{aa'}^{\left(I\right)}\left(\boldsymbol{\eta}^{\left(j\right)},\omega\right)+\left(1-\delta_{j,j'}\right)\alpha_{aa'}^{\left(II\right)}\left(\boldsymbol{\eta}^{\left(j\right)},\boldsymbol{\eta}^{\left(j'\right)},\omega\right)\label{eq: model for microscopic polarizability}
\end{eqnarray}
Essentially, this susceptibility kernel describes a lattice periodic
arrangement of point dipoles \citep{Laue1931,Ewald1938,Laue1960}.
The diagonal terms $j=j'$ refer to the afore mentioned effects of
induced electronic polarization of \emph{single} atoms (ions, molecules)
at the positions $\boldsymbol{\eta}^{\left(j\right)}$ inside the
unit cell $C_{\Lambda}$ around the lattice vector $\mathbf{R}=\mathbf{0}$.
Based on the functional form of the frequency dependence of the individual
microscopic polarizability (\ref{eq: atom polarizability}) of a single
atom (ion, molecule), in actual fact being equivalent to a Lorentz-oscillator
model, we write now a simple phenomenological \emph{ansatz} with fitting
parameters $\alpha_{0}^{\left(j\right)}$and $\omega_{0}^{\left(j\right)}$(and
possibly also including a small life-time parameter $\tau^{\left(j\right)}=1/\gamma^{\left(j\right)}>0$
representing spontaneous emission, see supplemental material \citep{Supplementary}):
\begin{equation}
\alpha_{aa'}^{\left(I\right)}\left(\boldsymbol{\eta}^{\left(j\right)},\omega\right)=\delta_{a,a'}\frac{\alpha_{0}^{\left(j\right)}}{1-\left[\frac{\omega+\frac{i}{2}\gamma^{\left(j\right)}}{\omega_{0}^{\left(j\right)}}\right]^{2}}\label{eq: isotropic Lorentz polarizability diagonal}
\end{equation}
If $\omega\ll\omega_{0}^{\left(j\right)}$ the atom polarizability
is well approximated by its static value $\alpha_{0}^{\left(j\right)}$.
For most frequencies, except if $\omega$ approaches a transition
frequency $\omega_{0}^{\left(j\right)}$ near to an absorption band,
we may assume $\gamma^{\left(j\right)}\rightarrow0^{+}$.

In case there are two or more basis atoms in the Wigner-Seitz cell
qualitatively new effects need to be considered. Off diagonal terms
$j\neq j'$ exist for $M\geq2$ and designate mutual influences of
atoms (ions, molecules) positioned at different sites $\boldsymbol{\eta}^{\left(j\right)}$
and $\boldsymbol{\eta}^{\left(j'\right)}$ inside the unit cell. On
one hand, in reaction to the presence of a propagating electromagnetic
wave the overlap integral(s) determining the sharing of electron pairs
between neighbouring atoms undergo (slight) changes. On the other
hand, in crystals like $NaCl$, $CsI$, $RbCl$ etc., the formation
of \emph{ions} needs to be taken into account. Besides the effect
of induced electronic polarization of \emph{single} atoms (ions),
attributed to a shift of the barycenter of the electrons bound to
individual atoms (ions) under the action of the local field on-site
$\mathbf{R}^{\left(j\right)}$, an additional shift of the position
of the positive ions relative to the position of the negative ions
concurs, thus leading to ionic\emph{ displacement }polarizability.
This effect is typically noticeable in the electromagnetic response
to radiation with frequency $\omega$ of order of characteristic \emph{lattice
vibration} frequencies $\omega_{ph}$, de facto being mainly of concern
for radiation in the infrared. 

Conceiving also the off-diagonal polarizabilities $j\neq j'$ not
as input from microscopic theory, but in the phenomenological guise
of a Lorentz-oscillator model for oppositely charged ion pairs \citep{Ashcroft1981},

\begin{equation}
\alpha_{aa'}^{\left(II\right)}\left(\boldsymbol{\eta}^{\left(j\right)},\boldsymbol{\eta}^{\left(j'\right)},\omega\right)=\delta_{a,a'}\frac{\alpha_{0}^{\left(j,j'\right)}}{1-\left[\frac{\omega+\frac{i}{2}\gamma^{\left(j,j'\right)}}{\omega_{0}^{\left(j,j'\right)}}\right]^{2}}\:,\label{eq: isotropic Lorentz polarizability off diagonal}
\end{equation}
with fitting parameters $\alpha_{0}^{\left(j,j'\right)}$, $\omega_{0}^{\left(j,j'\right)}$
and small damping $\gamma^{\left(j,j'\right)}$, we find the well
known chromatic dispersion of the index of refraction $n\left(\omega\right)$
is nicely reproduced by our theory of the dielectric tensor $\varepsilon_{\Lambda}\left(\mathbf{q},\omega\right)$
for a variety of dielectric crystals, see Table \ref{results} in
section \ref{sec:Macroscopic-Electric-Field-and-dielectric-function}.
While the electronic polarizabilities (\ref{eq: isotropic Lorentz polarizability diagonal})
of single atoms are often well approximated by their static value,
the ionic displacement polarizabilities (\ref{eq: isotropic Lorentz polarizability off diagonal})
have a characteristic dependence on frequency in the infrared. Note,
that we refrain adding the contributions of the electronic polarizabilities
of single atoms to the contributions of the ionic displacement polarizabilities,
there being no justification for this \citep{Ashcroft1981}.

For example, in the case of $CsI$ with $M=2$ ions in the unit cell,
our phenomenological approach in (\ref{eq: phenomenological dielectric kernel})
brings into altogether $6$ parameters. These are the two transition
frequencies $\omega_{0}^{\left(1\right)}$, $\omega_{0}^{\left(2\right)}$
in the optical regime to model the \emph{induced electronic polarization}
of the two ions together with two values $\alpha_{0}^{\left(1\right)}$,
$\alpha_{0}^{\left(2\right)}$ for the individual \emph{static} polarizabilities
of the respective ions, and a third resonance frequency $\omega_{0}^{\left(1,2\right)}$
with associated value $\alpha_{0}^{\left(1,2\right)}$shaping the
strength of the ionic displacement polarizabilty, that frequency $\omega_{0}^{\left(1,2\right)}$
being characteristic of lattice vibrational frequencies $\omega_{ph}$.
Such a distinction is justifiable if the time scale for electronic
polarization of individual ions is much faster than the time scale
for the lattice vibrations: $\omega_{ph}\simeq\omega_{0}^{\left(1,2\right)}\ll\min\left(\omega_{0}^{\left(1\right)},\omega_{0}^{\left(2\right)}\right)$.
With the model (\ref{eq: phenomenological dielectric kernel}) we
then successfully reproduced experimental data for the chromatic dispersion
of the refraction index $n\left(\omega\right)$, exemplarily for $CsI$
and $RbCl$, over a wide frequency interval ranging from ultraviolet
to infrared, utilizing only these $6$ parameters instead of $17$
parameters as required by a Sellmeier fit, see Fig.\ref{Sellmeier}
and Table \ref{fit}.

Restricting to optical frequencies well above the infrared range (but
always well below atomic excitation energies), then mostly the electrons
bound around individual ions (atoms) will react to the electromagnetic
fields. In this case the effect of induced electric polarization is
predominant and the effects of ionic polarisation, as represented
by the off diagonal contributions $j\neq j'$ in (\ref{eq: model for microscopic polarizability}),
can be considered as small, so that 
\begin{eqnarray}
\omega & \gg & \omega_{ph}\nonumber \\
\alpha_{aa'}\left(\boldsymbol{\eta}^{\left(j\right)},\boldsymbol{\eta}^{\left(j'\right)},\omega\right) & = & \delta_{j,j'}\alpha_{aa'}^{\left(I\right)}\left(\boldsymbol{\eta}^{\left(j\right)},\omega\right)\:.\label{eq: high frequency limit of optical polarizability}
\end{eqnarray}
It should be noted that retaining in (\ref{eq: high frequency limit of optical polarizability})
the possibility of non zero off diagonal Cartesian terms $a\neq a'$
in the Lorentz-oscillator model of atomic polarizabilities $\tilde{\alpha}_{aa'}\left(\boldsymbol{\eta}^{\left(j\right)},\omega\right)$
then enables the study of crystals composed of anisotropic polarizable
ionic or molecular subunits in the elementary cell. For instance,
the study of the influence of an external \emph{static} magnetic induction
field $\mathbf{B}_{0}$ or \emph{static} electric field $\mathbf{E}_{0}$
on the propagation of light in crystals also requires to retain non
zero off diagonal Cartesian components $a\neq a'$ in (\ref{eq: high frequency limit of optical polarizability}).
As important examples we mention the \emph{magneto-optical} Faraday
effect and the \emph{electro-optical} Pockels effect \citep{Agranovich1984}.

Obviously, under a translation by a lattice vector $\mathbf{R}\in\Lambda$
the dielectric kernel (\ref{eq: phenomenological dielectric kernel})
remains invariant: 
\begin{equation}
\tilde{\chi}_{a,a'}\left(\mathbf{r}+\mathbf{R},\mathbf{r}'+\mathbf{R},\omega\right)=\tilde{\chi}_{a,a'}\left(\mathbf{r},\mathbf{r}',\omega\right)\label{eq: periodicity of dielectric kernel}
\end{equation}
The dielectric kernel considered at fixed position $\mathbf{r}\in\Omega_{P}$
as a function of $\mathbf{r}'-\mathbf{r}$ will typically undergo
already discernible variations traversing a short route of order of
the interatomic distance $a$. In what follows we show, that the microscopic
local electric field amplitude $\tilde{\mathbf{E}}\left(\mathbf{r},\omega\right)$
then also displays discernible spatial variations on that same short
length scale $a$, even though the primary incident light signal $\tilde{\mathbf{E}}_{ext}\left(\mathbf{r},\omega\right)$
was discernibly varying only on the much longer length scale set by
the wavelength $\lambda\gg a$ of light in free space.

\subsection*{Non-Standard Bloch Functions }

Because of the periodicity (\ref{eq: periodicity of dielectric kernel})
it is an obvious choice to expand the microscopic local electric field
in a complete basis of Bloch eigenstates of the translation operator
$\hat{T}_{\mathbf{\mathbf{R}}}$ shifting the argument of any function
$f\left(\mathbf{r}\right)$ according to
\begin{equation}
\hat{T}_{\mathbf{\mathbf{R}}}f\left(\mathbf{r}\right)=f\left(\mathbf{r}+\mathbf{R}\right)\label{eq: shift operator}
\end{equation}
 with $\mathbf{R}\in\Lambda$ a Bravais lattice vector. Routinely
in problems with a lattice periodic potential an expansion provided
by the orthonormal and complete basis system of plane waves constructed
from eigenfunctions of the \emph{momentum} operator is deployed
\begin{equation}
\left\{ e^{i(\mathbf{k}+\mathbf{G})\cdot\mathbf{r}}\right\} _{\mathbf{k}\in C_{\Lambda}^{-1},\mathbf{G}\in\Lambda^{-1}}\;.\label{eq: basis system plane waves}
\end{equation}
Here $\Lambda^{-1}$ denotes the reciprocal lattice conjugate to the
lattice $\Lambda$ and $C_{\Lambda}^{-1}$ is the Brillouin zone.
Making use of $e^{i\mathbf{G}\cdot\mathbf{R}}=1$ for $\mathbf{R}\in\Lambda$
and any reciprocal lattice vector $\mathbf{G}\in\Lambda^{-1}$ one
readily confirms
\begin{equation}
\hat{T}_{\mathbf{\mathbf{R}}}e^{i(\mathbf{k}+\mathbf{G})\cdot\mathbf{r}}=e^{i\mathbf{k}\cdot\mathbf{R}}e^{i(\mathbf{k}+\mathbf{G})\cdot\mathbf{r}}.
\end{equation}
The eigenvalues $e^{i\mathbf{k}\cdot\mathbf{R}}$ associated with
the eigenfunctions $e^{i(\mathbf{k}+\mathbf{G})\cdot\mathbf{r}}$
of the translation operator $\hat{T}_{\mathbf{\mathbf{R}}}$ being
highly degenerate, any function of the form 
\begin{eqnarray}
f_{\mathbf{k}}\left(\mathbf{r}\right) & = & e^{i\mathbf{k}\cdot\mathbf{r}}u\left(\mathbf{r}\right)\\
u\left(\mathbf{r}\right) & = & \sum_{\mathbf{G}\in\Lambda^{-1}}u_{\mathbf{G}}e^{i\mathbf{G}\cdot\mathbf{r}}=u\left(\mathbf{r}+\mathbf{R}\right)\:,
\end{eqnarray}
(Fourier coefficients denoted as $u_{\mathbf{G}}$), is an eigenfunction
of the translation operator(s) $\hat{T}_{\mathbf{\mathbf{R}}}$: 
\begin{equation}
\hat{T}_{\mathbf{\mathbf{R}}}f_{\mathbf{k}}\left(\mathbf{r}\right)=e^{i\mathbf{k}\cdot\mathbf{R}}f_{\mathbf{k}}\left(\mathbf{r}\right)
\end{equation}

Based on a rigorous theory of the microscopic electromagnetic response
kernel, Dolgov and Maximov \citep{Maksimov2012} presented for crystalline
systems a profound analysis of the dielectric function, representing
the microscopic kernels in (\ref{eq: connection dielectric kernel to response kernel})
as (infinite) matrices with respect to the basis system (\ref{eq: basis system plane waves}).
Adversely, the integral kernel $\mathcal{\tilde{G}}\circ\tilde{\chi}$
in (\ref{eq:local field integral equation}), when represented in
the basis of plane waves $\left\{ e^{i(\mathbf{k}+\mathbf{G})\cdot\mathbf{r}}\right\} _{\mathbf{k}\in C_{\Lambda}^{-1},\mathbf{G}\in\Lambda^{-1}}$,
turns out to be a full up matrix $\left\langle \mathbf{G}+\mathbf{k}\right|\left[\mathcal{\tilde{G}}\circ\tilde{\chi}\right]_{aa'}\left|\mathbf{G}'+\mathbf{k}\right\rangle $,
labeled by a wavevector $\mathbf{k}\in C_{\Lambda^{-1}}$ and an infinite
set of reciprocal lattice vectors $\mathbf{G},\mathbf{G}'\in\Lambda^{-1}$.
In the (numerical) calculations thus the handling of large matrices
is required. 

An alternative set of eigenfunctions of the translation operator(s)
$\hat{T}_{\mathbf{\mathbf{R}}}$, so that the dielectric kernel when
represented in the new basis appears as a \emph{sparse} matrix, is
highly desirable. Observing, that any lattice periodic function $u\left(\mathbf{r}\right)$
can be generated from a \emph{fragment} $u^{\left(0\right)}\left(\mathbf{s}\right)$
that equals to $u\left(\mathbf{s}\right)$ inside the Wigner-Seitz
cell $C_{\Lambda}$ and is zero outside,
\begin{equation}
u\left(\mathbf{r}\right)=\sum_{\mathbf{R}'\in\Lambda}u^{\left(0\right)}\left(\mathbf{r}+\mathbf{R}'\right),\label{eq: generate lattice periodic function I-1}
\end{equation}
we may as well represent any such fragment $u^{\left(0\right)}\left(\mathbf{s}\right)\equiv\left\langle \mathbf{s}|u^{\left(0\right)}\right\rangle $
as a linear combination 
\begin{equation}
\left|u^{\left(0\right)}\right\rangle =\int_{C_{\Lambda}}d^{3}s'u^{\left(0\right)}\left(\mathbf{s}'\right)\left|\mathbf{s}'\right\rangle ,
\end{equation}
with the complete and orthonormal set of eigenstates $\left\{ \left|\mathbf{s}\right\rangle \right\} _{\mathbf{s}\in C_{\Lambda}}$
of the position operator $\hat{r}_{a}$ obeying to the well known
relations 
\begin{eqnarray}
\hat{r}_{a}\left|\mathbf{s}\right\rangle  & = & s_{a}\left|\mathbf{s}\right\rangle \nonumber \\
\int_{C_{\Lambda}}d^{3}s\left|\mathbf{s}\right\rangle \left\langle \mathbf{s}\right| & = & \hat{1}_{C_{\Lambda}}\nonumber \\
\left\langle \mathbf{r}|\mathbf{s}\right\rangle  & = & \delta^{\left(3\right)}\left(\mathbf{s}-\mathbf{r}\right).
\end{eqnarray}
So instead of expanding into the basis system of plane waves $\left\{ e^{i(\mathbf{k}+\mathbf{G})\cdot\mathbf{r}}\right\} _{\mathbf{k}\in C_{\Lambda}^{-1},\mathbf{G}\in\Lambda^{-1}}$
there emerges as an alternative expanding into the following system
of eigenfunctions of the translation operator $\hat{T}_{\mathbf{\mathbf{R}}}$:
\begin{eqnarray}
w\left(\mathbf{r};\mathbf{s},\mathbf{k}\right) & = & \frac{e^{i\mathbf{k}\cdot\mathbf{r}}}{\sqrt{\left|\Lambda_{P}\right|}}\sum_{\mathbf{R}'\in\Lambda}\left\langle \mathbf{r}|\mathbf{\mathbf{s}+\mathbf{R}'}\right\rangle \label{eq: basis w(r,s,k)}
\end{eqnarray}
By construction: 
\begin{equation}
\hat{T}_{\mathbf{\mathbf{R}}}w\left(\mathbf{r};\mathbf{s},\mathbf{k}\right)=w\left(\mathbf{r}+\mathbf{R};\mathbf{s},\mathbf{k}\right)=e^{i\mathbf{k}\cdot\mathbf{R}}w\left(\mathbf{r};\mathbf{s},\mathbf{k}\right)
\end{equation}
The set of states $\left\{ w\left(\mathbf{r};\mathbf{s},\mathbf{k}\right)\right\} _{\mathbf{k}\in C_{\Lambda^{-1}},\mathbf{\mathbf{s}}\in C_{\Lambda}}$
is for one thing labeled by a wavevector $\mathbf{k}\in C_{\Lambda^{-1}}$,
that ranges through the count $\left|\Lambda_{P}\right|$ of values
of wave vectors $\mathbf{k}$ in the first Brillouin zone $C_{\Lambda^{-1}}$,
consistent with the Born-von Karman periodic boundary conditions,
and for another thing it is labeled by position vectors $\mathbf{s}\in C_{\Lambda}$
that range within the Wigner-Seitz cell $C_{\Lambda}$ of the lattice
$\Lambda$. The system $\left\{ w\left(\mathbf{r};\mathbf{s},\mathbf{k}\right)\right\} _{\mathbf{k}\in C_{\Lambda^{-1}},\,\mathbf{\mathbf{s}}\in C_{\Lambda}}$
spans a \emph{complete} and \emph{orthonormal} basis system of eigenfunctions
of the translation operator(s) $\hat{T}_{\mathbf{\mathbf{R}}}$, for
a proof see supplemental material \citep{Supplementary}.

\subsection*{Solution of the Field-Integral Equations for a Dielectric Crystal}

Representing now the microscopic local electric field in the complete
and orthonormal basis system $\left\{ w\left(\mathbf{r};\mathbf{s},\mathbf{k}\right)\right\} _{\mathbf{k}\in C_{\Lambda^{-1}},\,\mathbf{\mathbf{s}}\in C_{\Lambda}}$
we write
\begin{equation}
\tilde{E}_{a}\left(\mathbf{r},\omega\right)=\sum_{\mathbf{k}\in C_{\Lambda^{-1}}}\int_{C_{\Lambda}}d^{3}s\:w\left(\mathbf{r};\mathbf{s},\mathbf{k}\right)\tilde{\mathfrak{e}}_{a}\left(\mathbf{s},\mathbf{k},\omega\right)\:.\label{eq:local electric field spanned in basis w(r,s,k)}
\end{equation}
Conversely, the expansion coefficients representing that field are
\begin{eqnarray}
\tilde{\mathfrak{e}}_{a}\left(\mathbf{s},\mathbf{k},\omega\right) & = & \int_{\Omega_{P}}d^{3}r'\left[w\left(\mathbf{r}';\mathbf{s},\mathbf{k}\right)\right]^{\dagger}\tilde{E}_{a}\left(\mathbf{r}',\omega\right)\:.\label{eq: expansion coefficient of  local field}
\end{eqnarray}
So, with 
\begin{eqnarray}
\tilde{\mathfrak{e}}_{ext,a}\left(\mathbf{s},\mathbf{k},\omega\right) & = & \int_{\Omega_{P}}d^{3}r'\left[w\left(\mathbf{r}';\mathbf{s},\mathbf{k}\right)\right]^{\dagger}\tilde{E}_{ext,a}\left(\mathbf{r}',\omega\right)\label{eq:expansion coefficients external field}
\end{eqnarray}
the field-integral equation (\ref{eq:local field integral equation})
is transformed into an equivalent integral equation determining the
expansion coefficients $\tilde{\mathfrak{e}}_{a}\left(\mathbf{s},\mathbf{k},\omega\right)$:
\begin{equation}
\tilde{\mathfrak{e}}_{a}\left(\mathbf{s},\mathbf{k},\omega\right)=\tilde{\mathfrak{e}}_{ext,a}\left(\mathbf{s},\mathbf{k},\omega\right)+\sum_{\mathbf{k}'\in C_{\Lambda^{-1}}}\int_{C_{\Lambda}}d^{3}s'\sum_{a'}\left\langle \mathbf{s},\mathbf{k}\right|\left[\mathcal{\tilde{G}}\circ\tilde{\chi}\right]_{a,a'}\left|\mathbf{s}',\mathbf{k}'\right\rangle \tilde{\mathfrak{e}}_{a'}\left(\mathbf{s}',\mathbf{k}',\omega\right)\label{eq:local field integral equation-1}
\end{equation}
The matrix elements of the dielectric kernel in the basis (\ref{eq: basis w(r,s,k)})
are readily evaluated, see supplemental material \citep{Supplementary}:
\begin{eqnarray}
\left\langle \mathbf{s},\mathbf{k}\right|\left[\mathcal{\tilde{G}}\circ\tilde{\chi}\right]_{a,a'}\left|\mathbf{s}',\mathbf{k}'\right\rangle  & = & \int_{\Omega_{P}}d^{3}r\int_{\Omega_{P}}d^{3}r'\left[w\left(\mathbf{r};\mathbf{s},\mathbf{k}\right)\right]^{\dagger}\left[\mathcal{\tilde{G}}\circ\tilde{\chi}\right]_{a,a'}\left(\mathbf{r},\mathbf{r}',\omega\right)w\left(\mathbf{r}';\mathbf{s}',\mathbf{k}'\right)\label{eq:matrix element dielectric kernel}\\
 & = & \frac{1}{\varepsilon_{0}}\sum_{j',j''}\sum_{a''}\left[\zeta_{\varLambda}(\mathbf{s}-\boldsymbol{\eta}^{\left(j''\right)},\mathbf{k},\omega)\right]_{a,a''}\alpha_{a'',a'}^{j'',j'}\left(\mathbf{k},\omega\right)\delta^{\left(3\right)}\left(\mathbf{s}'-\boldsymbol{\eta}^{\left(j'\right)}\right)\delta_{\mathbf{k},\mathbf{k}'}\nonumber 
\end{eqnarray}
Here we introduced notation such that 
\begin{equation}
\alpha_{a'',a'}^{\left(j'',j'\right)}\left(\mathbf{k},\omega\right)\equiv e^{-i\mathbf{k}\cdot\mathbf{\boldsymbol{\eta}}^{\left(j''\right)}}\alpha_{a'',a'}\left(\boldsymbol{\eta}^{\left(j''\right)},\boldsymbol{\eta}^{\left(j'\right)},\omega\right)e^{i\mathbf{k}\cdot\boldsymbol{\eta}^{\left(j'\right)}},\label{eq:block  matrix alpha}
\end{equation}
and $\tilde{\zeta}_{\varLambda}$ denotes a $3\times3$ matrix of
\emph{lattice sums} formed with the electromagnetic kernel (\ref{eq:3x3 matrix propagator}):
\begin{equation}
\left[\zeta_{\varLambda}(\mathbf{s},\mathbf{k},\omega)\right]_{a,a'}=\begin{cases}
\sum_{\mathbf{R}'\in\varLambda}e^{-i\mathbf{k}\cdot\left(\mathbf{s}+\mathbf{R}'\right)}\mathcal{\tilde{G}}_{a,a'}\left(\mathbf{s}+\mathbf{R}',\omega\right) & \textrm{if}\:\mathbf{s}\neq\mathbf{0}\\
\sum_{\mathbf{R}'\in\varLambda\setminus\left\{ \mathbf{\mathbf{0}}\right\} }e^{-i\mathbf{k}\cdot\mathbf{R}'}\mathcal{\tilde{G}}_{a,a'}\left(\mathbf{R}',\omega\right) & \textrm{if}\:\mathbf{s}=\mathbf{0}
\end{cases}\label{eq: lattice sums zeta}
\end{equation}
For any Bravais lattice vector $\mathbf{R}\in\varLambda$ there holds
\begin{equation}
\zeta_{\varLambda}(\mathbf{s}+\mathbf{R},\mathbf{k},\omega)=\zeta_{\varLambda}(\mathbf{s},\mathbf{k},\omega).\label{eq: periodicity property of lattice sum}
\end{equation}
The definition of the lattice sums (\ref{eq: lattice sums zeta})
by cases is to ensure no atom can get polarized by its self-generated
electromagnetic field. It is important to realize that 
\begin{equation}
\left[\zeta_{\varLambda}^{\left(0\right)}(\mathbf{k},\omega)\right]_{aa'}\equiv\left[\zeta_{\varLambda}(\mathbf{s}=\mathbf{0},\mathbf{k},\omega)\right]_{aa'}\neq\lim_{\left|\mathbf{\mathbf{s}}\right|\rightarrow0^{+}}\left[\zeta_{\varLambda}(\mathbf{s},\mathbf{k},\omega)\right]_{aa'}.\label{eq: zeta lattice sum null value}
\end{equation}
Instead there holds
\begin{equation}
\left[\zeta_{\varLambda}^{\left(0\right)}(\mathbf{k},\omega)\right]_{aa'}=\lim_{\left|\mathbf{\mathbf{s}}\right|\rightarrow0^{+}}\left[\zeta_{\varLambda}(\mathbf{s},\mathbf{k},\omega)-\mathcal{\tilde{G}}\left(\mathbf{s},\omega\right)\right]_{aa'}.\label{eq: lattice sums zeta for s=00003D0}
\end{equation}
A fast and precise numerical method for the calculation of the lattice
sums $\zeta_{\varLambda}(\mathbf{s},\mathbf{k},\omega)$ and $\zeta_{\varLambda}^{\left(0\right)}(\mathbf{k},\omega)$
we disclose in our supplemental material \citep{Supplementary}.

Insertion of (\ref{eq:matrix element dielectric kernel}) gives at
once the field amplitudes $\tilde{\mathfrak{e}}_{a}\left(\mathbf{s},\mathbf{k},\omega\right)$
in terms of a finite number of amplitudes $\tilde{\mathfrak{e}}_{a}^{\left(j\right)}\left(\mathbf{k},\omega\right)\equiv\tilde{\mathfrak{e}}_{a}\left(\boldsymbol{\eta}^{\left(j\right)},\mathbf{k},\omega\right)$,
with $j\in\left\{ 1,2,...,M\right\} $ counting the positions $\mathbf{s}=\boldsymbol{\eta}^{\left(j\right)}$
of the polarizable atoms (ions, molecules) inside the Wigner-Seitz
cell $C_{\Lambda}$ and $a\in\left\{ x,y,z\right\} $ denoting Cartesian
components: 
\begin{equation}
\tilde{\mathfrak{e}}_{a}\left(\mathbf{s},\mathbf{k},\omega\right)=\tilde{\mathfrak{e}}_{ext,a}\left(\mathbf{s},\mathbf{k},\omega\right)+\frac{1}{\varepsilon_{0}}\sum_{j',j''}\sum_{a',a''}\left[\zeta_{\varLambda}(\mathbf{s}-\boldsymbol{\eta}^{\left(j''\right)},\mathbf{k},\omega)\right]_{a,a''}\alpha_{a'',a'}^{\left(j'',j'\right)}\left(\mathbf{k},\omega\right)\tilde{\mathfrak{e}}_{a'}^{\left(j'\right)}\left(\mathbf{k},\omega\right)\label{eq:amplitudes of local field in basis w(r,s,k)}
\end{equation}
Taking subsequently for $j=1,2,...,M$ the limit $\mathbf{s}\rightarrow\boldsymbol{\eta}^{\left(j\right)}$,
then from (\ref{eq:amplitudes of local field in basis w(r,s,k)})
a $3M\times3M$ system of linear equations determining those amplitudes
$\tilde{\mathfrak{e}}_{a}^{\left(j\right)}\left(\mathbf{k},\omega\right)$
subject to the prescribed amplitudes $\tilde{\mathfrak{e}}_{ext,a}^{\left(j\right)}\left(\mathbf{k},\omega\right)$
of the external field is obtained: 
\begin{equation}
\tilde{\mathfrak{e}}_{a}^{\left(j\right)}\left(\mathbf{k},\omega\right)=\tilde{\mathfrak{e}}_{ext,a}^{\left(j\right)}\left(\mathbf{k},\omega\right)+\frac{1}{\varepsilon_{0}}\sum_{j',j''}\sum_{a',a''}\left[\zeta_{\varLambda}(\boldsymbol{\eta}^{\left(j\right)}-\boldsymbol{\eta}^{\left(j''\right)},\mathbf{k},\omega)\right]_{a,a''}\alpha_{a'',a'}^{\left(j'',j'\right)}\left(\mathbf{k},\omega\right)\tilde{\mathfrak{e}}_{a'}^{\left(j'\right)}\left(\mathbf{k},\omega\right)\label{eq: system of 3Mx3M equations for amplitudes of local field}
\end{equation}
This result clearly brings out the advantage of the basis system $\left\{ w\left(\mathbf{r};\mathbf{s},\mathbf{k}\right)\right\} _{\mathbf{k}\in C_{\Lambda^{-1}},\mathbf{\mathbf{s}}\in C_{\Lambda}}$
over the conventional basis system of plane waves $\left\{ e^{i(\mathbf{k}+\mathbf{G})\cdot\mathbf{r}}\right\} _{\mathbf{k}\in C_{\Lambda}^{-1},\mathbf{G}\in\Lambda^{-1}}$,
thus dispensing the consideration of large full up dielectric matrix
kernels $\left\langle \mathbf{G}+\mathbf{k}\right|\left[\mathcal{\tilde{G}}\circ\tilde{\chi}\right]_{aa'}\left|\mathbf{G}'+\mathbf{k}\right\rangle $
labeled by an infinite number of reciprocal lattice vectors $\mathbf{G},\mathbf{G}'\in\Lambda^{-1}$.
As an incidental remark see \footnote{The expansion basis system $\left\{ w\left(\mathbf{r};\mathbf{s},\mathbf{k}\right)\right\} _{\mathbf{k}\in C_{\Lambda^{-1}},\mathbf{\mathbf{s}}\in C_{\Lambda}}$
also could be used to advantage reducing the computational effort
determining the band structure of a massive particle moving in a periodic
array of \emph{point} scatterers, rewriting the Schr{\"o}dinger eigenvalue
problem as an integral equation, akin to the KKR method, for example
see \citep{Ashcroft1981}. }. 

The determination of the expansion coefficients $\tilde{\mathfrak{e}}_{a}\left(\mathbf{s},\mathbf{k},\omega\right)$
representing the local electric field amplitude via (\ref{eq:amplitudes of local field in basis w(r,s,k)})
requires solving (\ref{eq: system of 3Mx3M equations for amplitudes of local field}),
a \emph{small }sized $3M\times3M$ linear system of equations. Introducing
the $3M\times3M$ matrices 
\begin{equation}
\Gamma_{a,a'}^{\left(j,j'\right)}\left(\mathbf{k},\omega\right)=\sum_{j''}\sum_{a''}\left[\zeta_{\varLambda}(\boldsymbol{\eta}^{\left(j\right)}-\boldsymbol{\eta}^{\left(j''\right)},\mathbf{k},\omega)\right]_{a,a''}\frac{1}{\varepsilon_{0}}\alpha_{a'',a'}^{\left(j'',j'\right)}\left(\mathbf{k},\omega\right)\label{eq: Block matrix Gamma}
\end{equation}
and 
\begin{equation}
\left[I\right]_{a,a'}^{\left(j,j'\right)}=\delta_{j,j'}\delta_{a,a'}
\end{equation}
the explicit solution of (\ref{eq: system of 3Mx3M equations for amplitudes of local field})
reads 
\begin{equation}
\tilde{\mathfrak{e}}_{a}^{\left(j\right)}\left(\mathbf{k},\omega\right)=\sum_{j'}\sum_{a'}\left(\left[I-\Gamma\left(\mathbf{k},\omega\right)\right]^{-1}\right)_{a,a'}^{\left(j,j'\right)}\tilde{\mathfrak{e}}_{ext,a'}^{\left(j'\right)}\left(\mathbf{k},\omega\right).\label{eq: explicit amplitudes of local field}
\end{equation}
Finally, substituting the expansion coefficients (\ref{eq: explicit amplitudes of local field})
into (\ref{eq:amplitudes of local field in basis w(r,s,k)}) we then
obtain from (\ref{eq:local electric field spanned in basis w(r,s,k)})
the following explicit representation for the microscopic local electric
field: 
\begin{eqnarray}
\tilde{E}_{a}\left(\mathbf{r},\omega\right) & = & \sum_{\mathbf{k}\in C_{\Lambda^{-1}}}\int_{C_{\Lambda}}d^{3}s\:w\left(\mathbf{r};\mathbf{s},\mathbf{k}\right)\tilde{\mathfrak{e}}_{a}\left(\mathbf{s},\mathbf{k},\omega\right)\label{eq:microscopic local electric field}\\
 & = & \tilde{E}_{ext,a}\left(\mathbf{r},\omega\right)+\sum_{\mathbf{k}\in C_{\Lambda^{-1}}}\frac{e^{i\mathbf{k}\cdot\mathbf{r}}}{\sqrt{\left|\Lambda_{P}\right|}}\sum_{j',j''}\sum_{a',a''}\left[\zeta_{\varLambda}(\mathbf{r}-\boldsymbol{\eta}^{\left(j''\right)},\mathbf{k},\omega)\right]_{a,a''}\frac{1}{\varepsilon_{0}}\alpha_{a'',a'}^{\left(j'',j'\right)}\left(\mathbf{k},\omega\right)\tilde{\mathfrak{e}}_{a'}^{\left(j'\right)}\left(\mathbf{k},\omega\right)\nonumber 
\end{eqnarray}
Incidentally, the general result (\ref{eq:microscopic local electric field})
for the microscopic local electric field inside a crystal comprises
also the case of multiple beams of incident signals, all with frequency
$\omega$ but possibly with different wavevectors $\mathbf{k}$, in
this case represented by a multitude of expansion coefficients $\tilde{\mathfrak{e}}_{ext,a'}^{\left(j'\right)}\left(\mathbf{k},\omega\right)$
carrying different wave vectors $\mathbf{k}$, for example see \citep{Laue1960}. 

It should be emphasized that the well known dispersion relation $\omega=c\left|\mathbf{q}\right|$
of light in vacuum expresses the solvability condition for the external
field (\ref{eq:incident external signal}) solving the \emph{homogeneous}
Maxwell equations in free space. However, in reality external signals
represent solutions of the \emph{inhomogeneous} Maxwell equations
with a source current $\tilde{j}_{ext,a}\left(\mathbf{r},\omega\right)=\sum_{\mathbf{q}'}\bar{j}_{ext,a}\left(\mathbf{q}',\omega\right)e^{i\mathbf{q}'\cdot\mathbf{r}}$
flowing inside a source volume $\Omega_{S}$, the latter being usually
positioned at a large (but finite) distance to the probe volume $\Omega_{P}$,
see Eq. (57) in supplemental material \citep{Supplementary}. So a
typical external signal $\tilde{E}_{ext,a}\left(\mathbf{r},\omega\right)$
at \emph{fixed} (circular) frequency $\omega$ is composed of a bunch
of mixed wave vectors $\mathbf{q}$' forming that signal: 
\begin{equation}
\tilde{E}_{ext,a}\left(\mathbf{r},\omega\right)=\sum_{\mathbf{q}'}\mathcal{\bar{E}}_{ext,a}\left(\mathbf{\mathbf{q}}',\omega\right)e^{i\mathbf{\mathbf{q}}'\cdot\mathbf{r}}
\end{equation}
This feature enables to regard $\omega$ and $\mathbf{q}$ from now
on as \emph{independent} variables. Making use of the superposition
principle it suffices to specialize to a single beam. For convenience
we consider now an external signal in the guise of a plane wave, with
(circular) frequency $\omega$ and wavevector $\mathbf{q}$ so that
\begin{equation}
\tilde{E}_{ext,a}\left(\mathbf{r},\omega\right)=\mathcal{\bar{E}}_{ext,a}\left(\mathbf{q},\omega\right)e^{i\mathbf{q}\cdot\mathbf{r}}\:.\label{eq:incident external signal}
\end{equation}

Assuming then $\mathbf{q}\in C_{\Lambda^{-1}}$, certainly not a strong
limitation for optical signals propagating in crystalline materials,
it follows at once from the defining equation (\ref{eq:expansion coefficients external field})
that the expansion coefficients of the external signal are independent
on the value of the mode index $\mathbf{s}\in C_{\Lambda}$. All the
more the coefficients $\tilde{\mathfrak{e}}_{ext,a'}^{\left(j'\right)}\left(\mathbf{k},\omega\right)$
then being independent on $\boldsymbol{\eta}^{\left(j\right)}$ there
holds 
\begin{equation}
\tilde{\mathfrak{e}}_{ext,a}\left(\mathbf{s},\mathbf{k},\omega\right)=\sqrt{\left|\Lambda_{P}\right|}\delta_{\mathbf{k},\mathbf{q}}\mathcal{\bar{E}}_{ext,a}\left(\mathbf{q},\omega\right)\equiv\tilde{\mathfrak{e}}_{ext,a'}^{\left(j'\right)}\left(\mathbf{k},\omega\right)\:.\label{eq: amplitude of external fielod}
\end{equation}
Insertion of (\ref{eq: amplitude of external fielod}) into (\ref{eq: explicit amplitudes of local field})
then reduces (\ref{eq:microscopic local electric field}) to 
\begin{eqnarray}
\tilde{E}_{a}\left(\mathbf{r},\omega\right) & = & \mathcal{\bar{E}}_{ext,a}\left(\mathbf{q},\omega\right)e^{i\mathbf{q}\cdot\mathbf{r}}\label{eq:microscopic local field for incident plane wave with small q}\\
 &  & +e^{i\mathbf{q}\cdot\mathbf{r}}\sum_{j',j'',j'''}\sum_{a',a'',a'''}\left[\zeta_{\varLambda}(\mathbf{r}-\boldsymbol{\eta}^{\left(j''\right)},\mathbf{q},\omega)\right]_{a,a''}\frac{1}{\varepsilon_{0}}\alpha_{a'',a'}^{\left(j'',j'\right)}\left(\mathbf{q},\omega\right)\left(\left[I-\Gamma\left(\mathbf{q},\omega\right)\right]^{-1}\right)_{a',a'''}^{\left(j',j'''\right)}\mathcal{\bar{E}}_{ext,a'''}\left(\mathbf{q},\omega\right).\nonumber 
\end{eqnarray}
Exploiting next the lattice periodicity of $\zeta_{\varLambda}(\mathbf{\mathbf{s}},\mathbf{q},\omega)$,
see (\ref{eq: periodicity property of lattice sum}), there holds
for $\mathbf{s}\neq\mathbf{0}$ the Fourier series representation
\begin{equation}
\left[\zeta_{\varLambda}(\mathbf{s},\mathbf{q},\omega)\right]_{a,a'}=\sum_{\mathbf{G}\in\Lambda^{-1}}e^{i\mathbf{\mathbf{G}}\cdot\mathbf{s}}\left[\bar{\zeta}_{\varLambda}(\mathbf{G},\mathbf{q},\omega)\right]_{a,a'},\label{eq: Fourier series representation lattice sum}
\end{equation}
with $\bar{\zeta}_{\varLambda}(\mathbf{G},\mathbf{q},\omega)$ denoting
the Fourier coefficients, see supplemental material \citep{Supplementary}:
\begin{eqnarray}
\left[\bar{\zeta}_{\varLambda}(\mathbf{G},\mathbf{q},\omega)\right]_{a,a'} & = & \frac{1}{\left|C_{\Lambda}\right|}\int_{C_{\Lambda}}d^{3}s\left[e^{-i\mathbf{\mathbf{G}}\cdot\mathbf{s}}\zeta_{\varLambda}(\mathbf{s},\mathbf{q},\omega)\right]_{a,a'}\label{eq: Fourier coefficients lattice sum}\\
 & = & \frac{1}{\left|C_{\Lambda}\right|}\mathcal{\bar{G}}_{a,a'}(\mathbf{q}+\mathbf{G},\omega)\nonumber 
\end{eqnarray}
Insertion of the Fourier series representation (\ref{eq: Fourier series representation lattice sum})
into (\ref{eq:microscopic local field for incident plane wave with small q})
leads finally to the following result for the microscopic local electric
field:
\begin{equation}
\tilde{E}_{a}\left(\mathbf{r},\omega\right)=\mathcal{\bar{E}}_{ext,a}\left(\mathbf{q},\omega\right)e^{i\mathbf{q}\cdot\mathbf{r}}+\sum_{a'',a'''}\frac{1}{\left|C_{\Lambda}\right|}\sum_{\mathbf{G}\in\Lambda^{-1}}e^{i\left(\mathbf{q}+\mathbf{G}\right)\cdot\mathbf{r}}\mathcal{\bar{G}}_{a,a''}(\mathbf{q}+\mathbf{G},\omega)\left[\bar{K}_{\Lambda}\left(\mathbf{G},\mathbf{q},\omega\right)\right]_{a'',a'''}\mathcal{\bar{E}}_{ext,a'''}\left(\mathbf{q},\omega\right).\label{eq:eq:microscopic local field Fourier series representation}
\end{equation}
Here we introduced the kernel 
\begin{equation}
\left[\bar{K}_{\Lambda}\left(\mathbf{G},\mathbf{q},\omega\right)\right]_{a'',a'''}\equiv\sum_{j',j'',j'''}e^{-i\mathbf{G}\cdot\boldsymbol{\eta}^{\left(j''\right)}}\sum_{a'}\frac{1}{\varepsilon_{0}}\alpha_{a'',a'}^{\left(j'',j'\right)}\left(\mathbf{q},\omega\right)\left(\left[I-\Gamma\left(\mathbf{q},\omega\right)\right]^{-1}\right)_{a',a'''}^{\left(j',j'''\right)},\label{eq: kernel K(G,q,om)}
\end{equation}
a quantity being directly connected to the microscopic atom-individual
polarizabilities, see (\ref{eq:block  matrix alpha}). 

\begin{figure}
\includegraphics[scale=0.8]{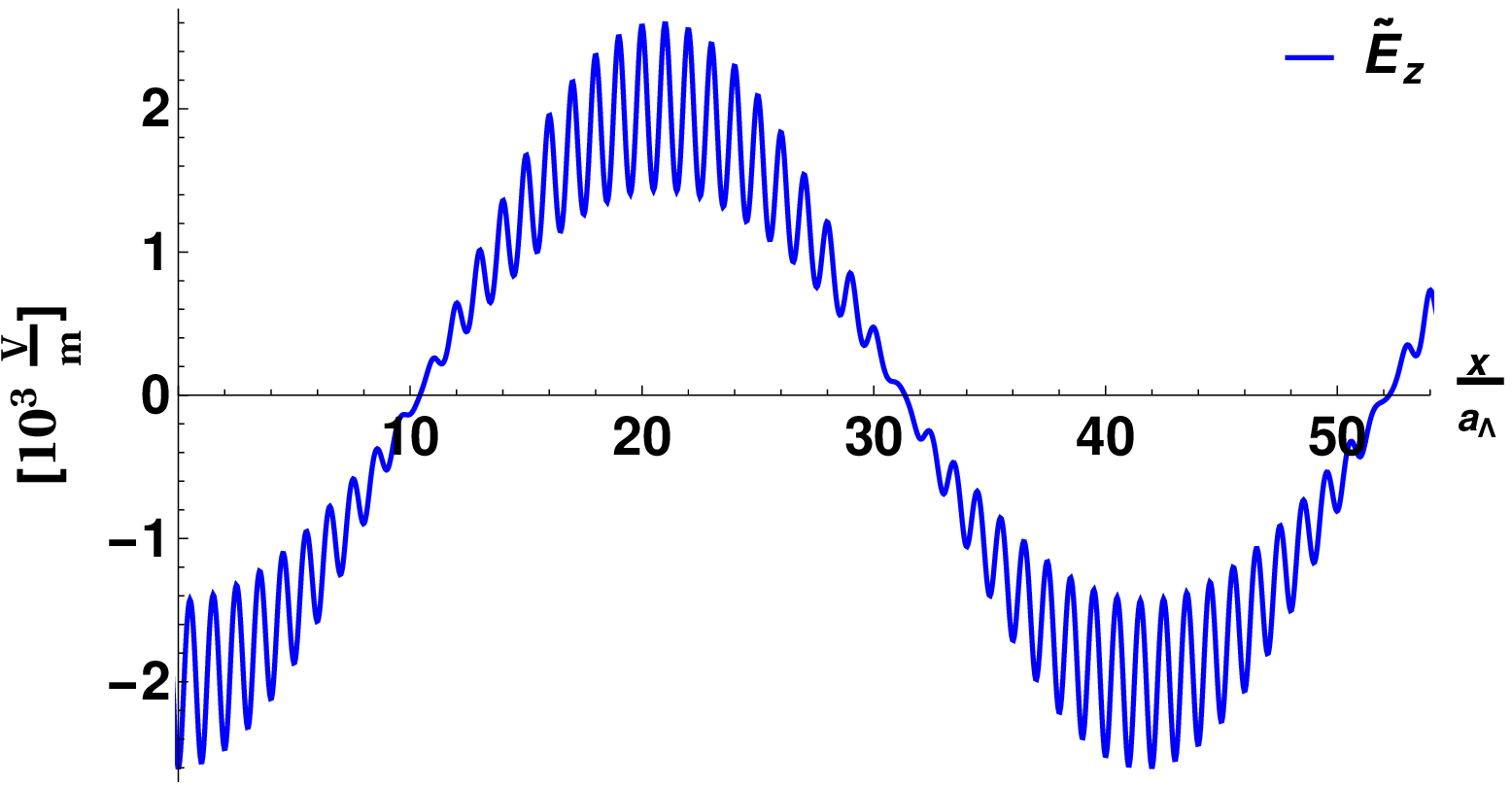}

\caption{\label{fig:short_wavelength_field_plot}The spatial variation of the
microscopic local electric field $\tilde{\mathbf{E}}\left(\mathbf{r},\omega\right)$
representing the solution (\ref{eq:eq:microscopic local field Fourier series representation})
to the inhomogenous field-integral equations (\ref{eq:local field integral equation})
for an\emph{ external} electric field amplitude $\mathbf{\tilde{E}}_{ext}\left(\mathbf{r},\omega_{ext}\right)=\mathbf{e}^{\left(z\right)}\exp\left(i\mathbf{q}\cdot\mathbf{r}\right)\left[\text{\ensuremath{\frac{V}{m}}}\right]$
in the guise of a plane wave, propagating in x - direction and linearly
polarized in z - direction. The plot visualizes the rapid spatial
variations of the microscopic local electric field traversing a simple
cubic crystal along a path $\mathbf{r}\left(x\right)=x\cdot\mathbf{e}^{\left(x\right)}+\frac{a_{\Lambda}}{2}\left(\mathbf{e}^{\left(y\right)}+\mathbf{e}^{\left(z\right)}\right)$
assuming for the external field a vacuum wavelength $\lambda=50\,nm$
in the extreme ultraviolet. A corresponding visualization that applies
for visible violet light with vacuum wavelength $\lambda=400\,nm$
is presented in Fig.\ref{fig:Local_Macro_Fields}.}

\end{figure}

Eq. (\ref{eq:eq:microscopic local field Fourier series representation})
constitutes the central result of this section. It explicitely determines
the microscopic \emph{local} electric field inside a crystal, that
is the electric field that polarizes atoms (ions, molecules) at their
positions $\mathbf{R}^{\left(j\right)}=\mathbf{R}+\boldsymbol{\eta}^{\left(j\right)}$
inside a unit cell of the lattice in reaction to an incident plane
wave (\ref{eq:incident external signal}). While the external field
$\tilde{E}_{ext,a}\left(\mathbf{r},\omega\right)$ displays spatial
variations on the length scale set by the wavelength $\lambda$ of
the incident optical signal in free space, the contributions of the
reciprocal lattice vectors $\mathbf{G}\neq\mathbf{0}$ to the Fourier
series in\emph{ }(\ref{eq:eq:microscopic local field Fourier series representation})
make it manifest that the\emph{ microscopic} field $\tilde{E}_{a}\left(\mathbf{r},\omega\right)$
displays on the back of the slowly varying envelope $e^{i\mathbf{q}\cdot\mathbf{r}}$
rapid spatial variations on the (atomic) scale set by the lattice
constant $a_{\Lambda}\ll\lambda$, see Fig. \ref{fig:short_wavelength_field_plot}.

\begin{figure}
\begin{minipage}[t]{1\columnwidth}%
\includegraphics[scale=0.6]{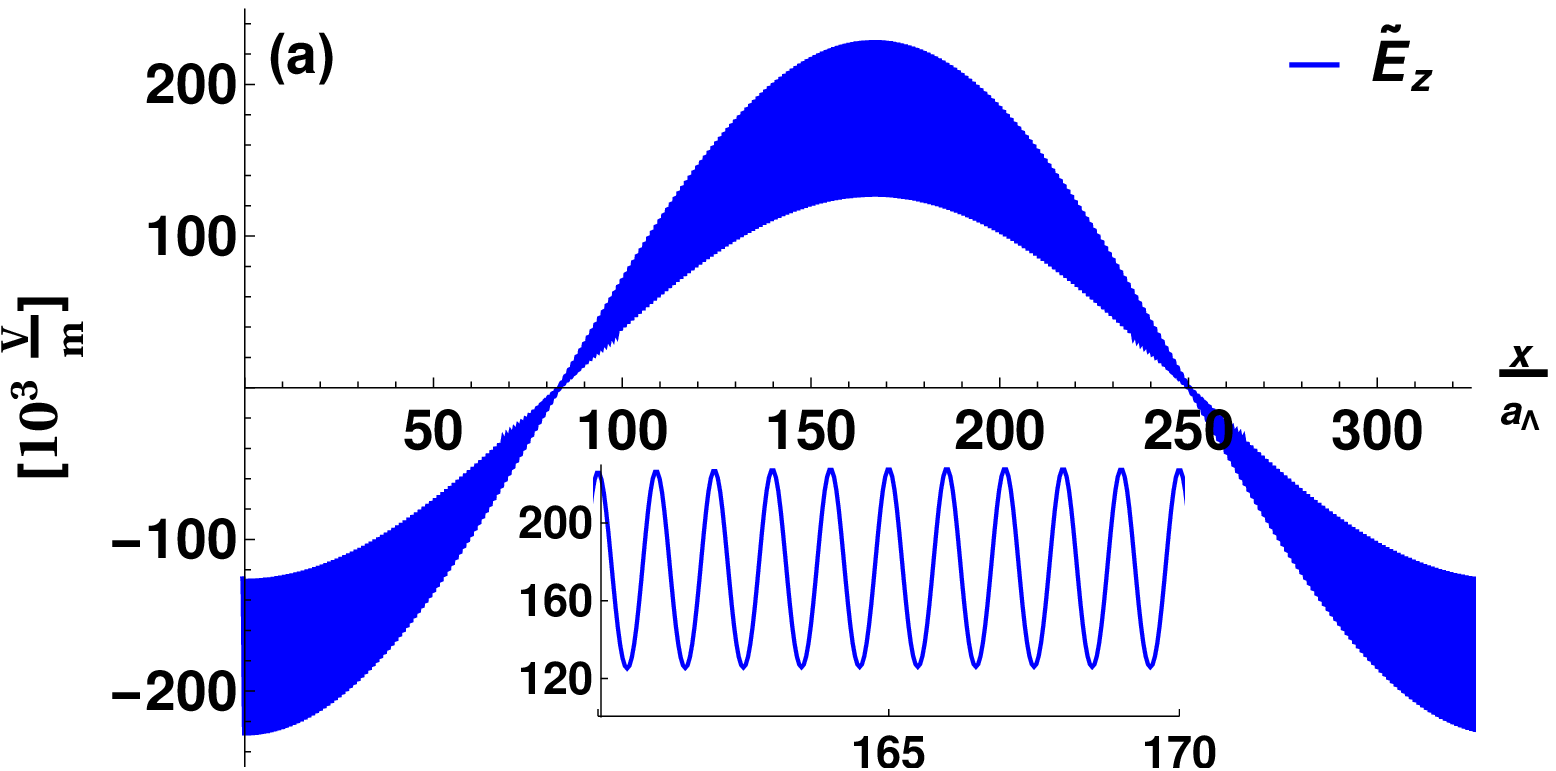}%
\end{minipage}

\vspace{0.5cm}

\begin{minipage}[t]{1\columnwidth}%
\includegraphics[bb=45bp 70bp 545bp 522bp,scale=0.25]{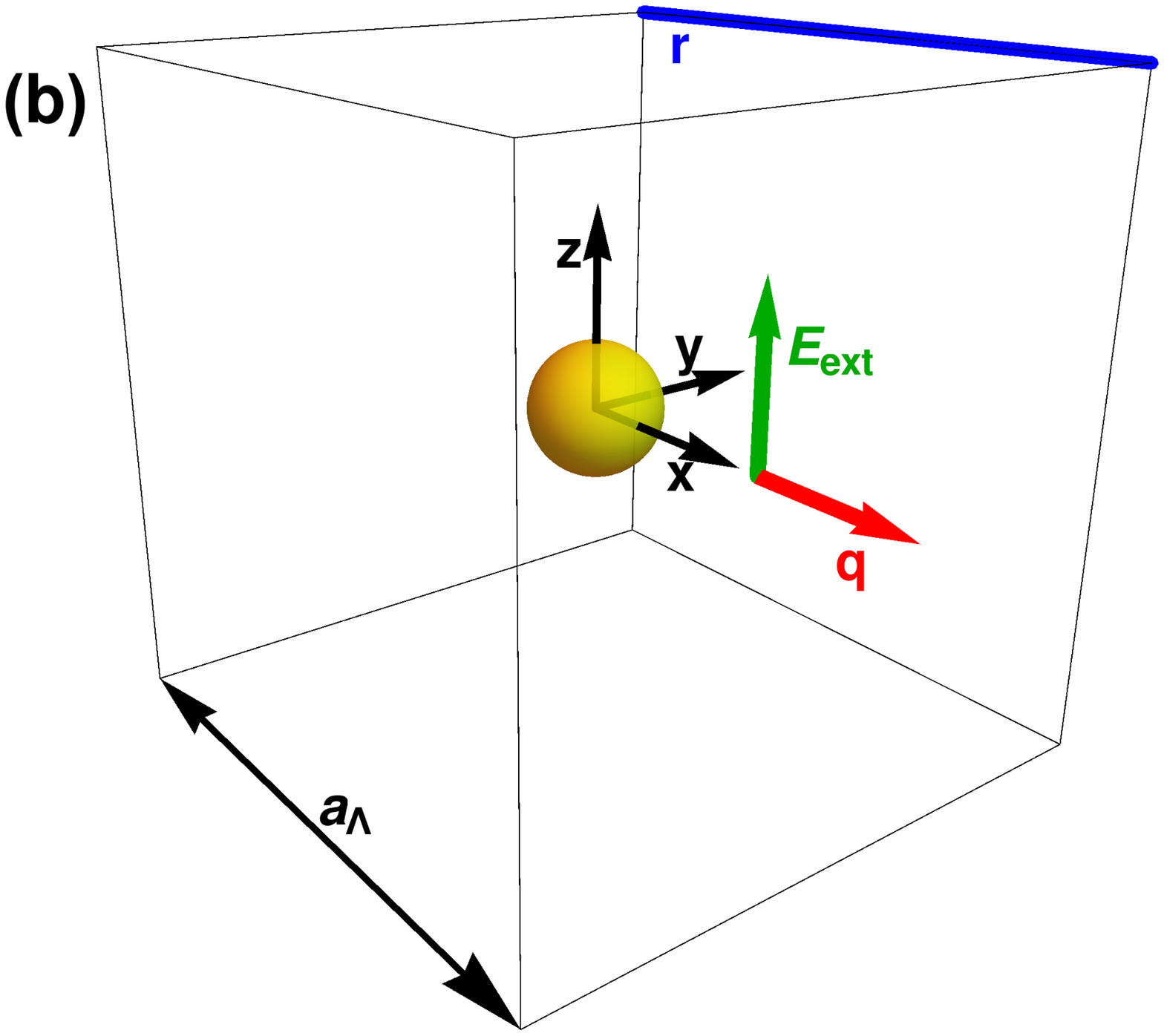}%
\end{minipage}

\vspace{0.5cm}

\begin{minipage}[t]{1\columnwidth}%
\includegraphics[scale=0.6]{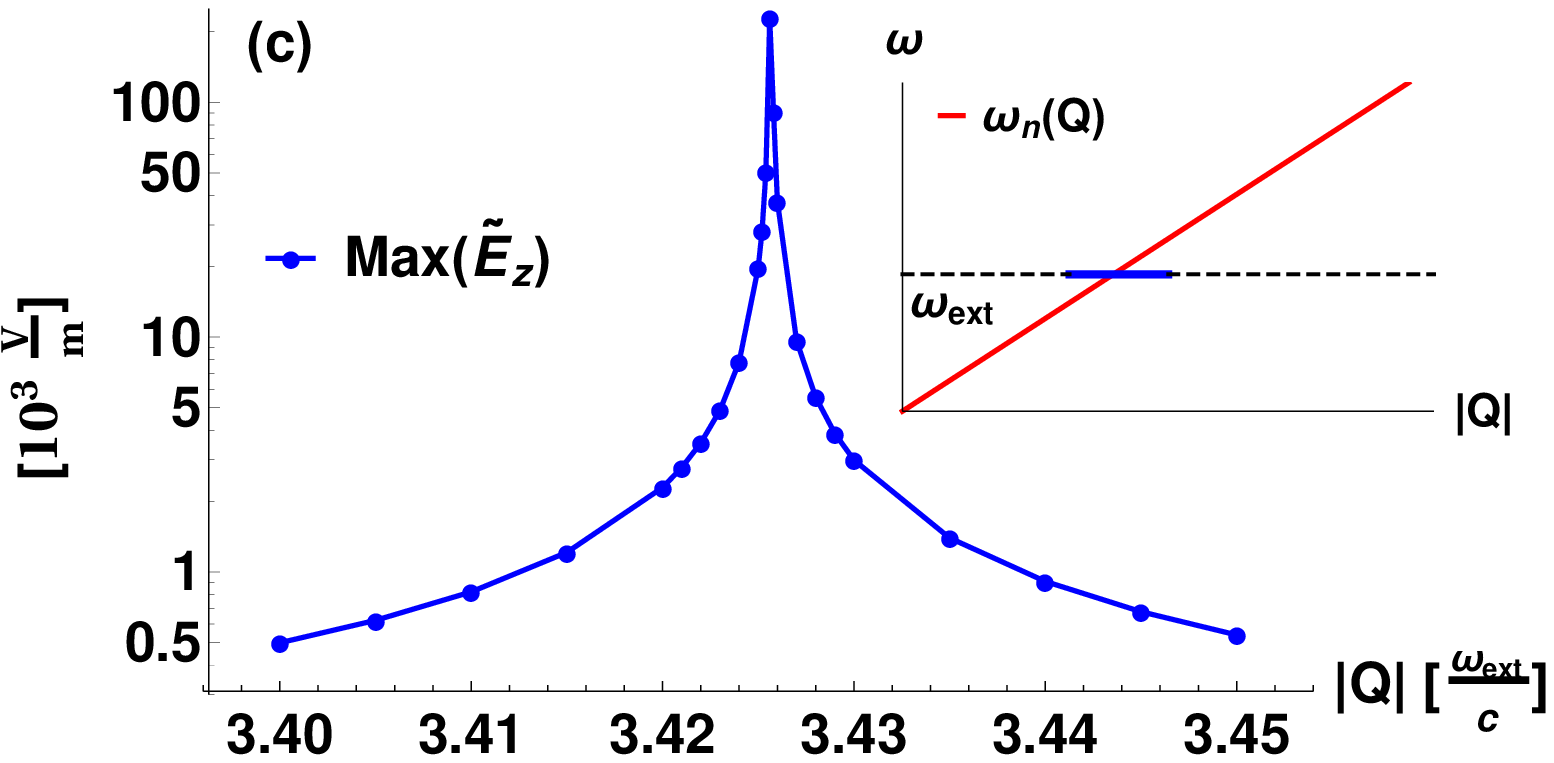}%
\end{minipage}

\vspace{0.5cm}

\begin{minipage}[t]{1\columnwidth}%
\includegraphics[scale=0.6]{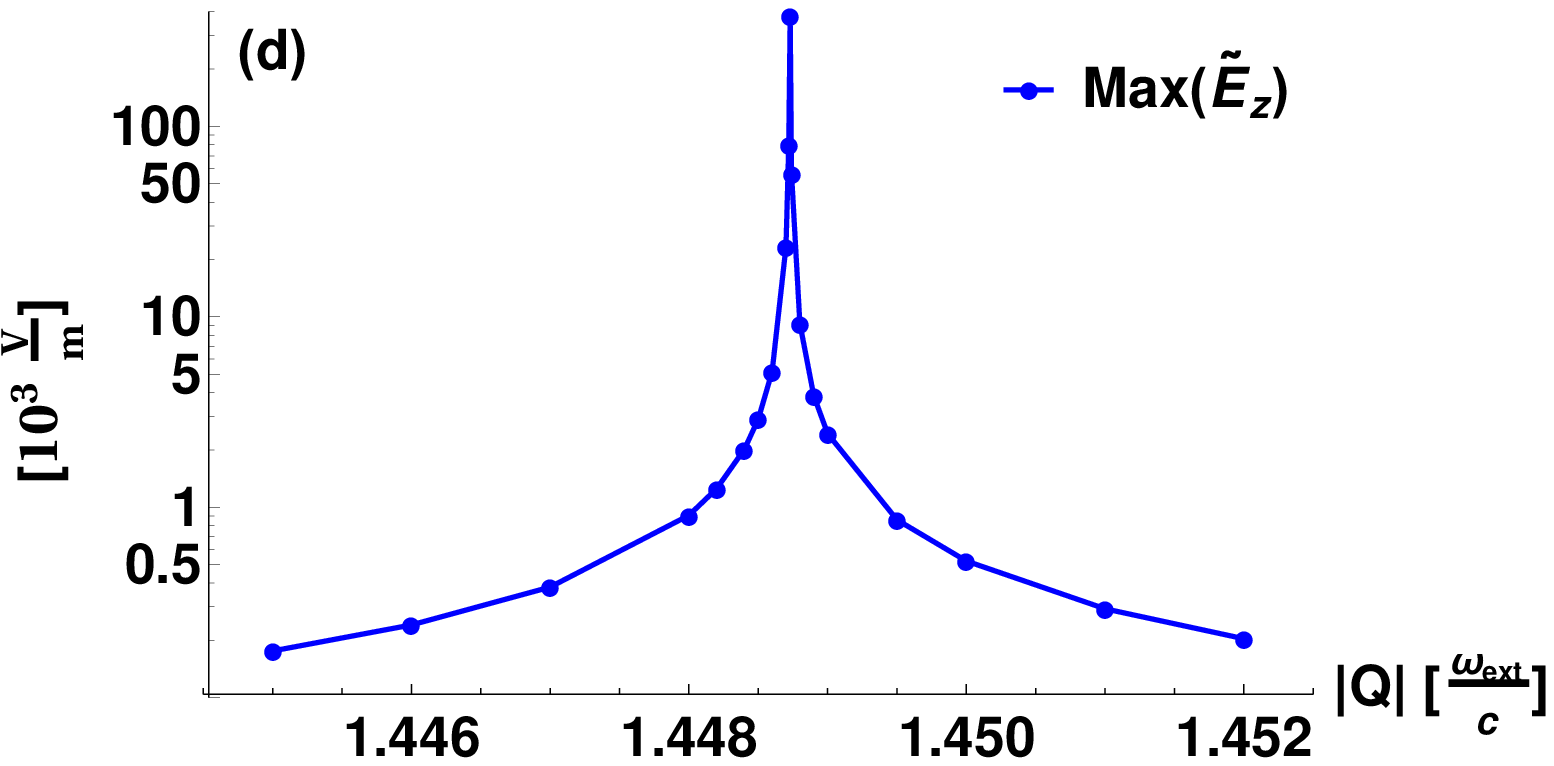}%
\end{minipage}

\caption{\label{fig:Local_Macro_Fields}(a) The spatial variation of the microscopic
local electric field $\tilde{\mathbf{E}}\left(\mathbf{r},\omega\right)$
representing the solution (\ref{eq:eq:microscopic local field Fourier series representation})
to the inhomogenous field-integral equations (\ref{eq:local field integral equation})
for an\emph{ external} field $\mathbf{\tilde{E}}_{ext}\left(\mathbf{r},\omega_{ext}\right)=\mathbf{e}^{\left(z\right)}\exp\left(i\mathbf{q}\cdot\mathbf{r}\right)\left[\text{\ensuremath{\frac{V}{m}}}\right]$
in the guise of a plane wave, propagating in x - direction and linearly
polarized in z - direction. The plot visualizes the rapid spatial
variations of the microscopic local electric field traversing a simple
cubic crystal along a path $\mathbf{r}\left(x\right)=x\cdot\mathbf{e}^{\left(x\right)}+\frac{a_{\Lambda}}{2}\left(\mathbf{e}^{\left(y\right)}+\mathbf{e}^{\left(z\right)}\right)$
assuming for the external field a vacuum wavelength $\lambda=400\,nm$
corresponding to visible violet light. The applied parameters in (a)
are: $a_{\Lambda}=3.5\text{\AA}$, $\frac{\alpha}{4\pi\varepsilon_{0}}=8{\text{\AA}}^{3}$,
$\omega_{\text{ext}}=\frac{2\pi c}{\lambda}$ , $n=3.4256$, $\mathbf{q}=\frac{\omega_{\text{ext}}}{c}n\mathbf{e}^{\left(x\right)}$.
The inset zooms to a smaller scale so that the spatial variations
of the microscopic local electric field are discernible. (b) Schematic
illustration of the path along which the spatial variations of the
microscopic local electric field are displayed in (a). (c) Depiction
of the maximal electric field strenght $\max_{x}\tilde{E}_{z}\left(\mathbf{r}\left(x\right),\omega_{ext}\right)$
along the path $\mathbf{r}\left(x\right)$ as displayed in (b), varying
the modulus $Q$ of the wave vector $\mathbf{Q}=Q\,\mathbf{e}^{\left(x\right)}$
of the external field $\mathbf{\tilde{E}}_{ext}\left(\mathbf{r},\omega_{ext}\right)=\mathbf{e}^{\left(z\right)}\exp\left(i\mathbf{Q}\cdot\mathbf{r}\right)\left[\text{\ensuremath{\frac{V}{m}}}\right]$.
As is clearly on view, only solutions to the field-integral equations
with wave vector $\left|\mathbf{Q}-\mathbf{q}\right|<\delta Q$ meeting
the conditions of tolerance set by the solvability condition of the
homogeneous field-integral equations, $\omega_{ext}=\omega_{n}\left(\mathbf{Q}\right)$
as shown in the inset of (c), may propagate with sufficient intensity
inside the crystal. By increasing the lattice constant to a value
of $a_{\Lambda}=5\text{\AA}$, the density of polarizable atoms as
well as the witdh $\delta Q$ of the distribution are decreased considerably,
see (d).}
\end{figure}

The result (\ref{eq:microscopic local field for incident plane wave with small q})
also reveals that the strength of the microscopic local field amplitude
inside the crystal strongly depends on the choice of the propagation
direction $\mathbf{\hat{q}}=\mathbf{\frac{q}{\left|\mathbf{q}\right|}}$,
a feature being directly connected to the photonic band structure
implicitely encoded in the eigenvalues of the matrix $\Gamma\left(\mathbf{q},\omega\right)$.
The huge size of the induced electric field strength, represented
by the difference $\tilde{E}_{a}\left(\mathbf{r},\omega\right)-\tilde{E}_{ext,a}\left(\mathbf{r},\omega\right)$,
can indeed be prorated to the predominant longitudinal character of
the microscopic local electric field (\ref{eq:microscopic local field for incident plane wave with small q})
inside a probe volume with a high density of polarizable atoms (ions,
molecules). This is intuitively accessible in view of the quasi static
electric field inside a material probe at a point $\mathbf{r}\in\Omega_{P}$
originating from nearby positioned induced atomic dipoles. 

\begin{figure}
\begin{minipage}[t]{1\columnwidth}%
\includegraphics[scale=0.7]{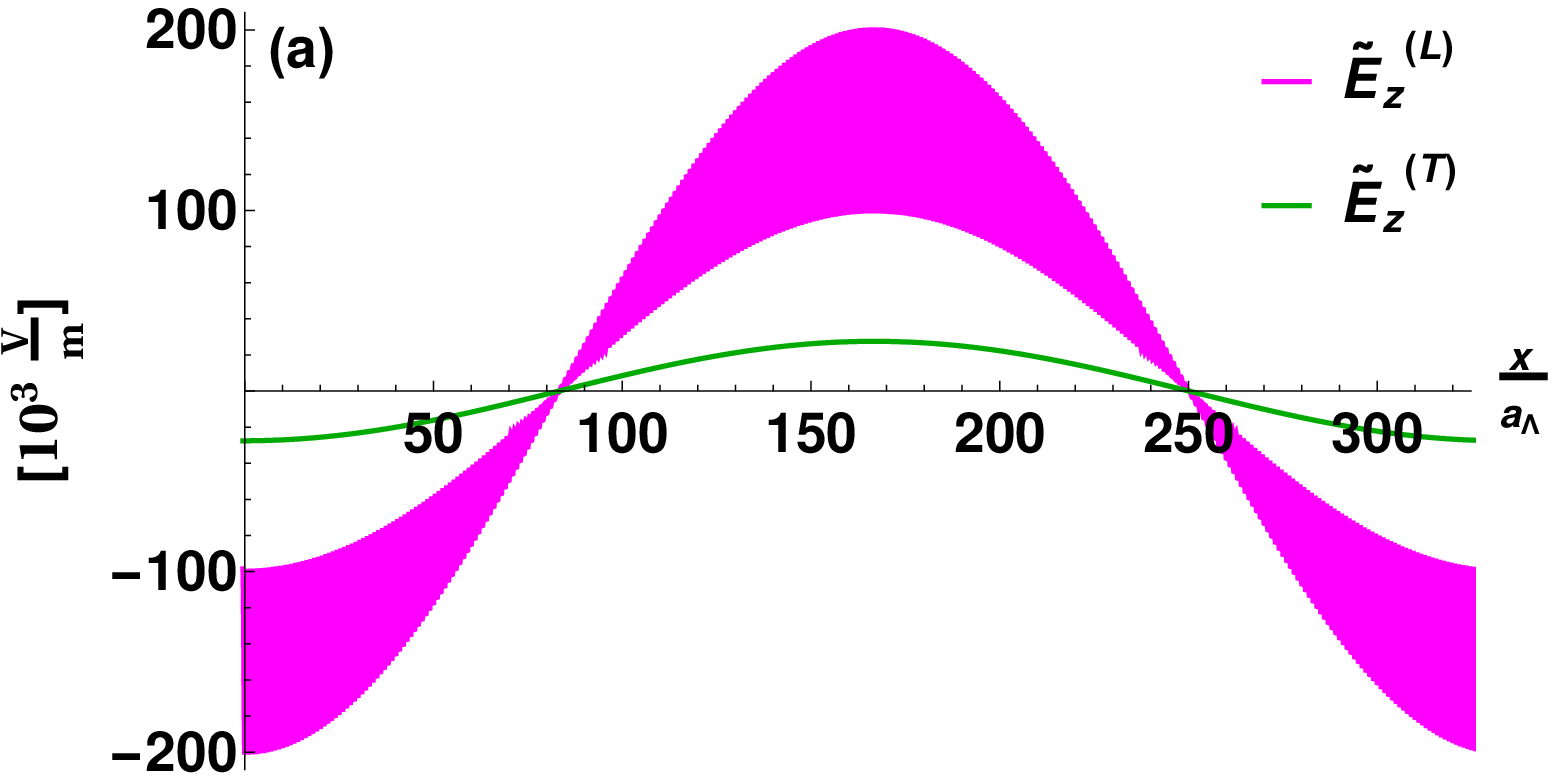}%
\end{minipage}

\vspace{1cm}

\begin{minipage}[t]{1\columnwidth}%
\includegraphics[scale=0.7]{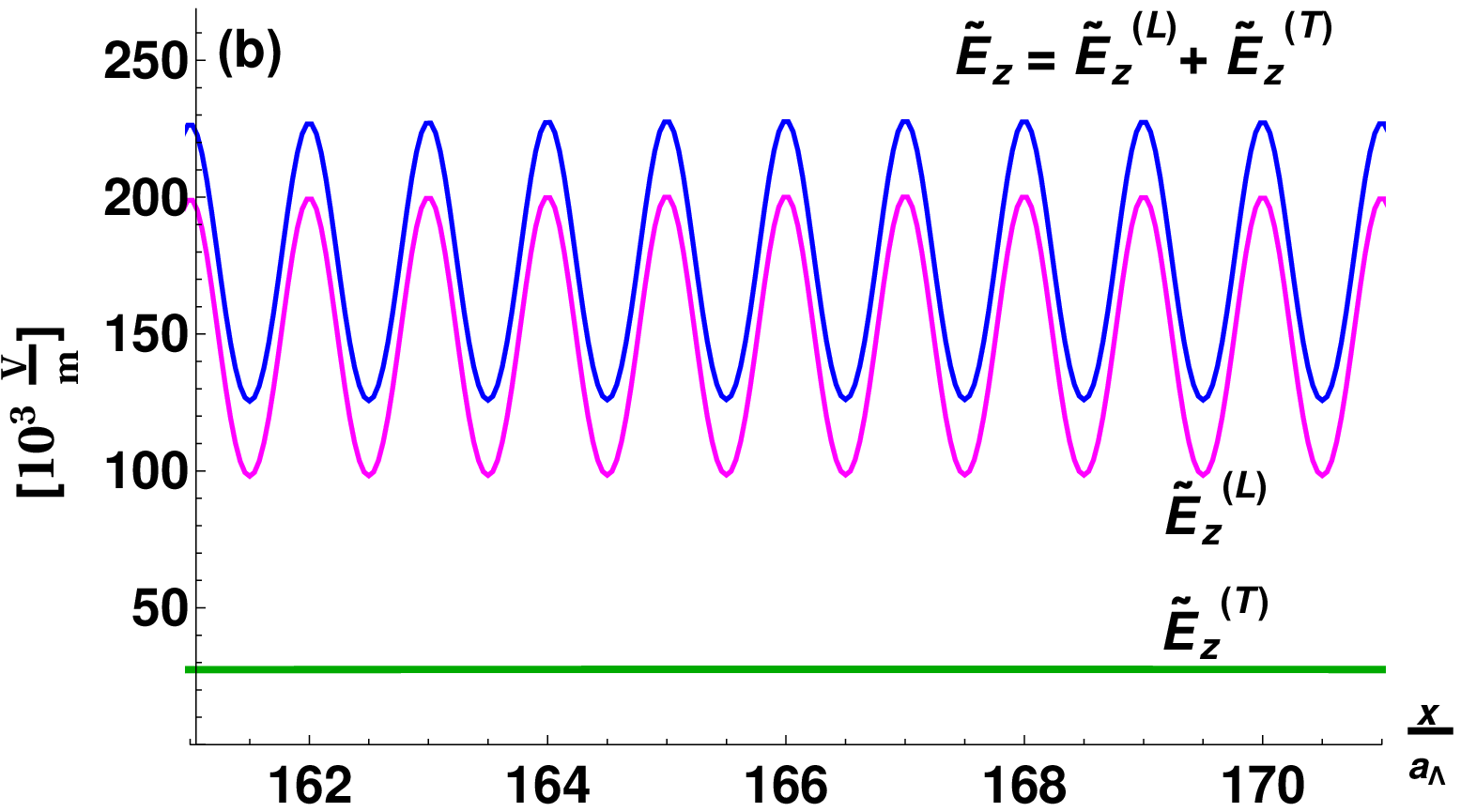}%
\end{minipage}

\caption{\label{fig:local_field_projections}(a) The spatial variation of the
z-component $\tilde{E}_{z}^{(L)}\left(\mathbf{r},\omega\right)$ of
the longitudinal part of the microscopic local electric field (magenta)
and the spatial variation of the z-component $\tilde{E}_{z}^{(T)}\left(\mathbf{r},\omega\right)$
of the transversal part of the microscopic local electric field (green)
corresponding to the same parameters as in Fig. \ref{fig:Local_Macro_Fields}.
The local field $\tilde{E}_{z}\left(\mathbf{r},\omega\right)=\tilde{E}_{z}^{\left(L\right)}\left(\mathbf{r},\omega\right)+\tilde{E}_{z}^{\left(T\right)}\left(\mathbf{r},\omega\right)$
is for a crystal of high particle density (high refractive index),
as depicted in (b), dominated by its static longitudinal (dipole)
part $\tilde{E}_{z}^{\left(L\right)}\left(\mathbf{r},\omega\right)$,
while the tranversal part $\tilde{E}_{z}^{\left(T\right)}\left(\mathbf{r},\omega\right)$
is distinctly smaller. }

\end{figure}

\begin{figure}
\begin{minipage}[t]{1\columnwidth}%
\includegraphics[scale=0.55]{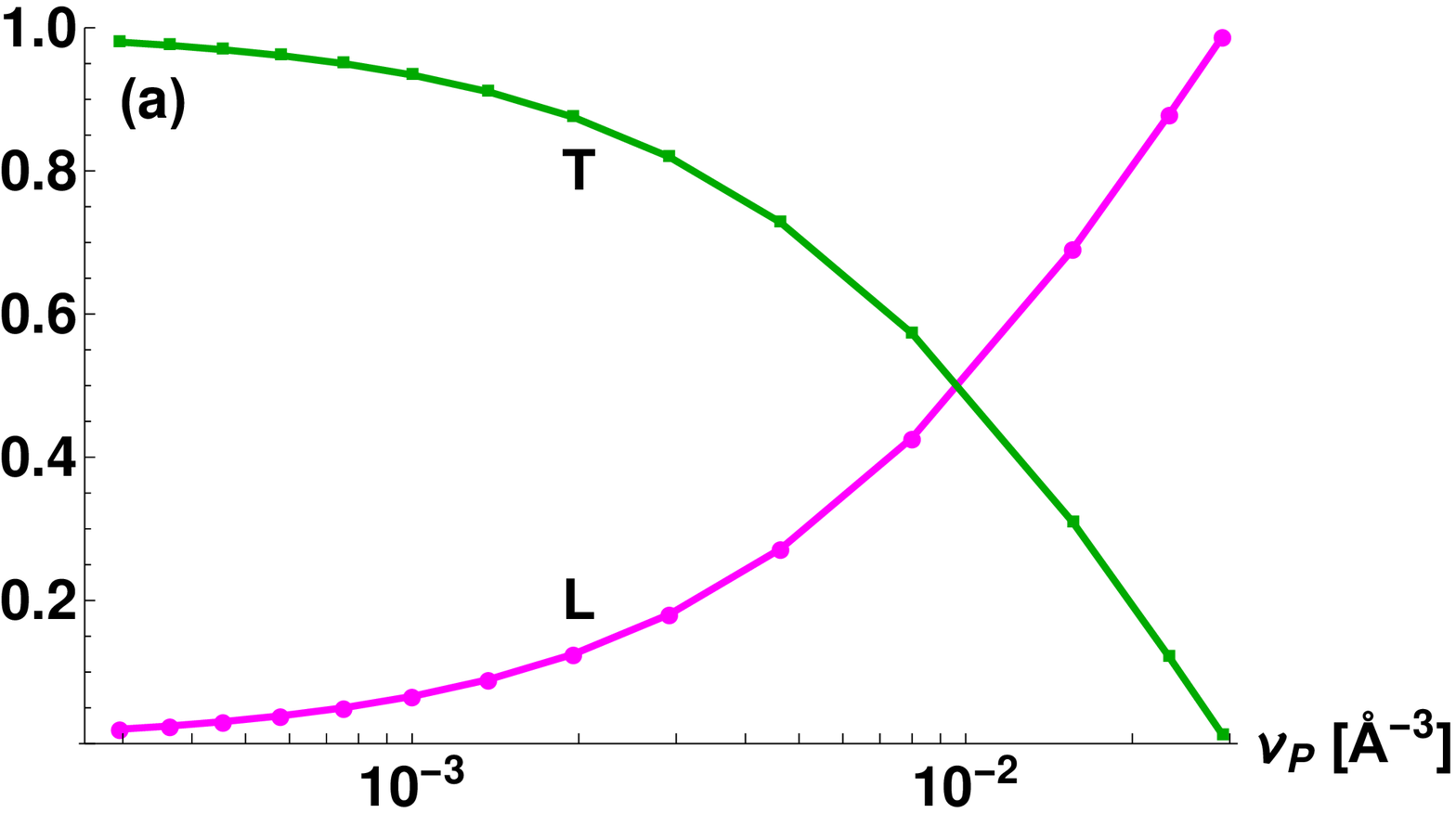}%
\end{minipage}

\vspace{1cm}

\begin{minipage}[t]{1\columnwidth}%
\includegraphics[scale=0.55]{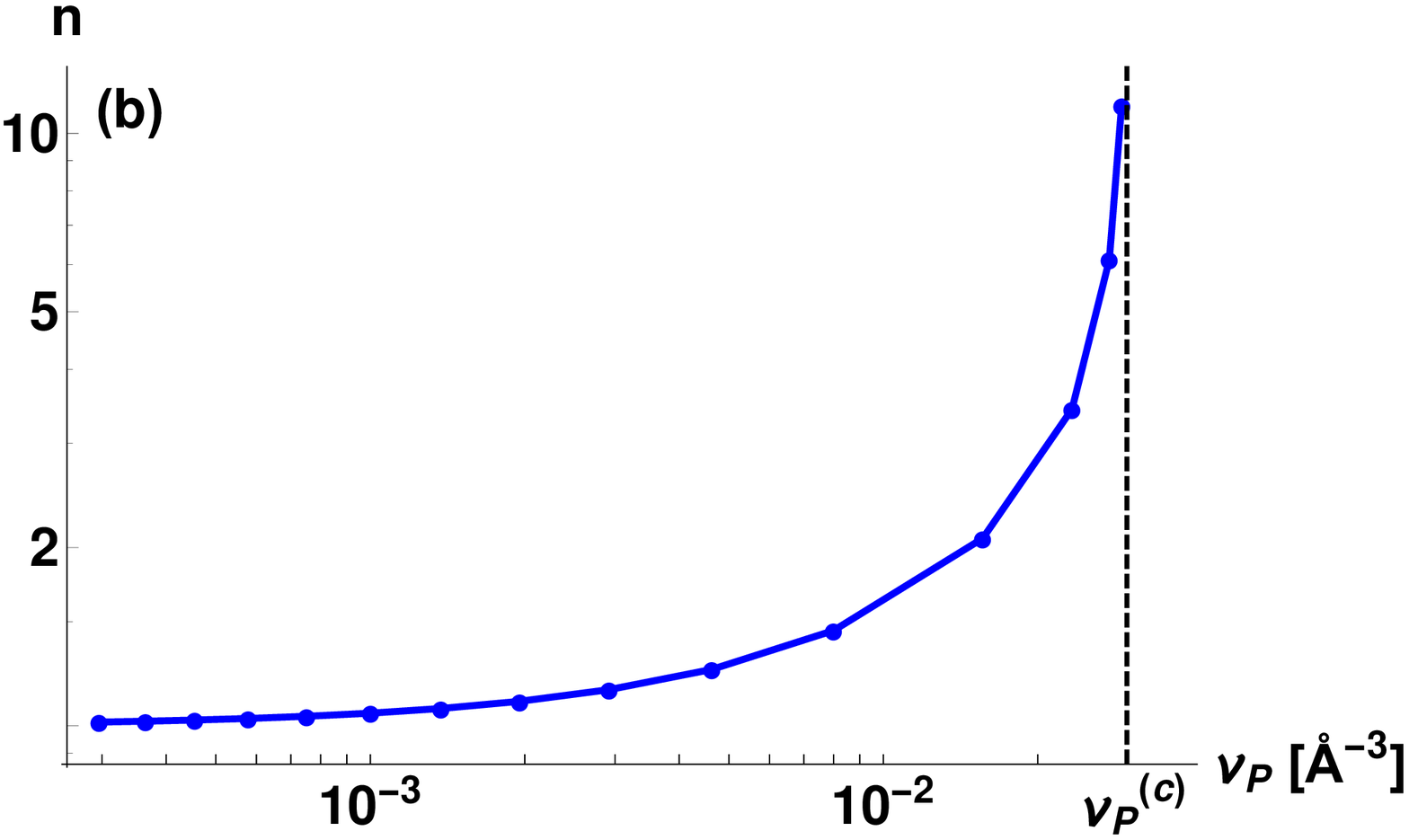}%
\end{minipage}

\caption{\label{fig:field_contributions_and_index}(a) Plot of the ratio of
maximum field strengths $\frac{\max_{x}\left|\tilde{E}_{z}^{(L,T)}\left(\mathbf{r}\left(x\right),\omega\right)\right|}{\max_{x}\left|\tilde{E}_{z}\left(\mathbf{r}\left(x\right),\omega\right)\right|}$
vs. particle density $\nu_{P}$ with parameters like in Fig.\ref{fig:Local_Macro_Fields}.
Near to the border of instability at $\nu=\nu_{P}^{\left(c\right)}$
the transversal (radiative) amplitude $\tilde{E}_{z}^{(T)}\left(\mathbf{r},\omega\right)$
is strongly suppressed while the longitudinal amplitude $\tilde{E}_{z}^{\left(L\right)}\left(\mathbf{r},\omega\right)$
becomes large. Conversely, at low particle density $\nu_{P}\rightarrow0$
the longitudinal amplitude $\tilde{E}_{z}^{\left(L\right)}\left(\mathbf{r},\omega\right)$
vanishes while the transversal amplitude essentially coincides with
the field amplitude of the external field. (b) The variation of the
index of refraction $n\left(\omega\right)$ vs. the density $\nu_{P}$
of polarizable atoms (ions, molecules) in a dielectric crystal. At
the borderline of stability the index of refraction $n\left(\omega\right)$
displays a singularity as $\nu_{P}$ approaches (from below) the critical
density $\nu_{P}^{\left(c\right)}$.}

\end{figure}

To elucidate the nature of the microscopic local electric field in
(\ref{eq:microscopic local field for incident plane wave with small q})
and (\ref{eq:eq:microscopic local field Fourier series representation})
respectively, let us decompose it into\emph{ longitudinal} and \emph{transversal
}parts. With (\ref{eq:eq:microscopic local field Fourier series representation})
we write now 
\begin{eqnarray}
\tilde{E}_{a}^{\left(L,T\right)}\left(\mathbf{r},\omega\right) & = & \sum_{a'}\int d^{3}r'\Pi_{a,a'}^{\left(L,T\right)}\left(\mathbf{r}'\right)\tilde{E}_{a'}\left(\mathbf{r}-\mathbf{r}',\omega\right)\\
 & = & \sum_{a'}\bar{\Pi}_{a,a'}^{\left(L,T\right)}\left(\mathbf{q}\right)\mathcal{\bar{E}}_{ext,a'}\left(\mathbf{q},\omega\right)e^{i\mathbf{q}\cdot\mathbf{r}}\nonumber \\
 &  & +\sum_{a',a'',a'''}\frac{1}{\left|C_{\Lambda}\right|}\sum_{\mathbf{G}\in\Lambda^{-1}}e^{i\left(\mathbf{q}+\mathbf{G}\right)\cdot\mathbf{r}}\bar{\Pi}_{a,a'}^{\left(L,T\right)}\left(\mathbf{q}+\mathbf{G}\right)\mathcal{\bar{G}}_{a',a''}(\mathbf{q}+\mathbf{G},\omega)\left[\bar{K}_{\Lambda}\left(\mathbf{G},\mathbf{q},\omega\right)\right]_{a'',a'''}\mathcal{\bar{E}}_{ext,a'''}\left(\mathbf{q},\omega\right).\nonumber 
\end{eqnarray}
Taking into account (\ref{eq: Fourier transform of 3x3 matrix propagator})
we have
\begin{eqnarray}
\sum_{a'}\bar{\Pi}_{a,a'}^{\left(L\right)}\left(\mathbf{q}+\mathbf{G}\right)\mathcal{\bar{G}}_{a',a''}(\mathbf{q}+\mathbf{G},\omega) & = & -\frac{\left(\mathbf{q}+\mathbf{G}\right)_{a}\left(\mathbf{q}+\mathbf{G}\right)_{a''}}{\left|\mathbf{q}+\mathbf{G}\right|^{2}}\\
\sum_{a'}\bar{\Pi}_{a,a'}^{\left(T\right)}\left(\mathbf{q}+\mathbf{G}\right)\mathcal{\bar{G}}_{a',a''}(\mathbf{q}+\mathbf{G},\omega) & = & \frac{\omega^{2}}{c^{2}}\frac{\delta_{a,a''}-\frac{\left(\mathbf{q}+\mathbf{G}\right)_{a}\left(\mathbf{q}+\mathbf{G}\right)_{a''}}{\left|\mathbf{q}+\mathbf{G}\right|^{2}}}{\left|\mathbf{\mathbf{q}}+\mathbf{G}\right|^{2}-\frac{\omega^{2}}{c^{2}}-i0^{+}}.
\end{eqnarray}
The searched-for longitudinal and transversal parts of the microscopic
local electric field are thus explicitely determined: 
\begin{eqnarray}
\tilde{E}_{a}^{\left(L\right)}\left(\mathbf{r},\omega\right) & = & \mathcal{\bar{E}}_{ext,a}^{\left(L\right)}\left(\mathbf{q},\omega\right)e^{i\mathbf{q}\cdot\mathbf{r}}\label{eq: longitudinal microscopic field}\\
 &  & -\sum_{a'',a'''}\frac{1}{\left|C_{\Lambda}\right|}\sum_{\mathbf{G}\in\Lambda^{-1}}e^{i\left(\mathbf{q}+\mathbf{G}\right)\cdot\mathbf{r}}\frac{\left(\mathbf{q}+\mathbf{G}\right)_{a}\left(\mathbf{q}+\mathbf{G}\right)_{a''}}{\left|\mathbf{q}+\mathbf{G}\right|^{2}}\left[\bar{K}_{\Lambda}\left(\mathbf{G},\mathbf{q},\omega\right)\right]_{a'',a'''}\mathcal{\bar{E}}_{ext,a'''}\left(\mathbf{q},\omega\right)\nonumber \\
\tilde{E}_{a}^{\left(T\right)}\left(\mathbf{r},\omega\right) & = & \mathcal{\bar{E}}_{ext,a}^{\left(T\right)}\left(\mathbf{q},\omega\right)e^{i\mathbf{q}\cdot\mathbf{r}}\label{eq:transversal microscopic field}\\
 &  & +\sum_{a'',a'''}\frac{1}{\left|C_{\Lambda}\right|}\sum_{\mathbf{G}\in\Lambda^{-1}}e^{i\left(\mathbf{q}+\mathbf{G}\right)\cdot\mathbf{r}}\frac{\omega^{2}}{c^{2}}\frac{\delta_{a,a''}-\frac{\left(\mathbf{q}+\mathbf{G}\right)_{a}\left(\mathbf{q}+\mathbf{G}\right)_{a''}}{\left|\mathbf{q}+\mathbf{G}\right|^{2}}}{\left|\mathbf{\mathbf{q}}+\mathbf{G}\right|^{2}-\frac{\omega^{2}}{c^{2}}-i0^{+}}\left[\bar{K}_{\Lambda}\left(\mathbf{G},\mathbf{q},\omega\right)\right]_{a'',a'''}\mathcal{\bar{E}}_{ext,a'''}\left(\mathbf{q},\omega\right)\nonumber 
\end{eqnarray}
\textcolor{black}{The spatial variations of the transversal amplitude
$\tilde{E}_{a}^{\left(T\right)}\left(\mathbf{r},\omega\right)$ and
the longitudinal amplitude $\tilde{E}_{a}^{\left(L\right)}\left(\mathbf{r},\omega\right)$
along the same path $\mathbf{r}\left(x\right)$ (and the same parameters
as in Fig. \ref{fig:Local_Macro_Fields}) is displayed in Fig. \ref{fig:local_field_projections}}.
It is important to realize that the longitudinal component $\tilde{E}_{a}^{\left(L\right)}\left(\mathbf{r},\omega\right)$
increases rapidly as the density of polarizable atoms (ions, molecules)
in the probe volume increases, see Fig.\ref{fig:field_contributions_and_index}.

\subsection*{Photonic Bandstructure\label{sec:Photonic-Bandstructure}}

\begin{figure}
\begin{minipage}[t]{1\columnwidth}%
\includegraphics[scale=0.4]{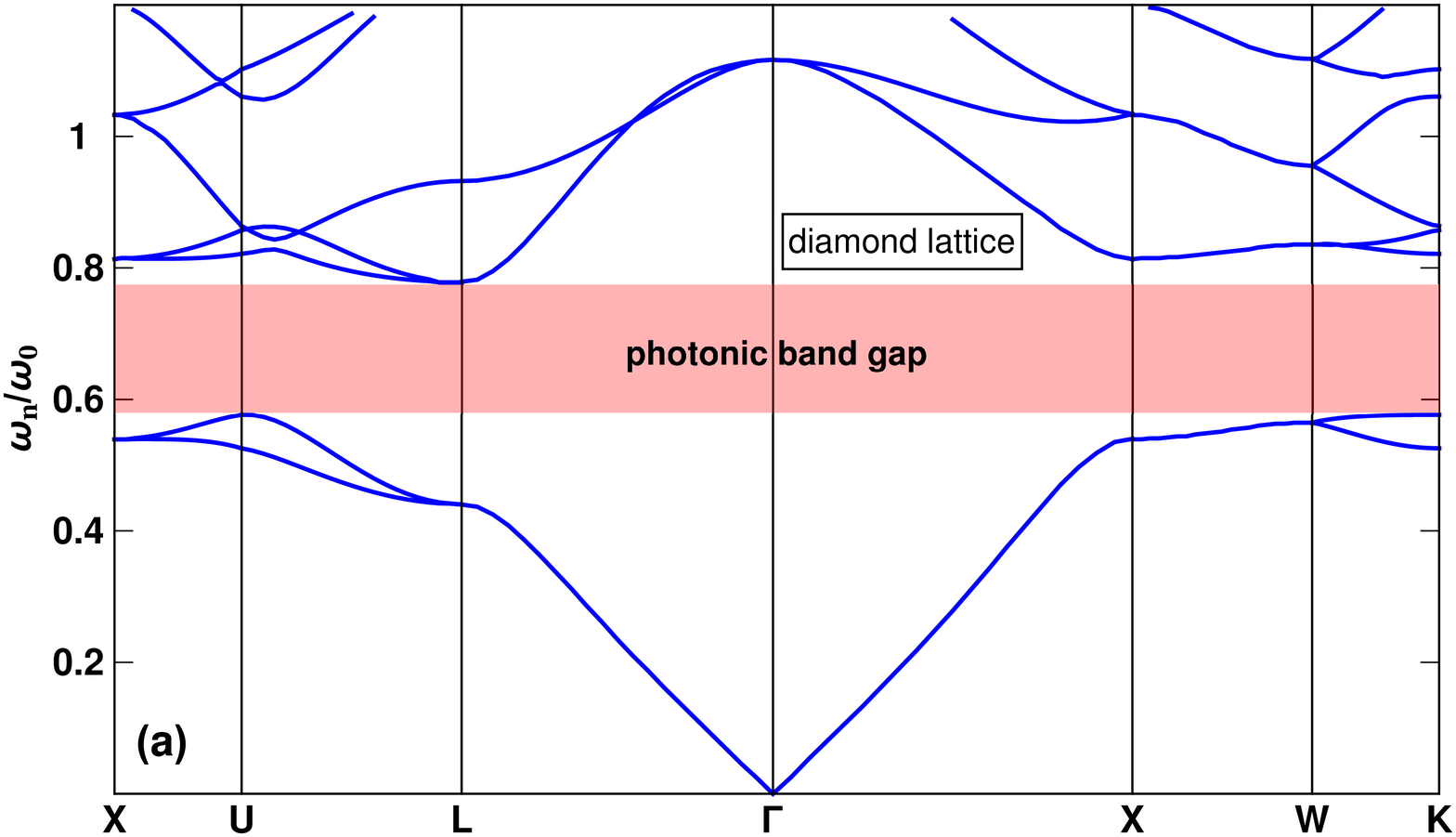}%
\end{minipage}

\vspace{1cm}

\begin{minipage}[t]{1\columnwidth}%
\includegraphics[scale=0.4]{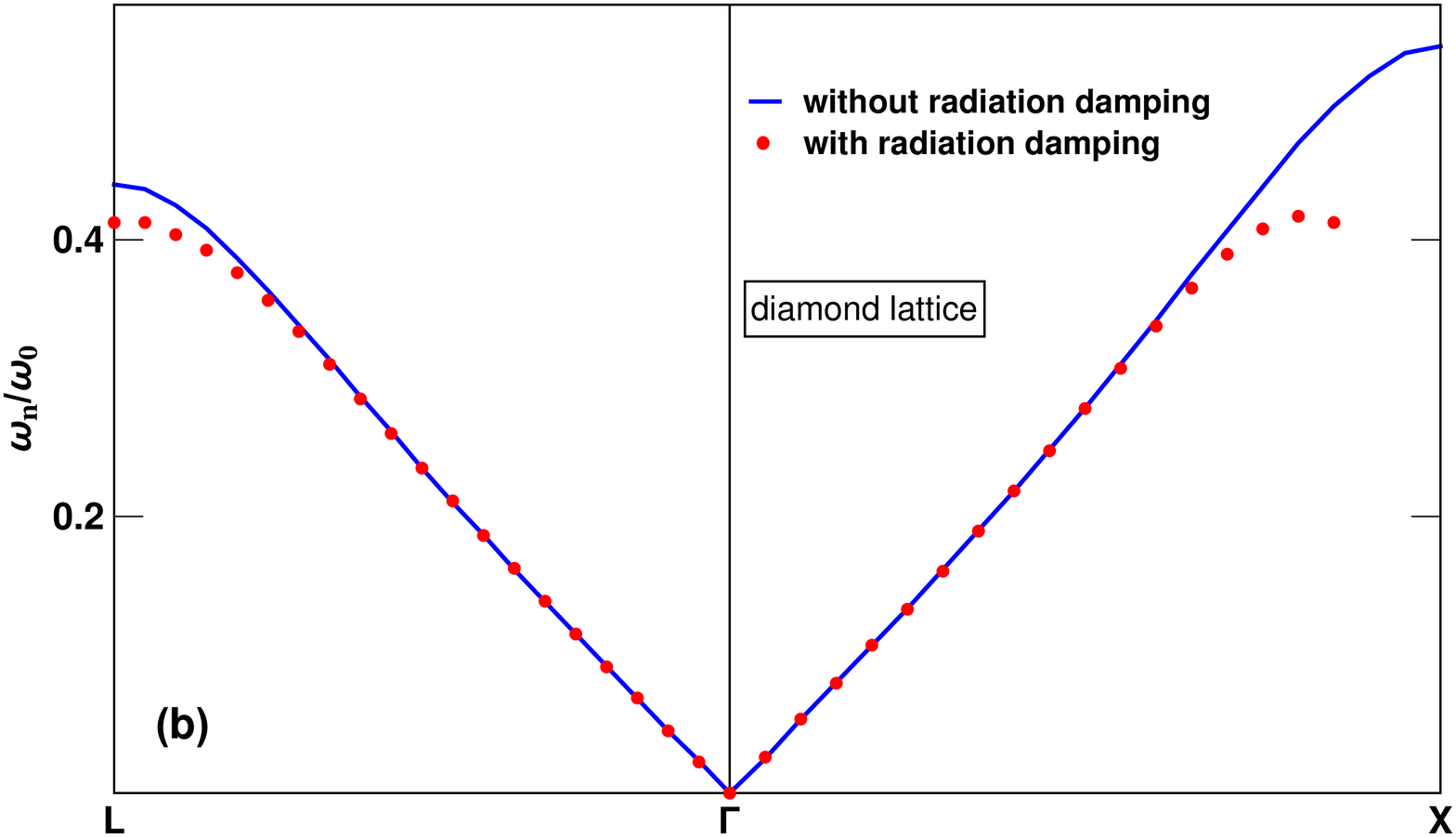}%
\end{minipage}

\caption{\label{fig:diamond-lattice}(a) Photonic band structure $\omega_{n}\left(\mathbf{q}\right)$
in units of $\omega_{0}=\frac{2\pi c}{a_{\Lambda}}$ as calculated
from (\ref{eq: eigenvalue problem photonic bandstructure}) for a
diamond lattice with wave vector $\mathbf{q}$ orientated along various
symmetry lines of the Brillouin zone choosing a lattice constant $a_{\Lambda}=6\text{\AA}$
and assuming a static electronic polarizability $\frac{\alpha_{0}}{4\pi\varepsilon_{0}}=3{\text{\AA}}^{3}$.
(b) Taking radiation damping into account, the meaning of the red
dots being explained in the supplemental material \citep{Supplementary},
the photonic band structure $\omega_{n}\left(\mathbf{q}\right)$ of
a dielectric crystalline material ceases to make sense approaching
the boundary of the Brillouin zone, where $\omega_{n}\left(\mathbf{q}\right)$
enters into the soft x-ray regime. }
\end{figure}

If no external field was incident, i.e. for $\tilde{\mathfrak{e}}_{ext,a}^{\left(j\right)}\left(\mathbf{q},\omega\right)\equiv0$,
a non trivial field amplitude $\tilde{\mathfrak{e}}_{a}^{\left(j\right)}\left(\mathbf{q},\omega\right)$
solving the homogenous system of equations (\ref{eq: system of 3Mx3M equations for amplitudes of local field})
is obviously identical to an eigenvector $\tilde{\mathfrak{v}}_{a,n}^{\left(j\right)}\left(\mathbf{q},\omega\right)$
associated with the \emph{special} eigenvalue
\begin{equation}
\gamma_{n}\left(\mathbf{q},\omega\right)=1\label{eq: eigenvalue unity}
\end{equation}
of the eigenvalue problem 
\begin{equation}
\sum_{j'}\sum_{a'}\Gamma_{a,a'}^{\left(j,j'\right)}\left(\mathbf{k},\omega\right)\tilde{\mathfrak{v}}_{a',n}^{\left(j'\right)}\left(\mathbf{q},\omega\right)=\gamma_{n}\left(\mathbf{q},\omega\right)\tilde{\mathfrak{v}}_{a,n}^{\left(j\right)}\left(\mathbf{q},\omega\right).\label{eq: eigenvalue problem photonic bandstructure}
\end{equation}
The dispersion relation of photons, i.e. the photonic bandstructure
$\omega_{n}\left(\mathbf{q}\right)$, can now be readily determined
for any number $M$ of basis atoms inside the unit cell $C_{\Lambda}$
by first solving (numerically) the eigenvalue problem (\ref{eq: eigenvalue problem photonic bandstructure})
for a given wave vector $\mathbf{q}\in C_{\varLambda^{-1}}$ as a
function of $\omega$, thus obtaining a family of $3M$ eigenvalue
curves $\gamma_{n}\left(\mathbf{q},\omega\right)$ vs. $\omega$,
and then solving (numerically) for $n=1,2,3,...3M$ the implicit equations
(\ref{eq: eigenvalue unity}) for the unknown $\omega$ so that 
\begin{equation}
\left[\gamma_{n}\left(\mathbf{q},\omega\right)-1\right]_{\omega\rightarrow\omega_{n}}=0.\label{eq: requirement determining photonic bandstructure}
\end{equation}
 Varying then the wavevector $\mathbf{q}$ along (widely) different
symmetry lines inside the Brillouin zone $C_{\Lambda^{-1}}$ various
pieces of the photonic bandstructure $\omega_{n}\left(\mathbf{q}\right)$
emerge, as is exemplarily displayed for the diamond lattice ($M=2$)
in Fig. \ref{fig:diamond-lattice}(a). Like electrons moving in a
periodic potential also electromagnetic waves propagating in a crystal
are governed by a bandstructure, for example \citep{Leung1990,Zhang1990,Soezueer1993}.
While the wave function for electrons (discarding spin-orbit forces
and Zeeman splitting ) is a scalar, propagating electromagnetic waves
are vectorfields. Incident lightsignals composed of frequencies within
an omni-directional band gap will be reflected from such a crystal
irrespective of the light source being polarized or unpolarized, which
is interesting for technical applications, for instance dielectric
mirrors, filters or antenna-substrates.

Further examples of photonic band structures calculated from (\ref{eq: eigenvalue problem photonic bandstructure})
and (\ref{eq: requirement determining photonic bandstructure}) with
our method, exemplarily for sc-, fcc- and bcc-lattices ($M=1$), are
given in supplemental material \citep{Supplementary}. Our results
comply well with calculations carried out within the frame of the
Fano-Hopfield model \citep{Klugkist2006,Antezza2009} in the case
of weak coupling of the oscillators. Earlier calculations for super-lattices
on the basis of the macroscopic Maxwell equations \citep{Leung1990,Zhang1990,Soezueer1993},
carried out in a basis of plane waves $e^{i\left(\mathbf{q}+\mathbf{G}\right)\mathbf{r}}$,
require for $\mathbf{q}$ fixed a large number $N$ of reciprocal
lattice vectors $\mathbf{G}$ to be taken into account, for accurate
computations typically $N\geq2000$ \citep{Hermann2001}, thus leading
to a huge $3N\times3N$ matrix eigenvalue problem for the modes and
mode freqencies. Unfortunately, these calculations suffer from an
inconsistency, as a number $N$ of spurious zero eigenvalue modes
needs explicit elimination by an ad hoc transversality constraint.
Because in our approach the extension of the scatterers is tiny compared
to the lattice constant, our results cease agreement with theirs in
certain details of the photonic bandstructure at high photon frequency,
while for optical frequencies and below our results are in full agreement
with theirs. Regarding the calculational cost of our approach: in
the case of sc-, fcc- and bcc-lattices with one atom in the unit cell
we solve (for each $\mathbf{q}$-vector) a $3\times3$ eigenvalue
problem, and in the case of diamond with two atoms in the unit cell
then a $6\times6$ eigenvalue problem. Parameters of interest to propagation
of light pulses, for instance group velocity and effective photon
mass, can be determined conveniently using $k\cdot p$-perturbation
theory \citep{Hermann2001}, a method often employed in solid state
electronic band structure theory \citep{Ashcroft1981}. 

Finally, it should be considered that higher bands $n>1$ of the photonic
band structure $\omega_{n}\left(\mathbf{q}\right)$ in real crystalline
materials are not credibly calculable. While the concept of a photonic
band structure exhibiting many band branches $\omega_{n}\left(\mathbf{q}\right)$
certainly applies for (artificial) superlattice structures with large
mesoscopic lattice constant $a_{\Lambda}\gg a$, it can be accepted
only with reserve for a real dielectric material. This is because
for a microscopic lattice constant $a_{\Lambda}\simeq a$ photon frequencies
above $\omega\geq\omega_{\Lambda}=c\times\frac{\pi}{a_{\Lambda}}$
are far and beyond the ultra-violet. In this case the effects of radiation
damping, as represented by the imaginary part of the lattice sums,
\begin{equation}
\mathtt{Im}\left[\zeta_{\varLambda}^{\left(0\right)}(\mathbf{q},\omega)\right]_{aa'}=-\frac{1}{6\pi}\left(\frac{\omega}{c}\right)^{3}\delta_{a,a'},\label{eq: Im Zeta_0}
\end{equation}
 can no longer be neglected. For a derivation of (\ref{eq: Im Zeta_0})
see supplemental material \citep{Supplementary}. In Fig.\ref{fig:diamond-lattice}(b)
the effect of taking into account radiation damping is exemplarily
shown for the diamond lattice ($M=2$). The corresponding plots revealing
the influence of radiation damping for the sc-, fcc- and bcc-lattices
($M=1$) we also present in \citep{Supplementary}. Of course, if
the wavelength of the external electromagnetic field is ultra short
then the point dipole ansatz (\ref{eq: phenomenological dielectric kernel})
for the dielectric susceptibility should be extended to include also
magnetic dipoles and electric quadrupoles on equal footing \citep{Raab2005}.

\section{The Dielectric Tensor of Macroscopic Electrodynamics \label{sec:Macroscopic-Electric-Field-and-dielectric-function}}

\begin{figure}
\includegraphics[scale=0.55]{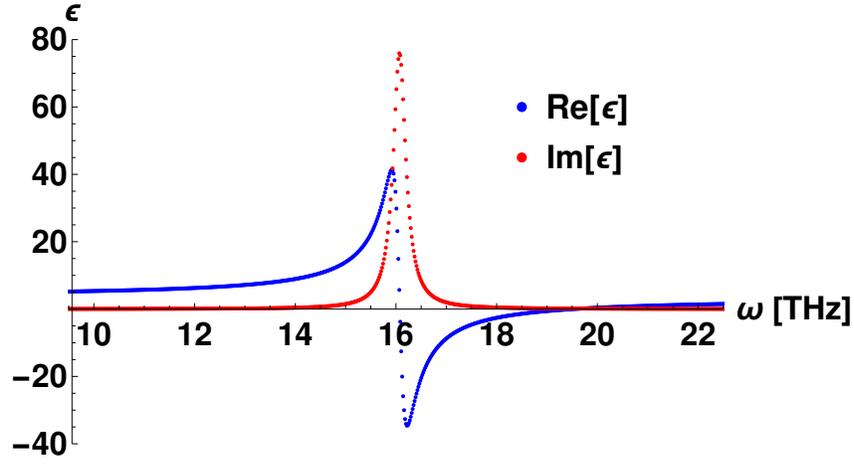}

\caption{\label{fig:Lyddane-Sachs-Teller}Plot of real and imaginary parts
of dielectric function $\varepsilon_{\Lambda}\left(\mathbf{q}=\mathbf{0},\omega\right)\equiv\varepsilon_{\Lambda}\left(\omega\right)$
as calculated from (\ref{eq: macroscopic dielectric function arbitrary M})
for $CsI$ with microscopic polarizabilities with the parameters from
Table \ref{fit}. The ionic polarizability with frequency dependence
given by (\ref{eq: isotropic Lorentz polarizability off diagonal})
also includes a small damping parameter $\gamma>0$. The roots of
$\text{Re}\left[\varepsilon_{\Lambda}\left(\omega\right)\right]$
are: $\omega_{T}=16.08$THz and $\omega_{L}=19.53$THz, corresponding
to frequencies of the transversel and longitudinal optical modes respectively.
The zeros of $\text{Re}\left[\varepsilon_{\Lambda}\left(\omega\right)\right]$
obey to the Lyddane-Sachs-Teller relation \citep{Lyddane1941}, connecting
the value of the static dielectric function $\varepsilon_{\Lambda}^{(0)}=4.45$
to its value $\varepsilon_{\Lambda}^{(\infty)}\equiv\varepsilon_{\Lambda}\left(\omega_{T}\cdot10^{2}\right)=3.05$
above and beyond all optical phonon frequencies: $\frac{\omega_{L}^{2}}{\omega_{T}^{2}}=\frac{\varepsilon_{\Lambda}^{(0)}}{\varepsilon_{\Lambda}^{(\infty)}}$
.}
\end{figure}

The \emph{macroscopic} electric field $\mathcal{\tilde{E}}_{a}\left(\mathbf{r},\omega\right)$
inside the probe we conceive as a low-pass filter applied to the Fourier-transformation
$\bar{E}_{a}\left(\mathbf{q},\omega\right)$ of the spatially rapidly
varying microscopic local electric field $\tilde{E}_{a}\left(\mathbf{r},\omega\right)$.
Introducing a cut-off wavenumber $q_{c}$ so that $q<q_{c}$ implies
$\mathbf{q}\in C_{\Lambda^{-1}}$, then
\begin{eqnarray}
\mathcal{\tilde{E}}_{a}\left(\mathbf{r},\omega\right) & = & \int\frac{d^{3}q}{\left(2\pi\right)^{3}}e^{i\mathbf{q}\cdot\mathbf{r}}\Theta\left(q_{c}-\left|\mathbf{q}\right|\right)\bar{E}_{a}\left(\mathbf{q},\omega\right)\label{eq: low pass filtered microscopic electric field}\\
\bar{E}_{a}\left(\mathbf{q},\omega\right) & = & \frac{1}{\left|\Omega_{P}\right|}\int_{\Omega_{P}}d^{3}re^{-i\mathbf{q}\cdot\mathbf{r}}\tilde{E}_{a}\left(\mathbf{r},\omega\right).\nonumber 
\end{eqnarray}
Likewise, the \emph{macroscopic} polarization $\mathcal{\tilde{P}}_{a}\left(\mathbf{r},\omega\right)$
inside the probe we conceive as a low-pass filter applied to the Fourier-transformation
$\bar{P}_{a}\left(\mathbf{q},\omega\right)$ of the microscopic polarization
$\tilde{P}_{a}\left(\mathbf{r},\omega\right)$, as defined in (\ref{eq:polarization vs local field}):
\begin{eqnarray}
\mathcal{\tilde{P}}_{a}\left(\mathbf{r},\omega\right) & = & \int\frac{d^{3}q}{\left(2\pi\right)^{3}}e^{i\mathbf{q\cdot}\mathbf{r}}\Theta\left(q_{c}-\left|\mathbf{q}\right|\right)\bar{P}_{a}\left(\mathbf{q},\omega\right)\label{eq: low pass filtered microscopic polarization}\\
\bar{P}_{a}\left(\mathbf{q},\omega\right) & = & \frac{1}{\left|\Omega_{P}\right|}\int_{\Omega_{P}}d^{3}re^{-i\mathbf{q}\cdot\mathbf{r}}\tilde{P}_{a}\left(\mathbf{r},\omega\right)\nonumber 
\end{eqnarray}
Thus in the long wavelength limit the Fourier components $\bar{\mathcal{E}}_{a}\left(\mathbf{q},\omega\right)$
of the macroscopic field $\tilde{\mathcal{E}}_{a}\left(\mathbf{r},\omega\right)$
coincide with those of the microscopic local field, and the Fourier
components $\mathcal{\bar{P}}_{a}\left(\mathbf{q},\omega\right)$
of the macroscopic polarization $\mathcal{\tilde{P}}_{a}\left(\mathbf{r},\omega\right)$
coincide with those of the microscopic polarization: 
\begin{eqnarray}
\mathcal{\bar{E}}_{a}\left(\mathbf{q},\omega\right) & = & \Theta\left(q_{c}-\left|\mathbf{q}\right|\right)\bar{E}_{a}\left(\mathbf{q},\omega\right)\\
\mathcal{\bar{P}}_{a}\left(\mathbf{q},\omega\right) & = & \Theta\left(q_{c}-\left|\mathbf{q}\right|\right)\bar{P}_{a}\left(\mathbf{q},\omega\right)
\end{eqnarray}
Restricting to $\left|\mathbf{q}\right|<q_{c}$ we now define the
dielectric $3\times3$ tensor $\left[\varepsilon_{\Lambda}\left(\mathbf{q},\omega\right)\right]_{aa'}$
of a crystalline dielectric by requiring the macroscopic polarization
being proportional to the macroscopic electric field:
\begin{equation}
\mathcal{\bar{P}}_{a}\left(\mathbf{q},\omega\right)=\varepsilon_{0}\sum_{a'\in\left\{ x,y,z\right\} }\left[\varepsilon_{\Lambda}\left(\mathbf{q},\omega\right)-I\right]_{aa'}\mathcal{\bar{E}}_{a'}\left(\mathbf{q},\omega\right)\label{eq: implicit definition dielectric function}
\end{equation}
Insertion of (\ref{eq:eq:microscopic local field Fourier series representation})
gives an explicit expression determining the Fourier amplitude of
the microscopic local electric field:
\begin{eqnarray}
\bar{E}_{a}\left(\mathbf{k},\omega\right) & = & \frac{1}{\left|\Omega_{P}\right|}\int_{\Omega_{P}}d^{3}re^{-i\mathbf{k}\cdot\mathbf{r}}\tilde{E}_{a}\left(\mathbf{r},\omega\right)\label{eq: Fourier transform microscopic local field}\\
 & = & \mathcal{\bar{E}}_{ext,a}\left(\mathbf{q},\omega\right)\delta_{\mathbf{k},\mathbf{q}}+\sum_{a'',a'''}\frac{1}{\left|C_{\Lambda}\right|}\sum_{\mathbf{G}'\in\Lambda^{-1}}\delta_{\mathbf{k},\mathbf{q}+\mathbf{G}'}\mathcal{\bar{G}}_{a,a''}(\mathbf{q}+\mathbf{G}',\omega)\left[\bar{K}_{\Lambda}\left(\mathbf{G}',\mathbf{q},\omega\right)\right]_{a'',a'''}\mathcal{\bar{E}}_{ext,a'''}\left(\mathbf{q},\omega\right)\nonumber 
\end{eqnarray}
 Decomposing $\mathbf{k}=\mathbf{k}_{0}+\mathbf{G}$ with $\mathbf{k}_{0}\in C_{\Lambda^{-1}}$
and $\mathbf{G}\in\Lambda^{-1}$ there holds for $\mathbf{q}\in C_{\Lambda^{-1}}$
and $\mathbf{G}'\in\Lambda^{-1}$ 
\begin{equation}
\delta_{\mathbf{k},\mathbf{q}+\mathbf{G}'}=\delta_{\mathbf{k}_{0},\mathbf{q}}\delta_{\mathbf{G},\mathbf{G}'}.
\end{equation}
Therefore
\begin{eqnarray}
\bar{E}_{a}\left(\mathbf{k},\omega\right) & = & \delta_{\mathbf{k}_{0},\mathbf{q}}\bar{E}_{a}\left(\mathbf{q}+\mathbf{G},\omega\right)\\
\bar{E}_{a}\left(\mathbf{q}+\mathbf{G},\omega\right) & = & \mathcal{\bar{E}}_{ext,a}\left(\mathbf{q},\omega\right)\delta_{\mathbf{G},\mathbf{0}}+\sum_{a'',a'''}\frac{1}{\left|C_{\Lambda}\right|}\mathcal{\bar{G}}_{a,a''}(\mathbf{q}+\mathbf{G},\omega)\left[\bar{K}_{\Lambda}\left(\mathbf{G},\mathbf{q},\omega\right)\right]_{a'',a'''}\mathcal{\bar{E}}_{ext,a'''}\left(\mathbf{q},\omega\right)\:.\label{eq:Fourier transform microscopic local field II}
\end{eqnarray}
Let us abbreviate for $\mathbf{G}=\mathbf{0}$:
\begin{equation}
\left[\bar{K}_{\Lambda}(\mathbf{q},\omega)\right]_{a,a''}\equiv\left[\bar{K}_{\Lambda}(\mathbf{0},\mathbf{q},\omega)\right]_{a,a''}=\sum_{1\leq j,j''\leq M}\left(\frac{1}{\varepsilon_{0}}\alpha\left(\mathbf{q},\omega\right)\circ\left[I-\Gamma\left(\mathbf{q},\omega\right)\right]^{-1}\right)_{a,a''}^{\left(j,j''\right)}\label{eq:  kernel K(G=00003D0,q,om)}
\end{equation}
Then for $\mathbf{q}\in C_{\Lambda^{-1}}$: 
\begin{eqnarray}
\bar{E}_{a}\left(\mathbf{q},\omega\right) & = & \sum_{a'}\left[I+\frac{1}{\left|C_{\Lambda}\right|}\bar{\mathcal{G}}(\mathbf{q},\omega)\circ\bar{K}_{\Lambda}(\mathbf{q},\omega)\right]_{a,a'}\mathcal{\bar{E}}_{ext,a'}\left(\mathbf{q},\omega\right)\label{eq: Fourier transform microscopic E- field}
\end{eqnarray}
On the other hand there holds keeping in mind the restriction $\mathbf{q}\in C_{\Lambda^{-1}}$:
\begin{eqnarray}
\bar{P}_{a}\left(\mathbf{q},\omega\right) & = & \varepsilon_{0}\frac{1}{\left|\Omega_{P}\right|}\int_{\Omega_{P}}d^{3}re^{-i\mathbf{q}\cdot\mathbf{r}}\int_{\Omega_{P}}d^{3}r'\sum_{a'}\tilde{\chi}_{aa'}\left(\mathbf{r},\mathbf{r}',\omega\right)\tilde{E}_{a'}\left(\mathbf{r}',\omega\right)\label{eq: Fourier transform microscopic polarization}\\
 & = & \frac{\varepsilon_{0}}{\left|C_{\Lambda}\right|}\sum_{a'}\left[\bar{K}_{\Lambda}(\mathbf{q},\omega)\right]_{a,a'}\mathcal{\bar{E}}_{ext,a'}\left(\mathbf{q},\omega\right)\nonumber 
\end{eqnarray}
Restricting to the long wavelength limit $\left|\mathbf{q}\right|<q_{c}$
and thus identifying $\mathcal{\bar{P}}_{a}\left(\mathbf{q},\omega\right)=\bar{P}_{a}\left(\mathbf{q},\omega\right)$
and $\mathcal{\bar{E}}_{a}\left(\mathbf{q},\omega\right)=\bar{E}_{a}\left(\mathbf{q},\omega\right)$,
and then combining (\ref{eq: Fourier transform microscopic E- field})
and (\ref{eq: implicit definition dielectric function}), a conditional
equation determining the macroscopic dielectric tensor $\varepsilon_{\Lambda}\left(\mathbf{q},\omega\right)$
is found 
\begin{eqnarray*}
\bar{\mathcal{P}}_{a}\left(\mathbf{q},\omega\right) & = & \varepsilon_{0}\sum_{a'}\left[\varepsilon_{\Lambda}\left(\mathbf{q},\omega\right)-I\right]_{aa'}\mathcal{\bar{E}}_{a'}\left(\mathbf{q},\omega\right)\\
 & = & \varepsilon_{0}\sum_{a',a''}\left[\varepsilon_{\Lambda}\left(\mathbf{q},\omega\right)-I\right]_{aa'}\left[I+\frac{1}{\left|C_{\Lambda}\right|}\bar{\mathcal{G}}(\mathbf{q},\omega)\circ\bar{K}_{\Lambda}(\mathbf{q},\omega)\right]_{a',a''}\mathcal{\bar{E}}_{ext,a''}\left(\mathbf{q},\omega\right)\\
 & \stackrel{!}{=} & \frac{\varepsilon_{0}}{\left|C_{\Lambda}\right|}\sum_{a'}\left[\bar{K}_{\Lambda}(\mathbf{q},\omega)\right]_{a,a'}\mathcal{\bar{E}}_{ext,a'}\left(\mathbf{q},\omega\right).
\end{eqnarray*}
Insisting both lines should be identical for any external field amplitude
$\mathcal{\bar{E}}_{ext,a}\left(\mathbf{q},\omega\right)$ immediately
leads (with help of elementary matrix algebra) to the identification
\begin{equation}
\varepsilon_{\Lambda}\left(\mathbf{q},\omega\right)-I=\frac{1}{\left|C_{\Lambda}\right|}\bar{K}_{\Lambda}(\mathbf{q},\omega)\circ\left[I+\frac{1}{\left|C_{\Lambda}\right|}\bar{\mathcal{G}}(\mathbf{q},\omega)\circ\bar{K}_{\Lambda}(\mathbf{q},\omega)\right]^{-1}.\label{eq: macroscopic dielectric function arbitrary M}
\end{equation}
This is a central result. The macroscopic dielectric tensor $\left[\varepsilon_{\Lambda}\left(\mathbf{q},\omega\right)\right]_{aa'}$
is solely determined by the lattice sums $\tilde{\zeta}_{\varLambda}(\mathbf{s},\mathbf{q},\omega)$
of the Bravais lattice $\Lambda$ under consideration and the individual
polarizations $\alpha_{a'',a'}\left(\boldsymbol{\eta}^{\left(j''\right)},\boldsymbol{\eta}^{\left(j'\right)},\omega\right)$
of the atoms (ions) inside the unit cell. As a test of the analytic
structure of the dielectric function $\varepsilon_{\Lambda}\left(\mathbf{q}=\mathbf{0},\omega\right)$
in the complex frequency domain we checked the Lyddane-Sachs-Teller
relation, see Fig. \ref{fig:Lyddane-Sachs-Teller}. In agreement with
general considerations under $\omega\rightarrow-\omega$ the real
part of $\varepsilon_{\Lambda}\left(\mathbf{q},\omega\right)$ is
an even function of $\omega$ and the imaginary part is an odd one. 

\subsection*{Macroscopic Electric Field}

\begin{figure}
\begin{minipage}[t]{1\columnwidth}%
\includegraphics[scale=0.8]{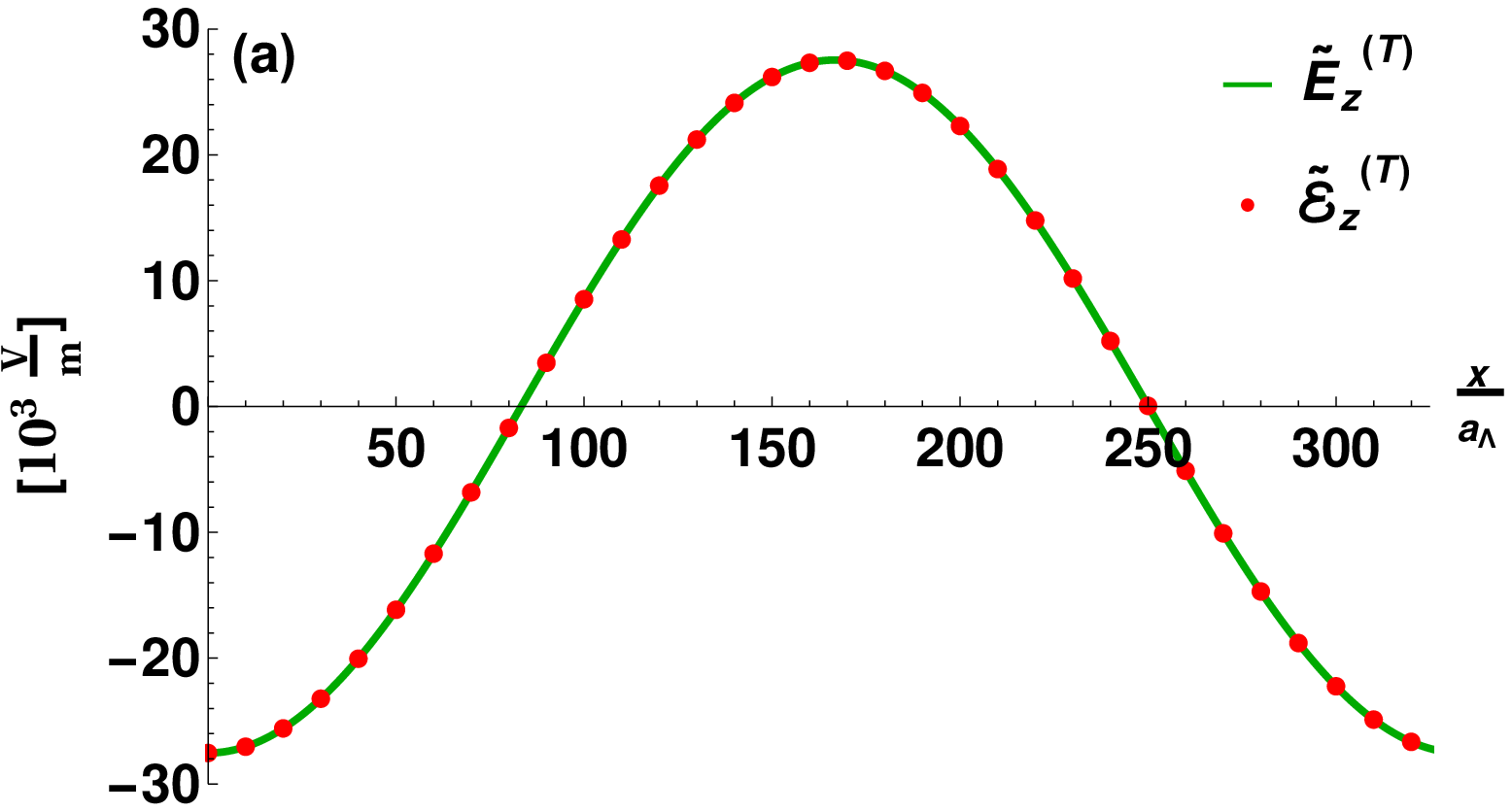}%
\end{minipage}

\vspace{1cm}

\begin{minipage}[t]{1\columnwidth}%
\includegraphics[scale=0.8]{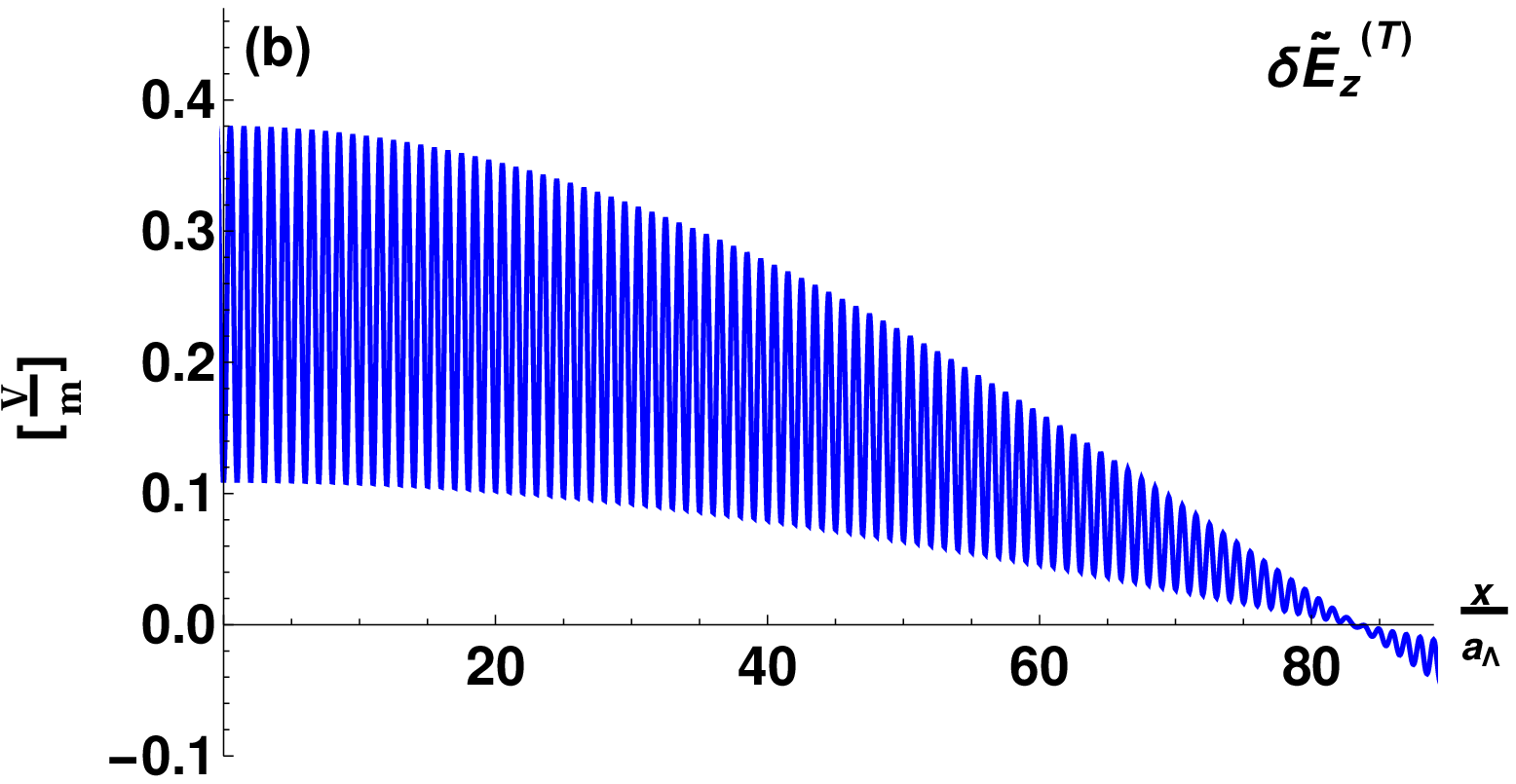}%
\end{minipage}

\caption{\label{fig:local_vs_macro_fields}(a) Plot of the transversal part
$\tilde{E}_{z}^{(T)}\left(\mathbf{r},\omega\right)$ of the microscopic
local electric field (green) and plot of amplitude $\mathcal{\tilde{E}}_{z}^{\left(T\right)}\left(\mathbf{r},\omega\right)$
of the transversal macroscopic field (red dots) along a path $\mathbf{r}\left(x\right)=x\cdot\mathbf{e}^{\left(x\right)}+\frac{a_{\Lambda}}{2}\left(\mathbf{e}^{\left(y\right)}+\mathbf{e}^{\left(z\right)}\right)$
with parameters as in Fig.\ref{fig:Local_Macro_Fields}, revealing
the transversal macroscopic field $\mathcal{\tilde{E}}_{z}^{\left(T\right)}\left(\mathbf{r},\omega\right)$
essentially coincides with the transversal part $E_{z}^{(T)}\left(\mathbf{r},\omega\right)$
of the microscopic local electric field.  (b) Plot of the residue
$\delta\tilde{E}_{z}^{(T)}\left(\mathbf{r},\omega\right)$ along the
same path $\mathbf{r}\left(x\right)$.}

\end{figure}

It follows from what has beeen said that the macroscopic electric
field amplitude is determined directly from the Fourier series representation
(\ref{eq:eq:microscopic local field Fourier series representation})
discarding all contributions of reciprocal lattice vectors $\mathbf{G}\neq\mathbf{0}$:
\begin{eqnarray}
\mathcal{\tilde{E}}_{a}\left(\mathbf{r},\omega\right) & = & \sum_{a'}\left(\delta_{a,a'}+\sum_{a''}\frac{1}{\left|C_{\Lambda}\right|}\mathcal{\bar{G}}_{a,a''}(\mathbf{q},\omega)\left[\bar{K}_{\Lambda}\left(\mathbf{q},\omega\right)\right]_{a'',a'}\right)\mathcal{\bar{E}}_{ext,a'}\left(\mathbf{q},\omega\right)e^{i\mathbf{q}\cdot\mathbf{r}}\label{eq:microscopic local field for incident plane wave with small q-1}
\end{eqnarray}
In position space then the longitudinal (L) and transversal (T) \emph{macroscopic}
electric field amplitudes read 
\begin{eqnarray}
\tilde{\mathcal{E}}_{a}^{\left(L\right)}\left(\mathbf{r},\omega\right) & = & \mathcal{\bar{E}}_{ext,a}^{\left(L\right)}\left(\mathbf{q},\omega\right)e^{i\mathbf{q}\cdot\mathbf{r}}-\frac{1}{\left|C_{\Lambda}\right|}\sum_{a',a''}\frac{q_{a}q_{a''}}{\left|\mathbf{q}\right|^{2}}\left[\bar{K}_{\Lambda}\left(\mathbf{q},\omega\right)\right]_{a'',a'}\mathcal{\bar{E}}_{ext,a'}\left(\mathbf{q},\omega\right)e^{i\mathbf{q}\cdot\mathbf{r}}\label{eq: longitudinal macroscopic field}\\
\tilde{\mathcal{E}}_{a}^{\left(T\right)}\left(\mathbf{r},\omega\right) & = & \mathcal{\bar{E}}_{ext,a}^{\left(T\right)}\left(\mathbf{q},\omega\right)e^{i\mathbf{q}\cdot\mathbf{r}}+\frac{1}{\left|C_{\Lambda}\right|}\sum_{a',a''}\frac{\omega^{2}}{c^{2}}\frac{\delta_{a,a''}-\frac{q_{a}q_{a''}}{\left|\mathbf{q}\right|^{2}}}{\left|\mathbf{\mathbf{q}}\right|^{2}-\frac{\omega^{2}}{c^{2}}-i0^{+}}\left[\bar{K}_{\Lambda}\left(\mathbf{q},\omega\right)\right]_{a'',a'}\mathcal{\bar{E}}_{ext,a'}\left(\mathbf{q},\omega\right)e^{i\mathbf{q}\cdot\mathbf{r}}\:.\label{eq:transversal macroscopic field}
\end{eqnarray}
Comparing now with (\ref{eq: longitudinal microscopic field}) and
(\ref{eq:transversal microscopic field}) we see at once that
\begin{eqnarray}
\tilde{E}_{a}^{\left(L\right)}\left(\mathbf{r},\omega\right) & = & \tilde{\mathcal{E}}_{a}^{\left(L\right)}\left(\mathbf{r},\omega\right)+\delta\tilde{E}_{a}^{\left(L\right)}\left(\mathbf{r},\omega\right)\\
\delta\tilde{E}_{a}^{\left(L\right)}\left(\mathbf{r},\omega\right) & = & -\sum_{a'',a'''}\frac{1}{\left|C_{\Lambda}\right|}\sum_{\mathbf{G}\neq\mathbf{0}}e^{i\left(\mathbf{q}+\mathbf{G}\right)\cdot\mathbf{r}}\frac{\left(\mathbf{q}+\mathbf{G}\right)_{a}\left(\mathbf{q}+\mathbf{G}\right)_{a''}}{\left|\mathbf{q}+\mathbf{G}\right|^{2}}\left[\bar{K}_{\Lambda}\left(\mathbf{G},\mathbf{q},\omega\right)\right]_{a'',a'''}\mathcal{\bar{E}}_{ext,a'''}\left(\mathbf{q},\omega\right)\nonumber \\
\nonumber \\
\tilde{E}_{a}^{\left(T\right)}\left(\mathbf{r},\omega\right) & = & \tilde{\mathcal{E}}_{a}^{\left(T\right)}\left(\mathbf{r},\omega\right)+\delta\tilde{E}_{a}^{\left(T\right)}\left(\mathbf{r},\omega\right)\\
\delta\tilde{E}_{a}^{\left(T\right)}\left(\mathbf{r},\omega\right) & = & \sum_{a'',a'''}\frac{1}{\left|C_{\Lambda}\right|}\sum_{\mathbf{G}\neq\mathbf{0}}e^{i\left(\mathbf{q}+\mathbf{G}\right)\cdot\mathbf{r}}\frac{\omega^{2}}{c^{2}}\frac{\delta_{a,a''}-\frac{\left(\mathbf{q}+\mathbf{G}\right)_{a}\left(\mathbf{q}+\mathbf{G}\right)_{a''}}{\left|\mathbf{q}+\mathbf{G}\right|^{2}}}{\left|\mathbf{\mathbf{q}}+\mathbf{G}\right|^{2}-\frac{\omega^{2}}{c^{2}}-i0^{+}}\left[\bar{K}_{\Lambda}\left(\mathbf{G},\mathbf{q},\omega\right)\right]_{a'',a'''}\mathcal{\bar{E}}_{ext,a'''}\left(\mathbf{q},\omega\right)\:.\nonumber 
\end{eqnarray}
Accordingly the microscopic local electric field and the macroscopic
electric field differ by the contributions of the sums over all reciprocal
lattice vectors $\mathbf{G}\neq\mathbf{0}$: 
\begin{eqnarray}
\tilde{E}_{a}\left(\mathbf{r},\omega\right) & = & \tilde{\mathcal{E}}_{a}\left(\mathbf{r},\omega\right)+\delta\tilde{E}_{a}\left(\mathbf{r},\omega\right)\label{eq: microscopic local electric field II}\\
\delta\tilde{E}_{a}\left(\mathbf{r},\omega\right) & = & \sum_{a'',a'}\frac{1}{\left|C_{\Lambda}\right|}\sum_{\mathbf{G}\neq\mathbf{0}}e^{i\left(\mathbf{q}+\mathbf{G}\right)\cdot\mathbf{r}}\mathcal{\bar{G}}_{a,a''}(\mathbf{q}+\mathbf{G},\omega)\left[\bar{K}_{\Lambda}\left(\mathbf{G},\mathbf{q},\omega\right)\right]_{a'',a'}\mathcal{\bar{E}}_{ext,a'}\left(\mathbf{q},\omega\right)\nonumber 
\end{eqnarray}
In Fig. \ref{fig:local_vs_macro_fields} we compare the spatial variation
of the transversal \emph{macroscopic} electric field amplitude (\ref{eq:transversal macroscopic field})
with the spatial variation of the transversal \emph{microscopic} local
electric field (\ref{eq:transversal microscopic field}) along a path
as shown in Fig.\ref{fig:Local_Macro_Fields}, assuming the \emph{external}
electric field was purely transversal, i.e. $\tilde{\mathcal{E}}_{ext,a}^{\left(L\right)}\left(\mathbf{r},\omega\right)=0$.
The residue $\delta\tilde{E}_{z}^{\left(T\right)}\left(\mathbf{r},\omega\right)$
turns out to be smaller by a factor $10^{-5}$ compared to the size
of the original amplitudes $\tilde{E}_{z}^{(T)}\left(\mathbf{r},\omega\right)$.

\subsection*{Macroscopic Magnetic Induction Field }

\begin{figure}
\begin{minipage}[t]{1\columnwidth}%
\includegraphics[scale=0.8]{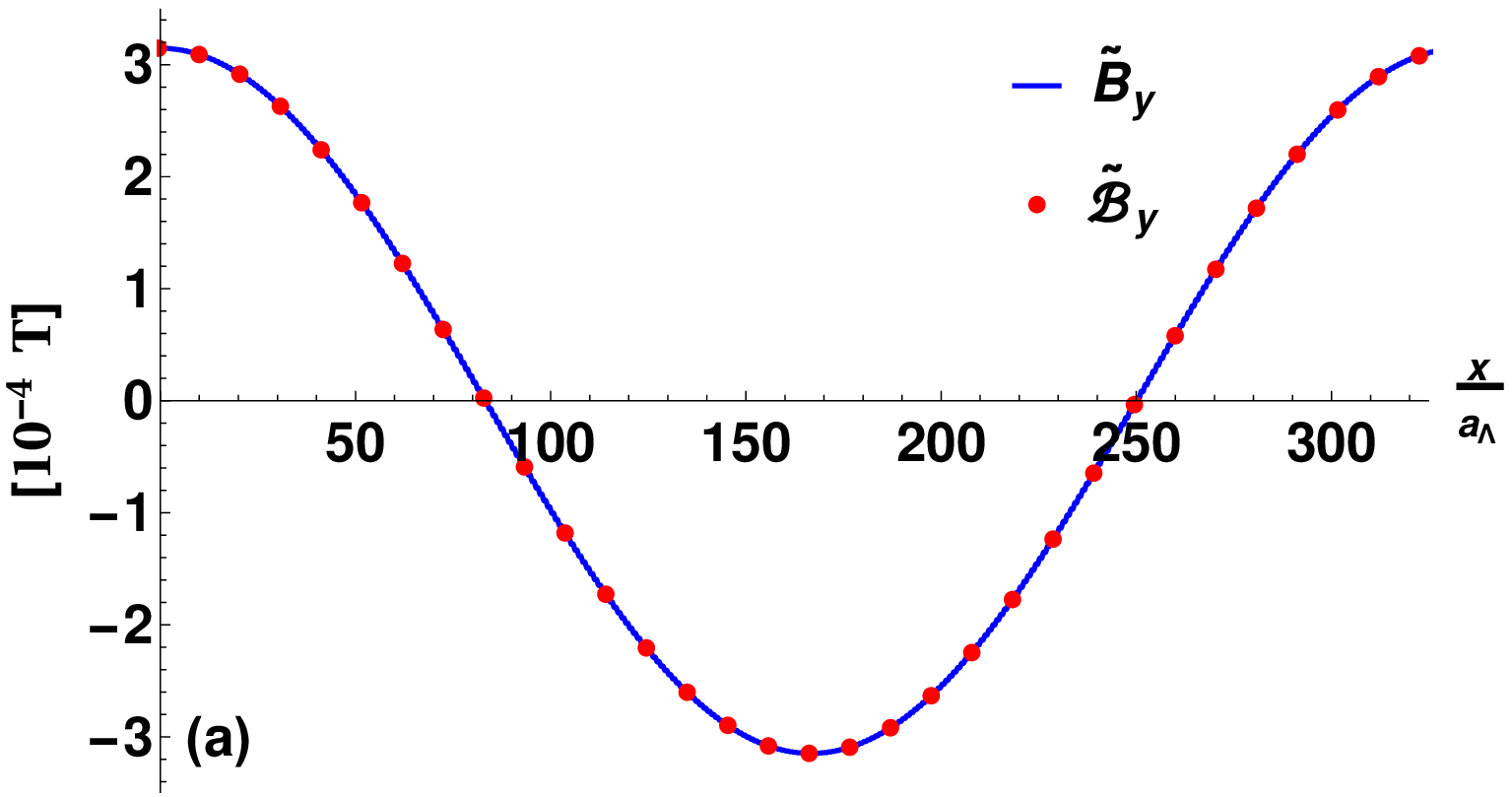}%
\end{minipage}

\vspace{1cm}

\begin{minipage}[t]{1\columnwidth}%
\includegraphics[scale=0.8]{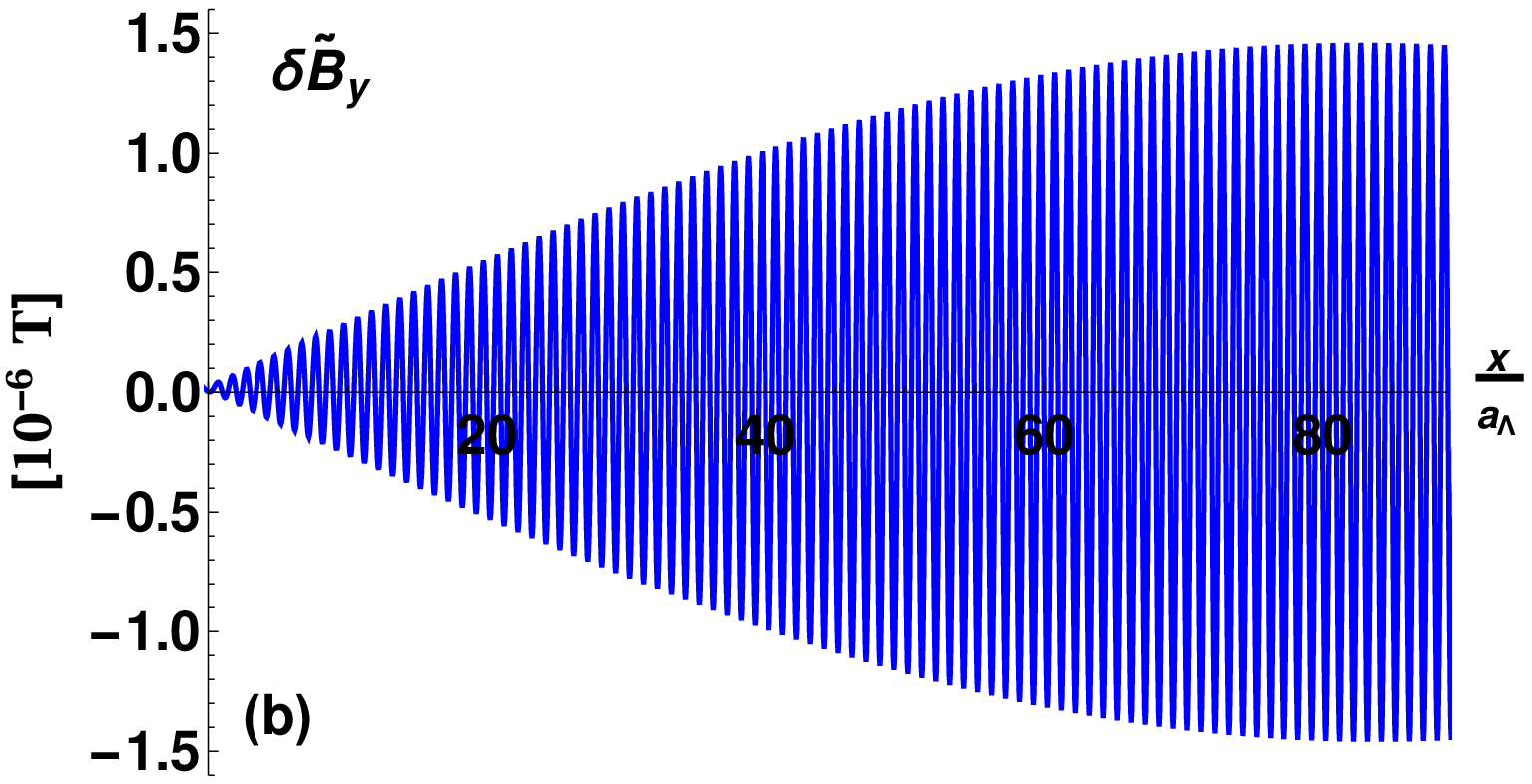}%
\end{minipage}

\caption{\label{fig:B-Field}(a) Plot of component $\tilde{B}_{y}\left(\mathbf{r},\omega\right)$
of the \emph{microscopic} local magnetic induction field amplitude
(blue) and plot of component $\mathcal{\tilde{B}}_{y}\left(\mathbf{r},\omega\right)$
of the \emph{macroscopic} magnetic induction field amplitude (red
dots) along a path $\mathbf{r}\left(x\right)=x\cdot\mathbf{e}^{\left(x\right)}+\frac{a_{\Lambda}}{2}\left(\mathbf{e}^{\left(y\right)}+\mathbf{e}^{\left(z\right)}\right)$
with parameters as in Fig.\ref{fig:Local_Macro_Fields}, revealing
that $\mathcal{\tilde{B}}_{y}\left(\mathbf{r},\omega\right)$ essentially
coincides with $\tilde{B}_{y}\left(\mathbf{r},\omega\right)$. (b)
Plot of the residue $\delta\tilde{B}_{y}\left(\mathbf{r},\omega\right)$
along the same path $\mathbf{r}\left(x\right)$.}
\end{figure}

The amplitude of the microscopic local magnetic induction field $\tilde{\mathbf{B}}\left(\mathbf{r},\omega\right)$
is of course directly connected to the amplitude of the local microscopic
electric field $\tilde{\mathbf{E}}\left(\mathbf{r},\omega\right)$
via Faraday's law:
\begin{eqnarray}
\tilde{\mathbf{B}}\left(\mathbf{r},\omega\right) & = & \frac{1}{i\omega}\mathbf{\boldsymbol{\nabla}}\wedge\tilde{\mathbf{E}}\left(\mathbf{r},\omega\right)
\end{eqnarray}
Insertion of the representation (\ref{eq: microscopic local electric field II})
for the microscopic local electric field amplitude, $\tilde{E}_{a}\left(\mathbf{r},\omega\right)=\tilde{\mathcal{E}}_{a}\left(\mathbf{r},\omega\right)+\delta\tilde{E}_{a}\left(\mathbf{r},\omega\right)$,
leads in this way immediately to
\begin{eqnarray}
\tilde{B}_{c}\left(\mathbf{r},\omega\right) & = & \sum_{b,a\in\left\{ x,y,z\right\} }\frac{1}{i\omega}\epsilon_{cba}\frac{\partial}{\partial r_{b}}\left\{ \tilde{\mathcal{E}}_{a}\left(\mathbf{r},\omega\right)+\delta\tilde{E}_{a}\left(\mathbf{r},\omega\right)\right\} \:.\label{eq: magnetic induction field}
\end{eqnarray}
Identifying now the macroscopic magnetic induction field amplitude
via
\begin{equation}
\tilde{\mathcal{B}}_{c}\left(\mathbf{r},\omega\right)=\sum_{b,a\in\left\{ x,y,z\right\} }\frac{1}{i\omega}\epsilon_{cba}\frac{\partial}{\partial r_{b}}\tilde{\mathcal{E}}_{a}\left(\mathbf{r},\omega\right)
\end{equation}
then 
\begin{equation}
\tilde{B}_{c}\left(\mathbf{r},\omega\right)=\tilde{\mathcal{B}}_{c}\left(\mathbf{r},\omega\right)+\delta\tilde{B}_{c}\left(\mathbf{r},\omega\right)\:,
\end{equation}
where the correction term representing the difference to the microscopic
magnetic induction field amplitude is
\begin{eqnarray}
\delta\tilde{B}_{c}\left(\mathbf{r},\omega\right) & = & \sum_{b,a\in\left\{ x,y,z\right\} }\frac{1}{i\omega}\epsilon_{cba}\frac{\partial}{\partial r_{b}}\delta\tilde{E}_{a}\left(\mathbf{r},\omega\right)\nonumber \\
 & = & \frac{\omega}{c^{2}}\sum_{c',c'',c'''}\frac{1}{\left|C_{\Lambda}\right|}\sum_{\mathbf{G}\neq\mathbf{0}}e^{i\left(\mathbf{q}+\mathbf{G}\right)\cdot\mathbf{r}}\epsilon_{cc'c''}\frac{q_{c'}+G_{c'}}{\left|\mathbf{\mathbf{q}+\mathbf{G}}\right|^{2}-\frac{\omega^{2}}{c^{2}}-i0^{+}}\left[\bar{K}_{\Lambda}\left(\mathbf{G},\mathbf{q},\omega\right)\right]_{c'',c'''}\mathcal{\bar{E}}_{ext,c'''}\left(\mathbf{q},\omega\right)\:.
\end{eqnarray}
 Like in the electric field case, the \emph{macroscopic} magnetic
induction field amplitude $\tilde{\mathcal{B}}_{c}\left(\mathbf{r},\omega\right)$
represents the low pass filtered \emph{microscopic} local magnetic
induction field amplitude $\tilde{B}_{c}\left(\mathbf{r},\omega\right)$.
The plot of the residue $\delta\tilde{B}_{y}\left(\mathbf{r},\omega\right)=\tilde{B}_{y}\left(\mathbf{r},\omega\right)-\tilde{\mathcal{B}}_{y}\left(\mathbf{r},\omega\right)$
is displayed in Fig.\ref{fig:B-Field}. Clearly, $\tilde{B}_{y}\left(\mathbf{r},\omega\right)$
and $\tilde{\mathcal{B}}_{y}\left(\mathbf{r},\omega\right)$ essentially
coincide. Note that in the electric field case the relative size of
the residue $\delta\tilde{E}_{z}^{\left(T\right)}\left(\mathbf{r},\omega\right)$
along the same path turned out to be even smaller, see Fig.\ref{fig:local_vs_macro_fields}.

\subsection*{Deducing the Differential Equations of Macroscopic Electrodynamics\label{sec:Deducing-the-Differential}}

Restricting to long wavelengths so that $\left|\mathbf{q}\right|<q_{c}$
let us rewrite (\ref{eq: Fourier transform microscopic E- field})
in the guise
\begin{eqnarray}
\bar{\mathcal{E}}_{a}\left(\mathbf{q},\omega\right)-\mathcal{\bar{E}}_{ext,a}\left(\mathbf{q},\omega\right) & = & \frac{1}{\left|C_{\Lambda}\right|}\sum_{a',a''}\bar{\mathcal{G}}_{a,a'}(\mathbf{q},\omega)\circ\left[\bar{K}_{\Lambda}(\mathbf{q},\omega)\right]_{a',a''}\mathcal{\bar{E}}_{ext,a''}\left(\mathbf{q},\omega\right)\label{eq: low pass filtered field integral equation}\\
 & = & \sum_{a',a''}\bar{\mathcal{G}}_{a,a'}(\mathbf{q},\omega)\circ\left[\varepsilon_{\Lambda}\left(\mathbf{q},\omega\right)-I\right]_{a',a''}\mathcal{\bar{E}}_{a''}\left(\mathbf{q},\omega\right)\:.\nonumber 
\end{eqnarray}
Multiplication on both sides with the inverse of the kernel (\ref{eq: Fourier transform of 3x3 matrix propagator})
gives
\begin{equation}
\sum_{a'}\left[\bar{\mathcal{G}}^{-1}(\mathbf{q},\omega)\right]_{a,a'}\left[\bar{\mathcal{E}}_{a'}\left(\mathbf{q},\omega\right)-\bar{\mathcal{E}}_{ext,a'}\left(\mathbf{q},\omega\right)\right]=\sum_{a'}\left[\varepsilon_{\Lambda}\left(\mathbf{q},\omega\right)-I\right]_{a,a'}\mathcal{\bar{E}}_{a'}\left(\mathbf{q},\omega\right)\:.\label{eq: intermediate I}
\end{equation}
In terms of the projection operators (\ref{eq:Fourier transform projection operators})
then
\begin{equation}
\frac{\omega^{2}}{c^{2}}\left[\bar{\mathcal{G}}^{-1}(\mathbf{q},\omega)\right]_{a,a'}=\left(\left|\mathbf{q}\right|^{2}-\frac{\omega^{2}}{c^{2}}\right)\bar{\Pi}_{a,a'}^{\left(T\right)}\left(\mathbf{q}\right)-\frac{\omega^{2}}{c^{2}}\bar{\Pi}_{a,a'}^{\left(L\right)}\left(\mathbf{q}\right)\:,\label{eq: inverse 3x3 propagator}
\end{equation}
so that (\ref{eq: intermediate I}) leads to
\begin{equation}
\sum_{a'}\left[\left|\mathbf{q}\right|^{2}\bar{\Pi}^{\left(T\right)}\left(\mathbf{q}\right)-\frac{\omega^{2}}{c^{2}}\varepsilon_{\Lambda}\left(\mathbf{q},\omega\right)\right]_{a,a'}\bar{\mathcal{E}}_{a'}\left(\mathbf{q},\omega\right)=\sum_{a'}\left[\left(\left|\mathbf{q}\right|^{2}-\frac{\omega^{2}}{c^{2}}\right)\bar{\Pi}_{a,a'}^{\left(T\right)}\left(\mathbf{q}\right)-\frac{\omega^{2}}{c^{2}}\bar{\Pi}_{a,a'}^{\left(L\right)}\left(\mathbf{q}\right)\right]\bar{\mathcal{E}}_{ext,a'}\left(\mathbf{q},\omega\right)\:.\label{eq:intermediate II}
\end{equation}

Identifying transversal and longitudinal components of the Fourier
amplitudes of the electric field,
\begin{eqnarray}
\sum_{a'}\bar{\Pi}_{a,a'}^{\left(L,T\right)}\left(\mathbf{q}\right)\bar{\mathcal{E}}_{ext,a'}\left(\mathbf{q},\omega\right) & = & \bar{\mathcal{E}}_{ext,a}^{\left(L,T\right)}\left(\mathbf{q},\omega\right)\label{eq: longitudinal and transversal projections of macroscopic electric field}\\
\sum_{a'}\bar{\Pi}_{a,a'}^{\left(L,T\right)}\left(\mathbf{q}\right)\bar{\mathcal{E}}_{a'}\left(\mathbf{q},\omega\right) & = & \bar{\mathcal{E}}_{a}^{\left(L,T\right)}\left(\mathbf{q},\omega\right)\nonumber 
\end{eqnarray}
let us reexpress the respective Fourier amplitudes of the \emph{external
}electric field in terms of the original sources inside the source
domain $\Omega_{S}$, namely the transversal external current distribution
$\bar{j}_{ext,a}^{(T)}\left(\mathbf{q},\omega\right)$ and the external
charge distribution $\bar{\varrho}_{ext}\left(\mathbf{q},\omega\right)$:
\begin{eqnarray}
\left(\left|\mathbf{q}\right|^{2}-\frac{\omega^{2}}{c^{2}}\right)\bar{\mathcal{E}}_{ext,a}^{\left(T\right)}\left(\mathbf{q},\omega\right) & = & \mu_{0}i\omega\bar{j}_{ext,a}^{(T)}\left(\mathbf{q},\omega\right)\label{eq: external transversal field}\\
\bar{\mathcal{E}}_{ext,a}^{\left(L\right)}\left(\mathbf{q},\omega\right) & = & -iq_{a}\bar{\phi}_{ext}\left(\mathbf{q},\omega\right)=-iq_{a}\frac{\bar{\varrho}_{ext}\left(\mathbf{q},\omega\right)}{\varepsilon_{0}\left|\mathbf{q}\right|^{2}}\label{eq: external longitudinal field}
\end{eqnarray}
So the right hand side in (\ref{eq:intermediate II}) reduces together
with the Fourier transformed relation (\ref{eq:external longitudinal electric field})
to
\begin{eqnarray}
\sum_{a'}\left[\left(\left|\mathbf{q}\right|^{2}-\frac{\omega^{2}}{c^{2}}\right)\bar{\Pi}_{a,a'}^{\left(T\right)}\left(\mathbf{q}\right)-\frac{\omega^{2}}{c^{2}}\bar{\Pi}_{a,a'}^{\left(L\right)}\left(\mathbf{q}\right)\right]\bar{\mathcal{E}}_{ext,a'}\left(\mathbf{q},\omega\right) & = & \left(\left|\mathbf{q}\right|^{2}-\frac{\omega^{2}}{c^{2}}\right)\bar{\mathcal{E}}_{ext,a}^{\left(T\right)}\left(\mathbf{q},\omega\right)-\frac{\omega^{2}}{c^{2}}\bar{\mathcal{E}}_{ext,a}^{\left(L\right)}\left(\mathbf{q},\omega\right)\nonumber \\
 & = & \mu_{0}i\omega\left[\bar{j}_{ext,a}^{(T)}\left(\mathbf{q},\omega\right)+\bar{j}_{ext,a}^{(L)}\left(\mathbf{q},\omega\right)\right]\\
 & = & \mu_{0}i\omega\bar{j}_{ext,a}\left(\mathbf{q},\omega\right)\:.
\end{eqnarray}
Consequently (\ref{eq:intermediate II}) assumes the guise 
\begin{eqnarray}
\sum_{a'}\left[\left|\mathbf{q}\right|^{2}\bar{\Pi}^{\left(T\right)}\left(\mathbf{q}\right)-\frac{\omega^{2}}{c^{2}}\varepsilon_{\Lambda}\left(\mathbf{q},\omega\right)\right]_{a,a'}\bar{\mathcal{E}}_{a'}\left(\mathbf{q},\omega\right) & = & \mu_{0}i\omega\bar{j}_{ext,a}\left(\mathbf{q},\omega\right)\:.\label{eq:intermediate III}
\end{eqnarray}

If the dependence on wavevector $\mathbf{q}$ of the dielectric tensor
in (\ref{eq:intermediate III}) can be ignored, we replace $\varepsilon_{\Lambda}\left(\mathbf{q},\omega\right)\rightarrow\varepsilon_{\Lambda}\left(\omega\right)$
and obtain then in position space the well known (so called) \emph{vector}
wave equation determining the \emph{macroscopic} electric field: 
\begin{eqnarray}
\nabla\wedge\left[\nabla\wedge\tilde{\mathcal{\boldsymbol{E}}}\left(\mathbf{r},\omega\right)\right]-\frac{\omega^{2}}{c^{2}}\varepsilon_{\Lambda}\left(\omega\right)\tilde{\mathcal{\boldsymbol{E}}}\left(\mathbf{r},\omega\right) & = & \mu_{0}i\omega\tilde{\mathbf{j}}_{ext}\left(\mathbf{r},\omega\right)\label{eq: vector wave equation for macroscopic electric field}
\end{eqnarray}
It should be noted, that here $\tilde{\mathcal{\boldsymbol{E}}}\left(\mathbf{r},\omega\right)$
still may be decomposed into divergence-free (transversal) and curl-free
(longitudinal) parts, $\tilde{\mathcal{\boldsymbol{E}}}\left(\mathbf{r},\omega\right)=\tilde{\mathcal{\boldsymbol{E}}}^{\left(T\right)}\left(\mathbf{r},\omega\right)+\tilde{\mathcal{\boldsymbol{E}}}^{\left(L\right)}\left(\mathbf{r},\omega\right)$.
Thus it is deceptive to interpret (\ref{eq: vector wave equation for macroscopic electric field})
as a \emph{wave equation} determining electromagnetic radiation as
propagating photons with speed determined by the eigenvalues of the
dielectric tensor $\varepsilon_{\Lambda}\left(\omega\right)$, unless
$\tilde{\mathcal{\boldsymbol{E}}}^{\left(L\right)}\left(\mathbf{r},\omega\right)\equiv\mathbf{0}$.

To find the differential equations for the transversal and longitudinal
parts of the macroscopic field amplitude let us first introduce block
matrix notation specifying transversal and longitudinal projections
of the dielectric tensor:
\begin{eqnarray}
\varepsilon_{a,a'}^{\left(A,B\right)}\left(\mathbf{q},\omega\right) & = & \sum_{b,b'}\bar{\Pi}_{a,b}^{\left(A\right)}\left(\mathbf{q}\right)\left[\varepsilon_{\Lambda}\left(\mathbf{q},\omega\right)\right]_{b,b'}\bar{\Pi}_{b',a'}^{\left(B\right)}\left(\mathbf{q}\right)\label{eq: projections of dielectric tensor}\\
A,B & \in & \left\{ L,T\right\} \nonumber 
\end{eqnarray}
Then the vector wave equation (\ref{eq:intermediate III}) separates
into two coupled equations for the respective transversal and longitudinal
Fourier amplitudes of the macroscopic field:
\begin{eqnarray}
\sum_{a'}\left(\left|\mathbf{q}\right|^{2}\delta_{a,a'}-\frac{\omega^{2}}{c^{2}}\varepsilon_{a,a'}^{\left(T,T\right)}\left(\mathbf{q},\omega\right)\right)\bar{\mathcal{E}}_{a'}^{\left(T\right)}\left(\mathbf{q},\omega\right)-\frac{\omega^{2}}{c^{2}}\sum_{a'}\varepsilon_{a,a'}^{\left(T,L\right)}\left(\mathbf{q},\omega\right)\bar{\mathcal{E}}_{a'}^{\left(L\right)}\left(\mathbf{q},\omega\right) & = & \mu_{0}i\omega\bar{j}_{ext,a}^{(T)}\left(\mathbf{q},\omega\right)\label{eq: coupled field equatios}\\
\sum_{a'}\varepsilon_{a,a'}^{\left(L,T\right)}\left(\mathbf{q},\omega\right)\bar{\mathcal{E}}_{a'}^{\left(T\right)}\left(\mathbf{q},\omega\right)+\sum_{a'}\varepsilon_{a,a'}^{\left(L,L\right)}\left(\mathbf{q},\omega\right)\bar{\mathcal{E}}_{a'}^{\left(L\right)}\left(\mathbf{q},\omega\right) & = & \bar{\mathcal{E}}_{ext,a}^{\left(L\right)}\left(\mathbf{q},\omega\right)\nonumber 
\end{eqnarray}

Choosing Eq. (\ref{eq: vector wave equation for macroscopic electric field})
as a starting point for the transport theory of radiation (light intensity)
inside a (possibly disordered) material appears according to what
has been said questionable, as the fluctuation contribution $\tilde{E}_{a}\left(\mathbf{r},\omega\right)-\tilde{\mathcal{E}}_{a}\left(\mathbf{r},\omega\right)\equiv\delta\tilde{E}_{a}\left(\mathbf{r},\omega\right)$,
see Eq.(\ref{eq: microscopic local electric field II}), is in this
case not included, despite the product $\delta\tilde{E}_{a}\left(\mathbf{r},\omega\right)$$\delta\tilde{E}_{b}\left(\mathbf{r}',\omega\right)$
apparently comprising a spatially slowly varying interference contribution.

\subsection*{Wave Equation with Renormalized Speed of Light}

If the external field was purely transversal, i.e. $\bar{\mathcal{E}}_{ext,a}^{\left(L\right)}\left(\mathbf{q},\omega\right)\equiv0$,
then the longitudinal component of the macroscopic field is readily
eliminated in (\ref{eq: coupled field equatios}), provided the inverse
of the longitudinal block $\varepsilon_{\Lambda}^{\left(L,L\right)}\left(\mathbf{q},\omega\right)$
exists:
\begin{equation}
\bar{\mathcal{E}}_{a}^{\left(L\right)}\left(\mathbf{q},\omega\right)=-\sum_{a'}\left[\left[\varepsilon^{\left(L,L\right)}\left(\mathbf{q},\omega\right)\right]^{-1}\circ\varepsilon^{\left(L,T\right)}\left(\mathbf{q},\omega\right)\right]_{a,a'}\bar{\mathcal{E}}_{a'}^{\left(T\right)}\left(\mathbf{q},\omega\right)
\end{equation}
Insertion leads to
\begin{equation}
\sum_{a'}\left(\left|\mathbf{q}\right|^{2}\delta_{a,a'}-\frac{\omega^{2}}{c^{2}}\left[\varepsilon^{\left(T,T\right)}\left(\mathbf{q},\omega\right)-\varepsilon^{\left(T,L\right)}\left(\mathbf{q},\omega\right)\circ\left[\varepsilon^{\left(L,L\right)}\left(\mathbf{q},\omega\right)\right]^{-1}\circ\varepsilon^{\left(L,T\right)}\left(\mathbf{q},\omega\right)\right]_{a,a'}\right)\bar{\mathcal{E}}_{a'}^{\left(T\right)}\left(\mathbf{q},\omega\right)=\mu_{0}i\omega\bar{j}_{ext,a}^{(T)}\left(\mathbf{q},\omega\right)\:.
\end{equation}
Introducing as an effective \emph{transversal} dielectric tensor 
\begin{eqnarray}
\varepsilon^{\left(T\right)}\left(\mathbf{q},\omega\right) & = & \varepsilon^{\left(T,T\right)}\left(\mathbf{q},\omega\right)-\varepsilon^{\left(T,L\right)}\left(\mathbf{q},\omega\right)\circ\left[\varepsilon^{\left(L,L\right)}\left(\mathbf{q},\omega\right)\right]^{-1}\circ\varepsilon^{\left(L,T\right)}\left(\mathbf{q},\omega\right)\:,\label{eq: transversal dielectric tensor}
\end{eqnarray}
 then
\begin{equation}
\sum_{a'}\left(\left|\mathbf{q}\right|^{2}\delta_{a,a'}-\frac{\omega^{2}}{c^{2}}\varepsilon_{a,a'}^{\left(T\right)}\left(\mathbf{q},\omega\right)\right)\bar{\mathcal{E}}_{a'}^{\left(T\right)}\left(\mathbf{q},\omega\right)=\mu_{0}i\omega\bar{j}_{ext,a}^{(T)}\left(\mathbf{q},\omega\right)\:.\label{eq: wave equation transverse macroscopic field Fourier space}
\end{equation}

If the dependence on wavevector $\mathbf{q}$ of the (transversal)
dielectric function can be ignored, then $\varepsilon_{ab}^{\left(T\right)}\left(\mathbf{q},\omega\right)\rightarrow\varepsilon_{ab}^{(T)}\left(\omega\right)$,
and equation (\ref{eq: wave equation transverse macroscopic field Fourier space})
corresponds in position space to a (scalar) wave equation determining
the Cartesian components of the transversal macroscopic electric field
amplitude $\tilde{\mathcal{\boldsymbol{E}}}^{\left(T\right)}\left(\mathbf{r},\omega\right)$
propagating inside a dielectric crystal, in full agreement with the
standard theory of the propagation of polarized light in transparent
dielectric materials:
\begin{equation}
\sum_{a'}\left(-\nabla^{2}\delta_{a,a'}-\frac{\omega^{2}}{c^{2}}\varepsilon_{aa'}^{(T)}\left(\omega\right)\right)\mathcal{\tilde{E}}_{a'}^{\left(T\right)}\left(\mathbf{r},\omega\right)=\mu_{0}i\omega\tilde{j}_{ext,a}^{\left(T\right)}\left(\mathbf{r},\omega\right)\label{eq: wave equation for transversal macroscopic field}
\end{equation}
With a choice of a coordinate frame such that the dielectric tensor
is diagonal in that frame, $\varepsilon_{aa'}^{(T)}\left(\omega\right)=\delta_{aa'}n_{a}^{2}\left(\omega\right)$,
one finds for light propagating along a high symmetry axis corresponding
to an eigenvector of that dielectric tensor the usual reduction of
the speed of light, $c\rightarrow c/n_{a}\left(\omega\right)$ characteristic
for in general birefringent crystalline dielectrics. Note that because
of chromatic dispersion of the index of refraction then those eigenvectors
may undergo a corresponding chromatic dispersion of axes as well,
yet this effect being observable only for monoclinic and triclinic
crystalline symmetry \citep{Born1999}. 

If the dependence on wavevector $\mathbf{q}$ of the (transversal)
dielectric function cannot be ignored, the Taylor expansion
\begin{equation}
\varepsilon_{ab}^{\left(T\right)}\left(\mathbf{q},\omega\right)=\varepsilon_{ab}^{\left(T\right)}\left(\omega\right)+i\sum_{c}\gamma_{abc}\left(\omega\right)q_{c}+\sum_{c,d}\alpha_{abcd}\left(\omega\right)q_{c}q_{d}+...\label{eq:Taylor expansion transversal dielectric tensor}
\end{equation}
 around $\mathbf{q}=\mathbf{0}$ in (\ref{eq: wave equation transverse macroscopic field Fourier space})
provides then insight into various optical phenomena connected to
retardation and non-locality of the dielectric tensor, in full agreement
with the phenomenological reasoning of Agranovich and Ginzburg \citep{Agranovich1984}:
the eigenvalues of the symmetric tensor $\varepsilon_{ab}^{(T)}\left(\omega\right)=\varepsilon_{ba}^{(T)}\left(\omega\right)$
describing chromatic dispersion of the index of refraction and birefringence,
the antisymmetric first order term $\gamma_{abc}\left(\omega\right)$
specifying rotary power (natural optical activity), the second order
terms $\alpha_{abcd}\left(\omega\right)$ shaping the intrinsic effects
of a spatial-dispersion-induced-birefringence. Of course, in a centrosymmetric
crystal there exists no natural optical activity: $\gamma_{abc}\left(\omega\right)\equiv0$.
The tensor $\alpha_{abcd}\left(\omega\right)$ originates from the
symmetry-breaking evoked by the finite $\mathbf{q}$-vector of the
light \citep{Agranovich1984}, it displays in general $3\times3\times3\times3=81$
components. But for crystals with cubic symmetry the number of independent
components of that tensor reduces substantially: for symmetry group
$T$ and $T_{h}$ there exist four independent components, for symmetry
group $T_{d},$$O_{h}$ and $O$ there exist three independent components
and for isotropic systems that number reduces to two \citep{Agranovich1984}.
The (weak) effects of a dispersion-induced-birefringence indeed give
as a matter of fact reason for concern regarding the image quality
of dielectric lenses made from $CaF_{2}$ and $BaF_{2}$, a topic
of prime importance designing modern lithographic optical systems
in the ultraviolet \citep{Burnett2002,Serebryakov2003}.

The dielectric tensor $\varepsilon_{\Lambda}\left(\mathbf{q},\omega\right)$
being a functional of the microscopic polarizabilities $\alpha\left(\boldsymbol{\eta}^{\left(j\right)},\omega\right)$
of atoms (ions, molecules) located at position $\boldsymbol{\eta}^{\left(j\right)}$
in the unit cell $C_{\Lambda}$, for example see (\ref{eq: atom polarizability}),
undergoes variations in proportion to changes of those atom individual
polarizabilities caused by external static fields. For instance, in
the presence of a \emph{static} external magnetic induction field
$\mathbf{B}_{0}$, the polarizability of a single atom displays now
a magnetic field induced anisotropy \citep{Baranova1979}, even though
the atom-individual polarizability in zero field $\alpha\left(\boldsymbol{\eta}^{\left(j\right)},\omega\right)$
was isotropic:
\begin{eqnarray}
\alpha_{aa'}\left(\boldsymbol{\eta}^{\left(j\right)},\omega;\mathbf{B}_{0}\right) & = & \left(\left[I+i\frac{\omega}{\left|e\right|}\alpha\left(\boldsymbol{\eta}^{\left(j\right)},\omega\right)\circ b\right]^{-1}\circ\alpha\left(\boldsymbol{\eta}^{\left(j\right)},\omega\right)\right)_{aa'}\label{eq: microscopic atom polarizability in static B-field}\\
b_{aa'} & = & \sum_{a''}\epsilon_{aa'a''}B_{a''}\nonumber 
\end{eqnarray}
 Such changes then reflect in corresponding changes of the dielectric
tensor. As a result the Taylor expansion of the transversal dielectric
tensor of the crystal now reads to leading order in the small quantities
$\mathbf{q}$ and $\mathbf{B}_{0}$: 
\begin{eqnarray}
\varepsilon_{ab}^{\left(T\right)}\left(\mathbf{q},\omega;\mathbf{B}_{0}\right) & = & \varepsilon_{ab}^{\left(T\right)}\left(\omega\right)+i\sum_{c}\beta_{abc}\left(\omega\right)B_{0,c}+i\sum_{c}\gamma_{abc}\left(\omega\right)q_{c}+...\label{eq: Taylor expansion transversal dielectric tensor in static B-field}
\end{eqnarray}
Note the symmetry 
\begin{equation}
\varepsilon_{ab}^{\left(T\right)}\left(-\mathbf{q},\omega;-\mathbf{B}_{0}\right)=\varepsilon_{ba}^{\left(T\right)}\left(\mathbf{q},\omega;\mathbf{B}_{0}\right)\:.
\end{equation}
Correspondingly, if the dependence of the transversal dielectric tensor
on wavevector $\mathbf{q}$ can be ignored, then light propagation
in the presence of a static field $\mathbf{B}_{0}$ inside a dielectric
crystal is governed by the wave equation (\ref{eq: wave equation for transversal macroscopic field}),
but with a dielectric tensor now dependent on that \emph{static} magnetic
field: 
\begin{equation}
\varepsilon_{ab}^{\left(T\right)}\left(\omega;\mathbf{B}_{0}\right)=\varepsilon_{ab}^{\left(T\right)}\left(\omega\right)+i\sum_{c}\beta_{abc}\left(\omega\right)B_{0,c}\label{eq: magnetic field dependence transversal dielectric tensor}
\end{equation}
The presence of the antisymmetric tensor $\beta_{abc}\left(\omega\right)=\sum_{c'}\epsilon_{abc'}\lambda_{c'c}\left(\omega\right)$
in (\ref{eq: magnetic field dependence transversal dielectric tensor})
and with that said in (\ref{eq: wave equation for transversal macroscopic field}),
now leads to left and right circularly polarized waves propagating
at slightly different speeds, thus giving rise to a magnetic field
induced \emph{circular} birefringence. If $\mathbf{B}_{0}\wedge\mathbf{q}=\mathbf{0}$,
i.e. if $\mathbf{B}_{0}$ is orientated parallel or anti-parallel
to the direction $\mathbf{\hat{q}}$ of light propagation, this is
the well known Faraday rotation effect of a light waves linear polarization
in a static magnetic field.

\subsection*{Electric-Field Screening}

Conversely, if the external current source was purely longitudinal,
i.e. $\bar{j}_{ext,a}^{(T)}\left(\mathbf{q},\omega\right)\equiv0$,
there follows directly from (\ref{eq: coupled field equatios}) upon
elimination of the transversal part $\bar{\mathcal{E}}_{a}^{\left(T\right)}\left(\mathbf{q},\omega\right)$
in favour of the field amplitude $\bar{\mathcal{E}}_{a}^{\left(L\right)}\left(\mathbf{q},\omega\right)$
now the relation 
\begin{equation}
\bar{\mathcal{E}}_{a}^{\left(L\right)}\left(\mathbf{q},\omega\right)=\sum_{a'}\left(\left[\varepsilon^{\left(L\right)}\left(\mathbf{q},\omega\right)\right]^{-1}\right)_{a,a'}\bar{\mathcal{E}}_{ext,a'}^{\left(L\right)}\left(\mathbf{q},\omega\right)\:,\label{eq:Longitudinal macroscopic field Fourier space}
\end{equation}
with the \emph{longitudinal} dielectric tensor 
\begin{equation}
\varepsilon^{\left(L\right)}\left(\mathbf{q},\omega\right)=\varepsilon^{\left(L,T\right)}\left(\mathbf{q},\omega\right)\circ\frac{\frac{\omega^{2}}{c^{2}}}{\left|\mathbf{q}\right|^{2}I-\frac{\omega^{2}}{c^{2}}\varepsilon^{\left(T,T\right)}\left(\mathbf{q},\omega\right)}\circ\varepsilon^{\left(T,L\right)}\left(\mathbf{q},\omega\right)+\varepsilon^{\left(L,L\right)}\left(\mathbf{q},\omega\right)\label{eq:longitudinal dielectric tensor}
\end{equation}
describing \emph{electric-field screening}. From equation (\ref{eq:Longitudinal macroscopic field Fourier space})
we readily infer 
\begin{equation}
\sum_{a,a'}iq_{a}\varepsilon_{aa'}^{\left(L\right)}\left(\mathbf{q},\omega\right)\bar{\mathcal{E}}_{a'}^{\left(L\right)}\left(\mathbf{q},\omega\right)=\sum_{a}iq_{a}\bar{\mathcal{E}}_{ext,a}^{\left(L\right)}\left(\mathbf{q},\omega\right)=\frac{\bar{\varrho}_{ext}\left(\mathbf{q},\omega\right)}{\varepsilon_{0}}\:.
\end{equation}
For optical frequencies and below it is (obviously) adequate to approximate
$\varepsilon^{\left(L\right)}\left(\mathbf{q},\omega\right)\simeq\varepsilon^{\left(L,L\right)}\left(\mathbf{q},\omega\right)$,
provided $\det\left[\left|\mathbf{q}\right|^{2}I-\frac{\omega^{2}}{c^{2}}\varepsilon^{\left(T,T\right)}\left(\mathbf{q},\omega\right)\right]\neq0$.
Then 
\begin{eqnarray}
\sum_{a,a'}iq_{a}\varepsilon_{aa'}^{\left(L\right)}\left(\mathbf{q},\omega\right)\bar{\mathcal{E}}_{a'}^{\left(L\right)}\left(\mathbf{q},\omega\right) & = & \sum_{a,a'}iq_{a}\varepsilon_{aa'}^{\left(L,L\right)}\left(\mathbf{q},\omega\right)\bar{\mathcal{E}}_{a'}^{\left(L\right)}\left(\mathbf{q},\omega\right)\nonumber \\
 & = & \sum_{a,a',b,b'}iq_{a}\Pi_{ab}^{\left(L\right)}\left(\mathbf{q}\right)\left[\varepsilon_{\Lambda}\left(\mathbf{q},\omega\right)\right]_{bb'}\Pi_{b'a'}^{\left(L\right)}\left(\mathbf{q}\right)\bar{\mathcal{E}}_{a'}^{\left(L\right)}\left(\mathbf{q},\omega\right)\nonumber \\
 & = & \sum_{b,b'}iq_{b}\left[\varepsilon_{\Lambda}\left(\mathbf{q},\omega\right)\right]_{bb'}\bar{\mathcal{E}}_{b'}^{\left(L\right)}\left(\mathbf{q},\omega\right)\:.
\end{eqnarray}
In position space, and representing the longitudinal electric field
$\mathcal{\tilde{E}}^{\left(L\right)}\left(\mathbf{r},\omega\right)=-\nabla\tilde{\phi}\left(\mathbf{r},\omega\right)$,
this conforms in the long wavelength limit $\mathbf{\left|q\right|}\rightarrow0$
to the usual Poisson type equation of electrostatics determining the
scalar potential $\tilde{\phi}\left(\mathbf{r},\omega\right)$ of
a given charge distribution $\tilde{\varrho}_{ext}\left(\mathbf{r},\omega\right)$:
\begin{equation}
\boldsymbol{-\nabla}\boldsymbol{\mathbf{\cdot}}\left[\varepsilon_{\Lambda}\left(\omega\right)\nabla\tilde{\phi}\left(\mathbf{r},\omega\right)\right]=\frac{\tilde{\varrho}_{ext}\left(\mathbf{r},\omega\right)}{\varepsilon_{0}}\label{eq: Poisson equation for longitudinal macroscopic field}
\end{equation}
Now it is manifest that it is the dielectric $3\times3$ tensor 
\begin{equation}
\varepsilon_{\Lambda}\left(\omega\right)\equiv\lim_{\left|\mathbf{q}\right|\rightarrow0}\varepsilon_{\Lambda}\left(\mathbf{q},\omega\right)
\end{equation}
that describes electric-field screening, like in electrostatics.

\subsection*{Results of Calculations for the Chromatic Dispersion, Natural Optical
Activity and Spatial-Dispersion-Induced Birefringence in Various Dielectric
Crystals\label{sec:chromatic-dispersion,-optical activity and spatial dispersion induced birefringence}}

\begin{figure*}
\begin{minipage}[t]{0.49\linewidth}%
 \includegraphics[width=1\linewidth]{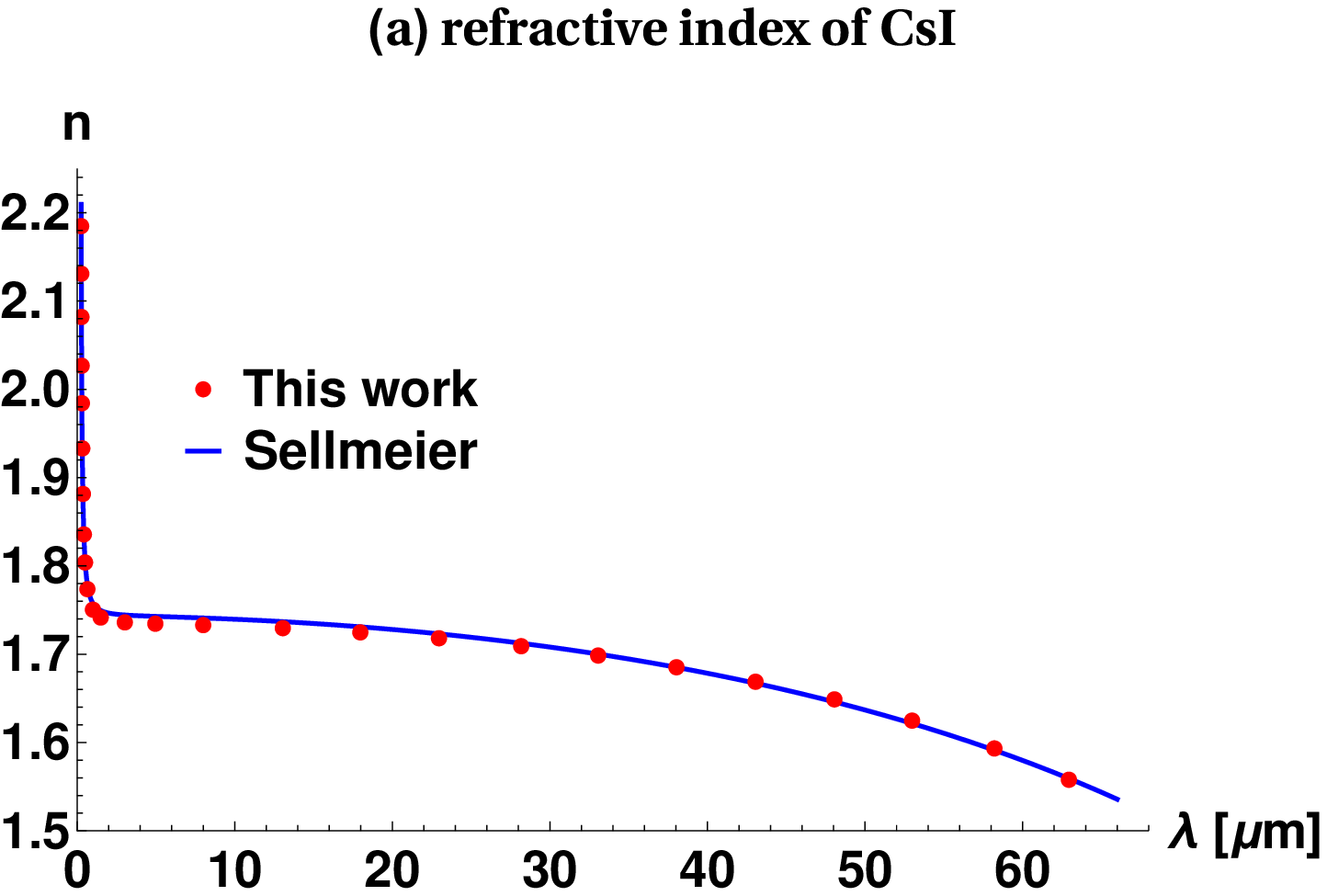} %
\end{minipage}%
\begin{minipage}[t]{0.49\linewidth}%
 \includegraphics[width=1\linewidth]{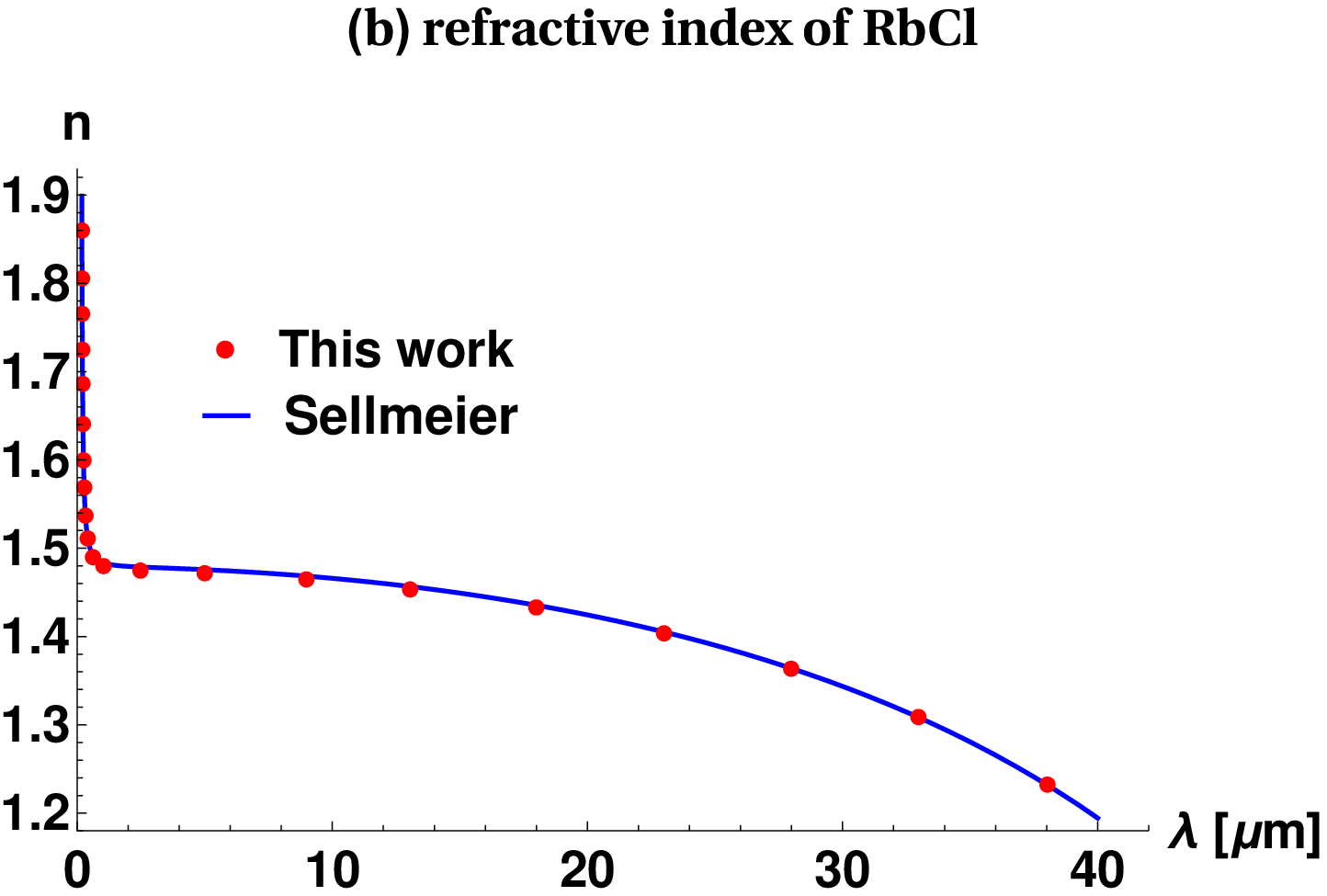} %
\end{minipage}\vspace{1cm}
\begin{minipage}[t]{0.49\linewidth}%
 \includegraphics[width=1\linewidth]{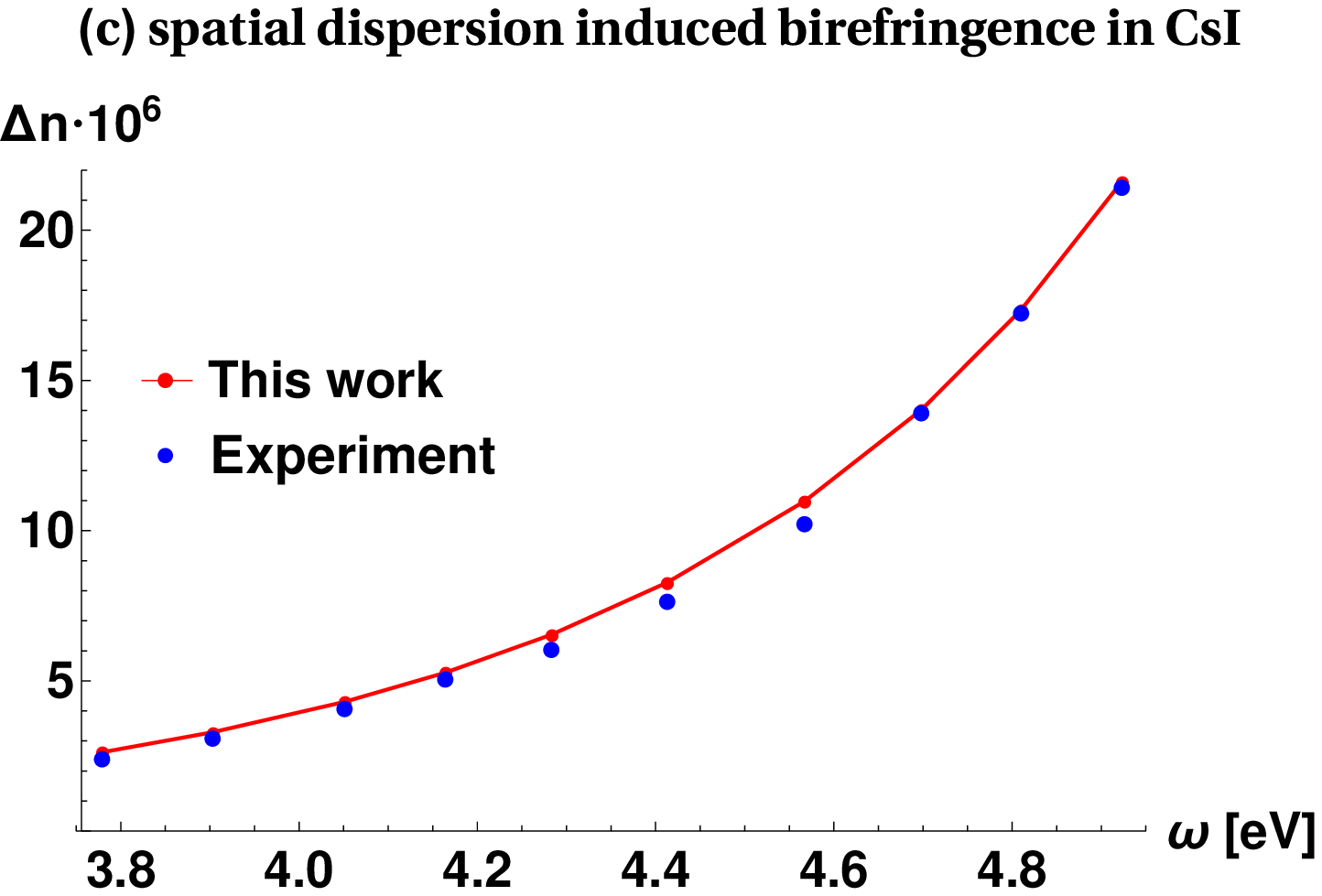} %
\end{minipage}%
\begin{minipage}[t]{0.49\linewidth}%
 \includegraphics[width=1\linewidth]{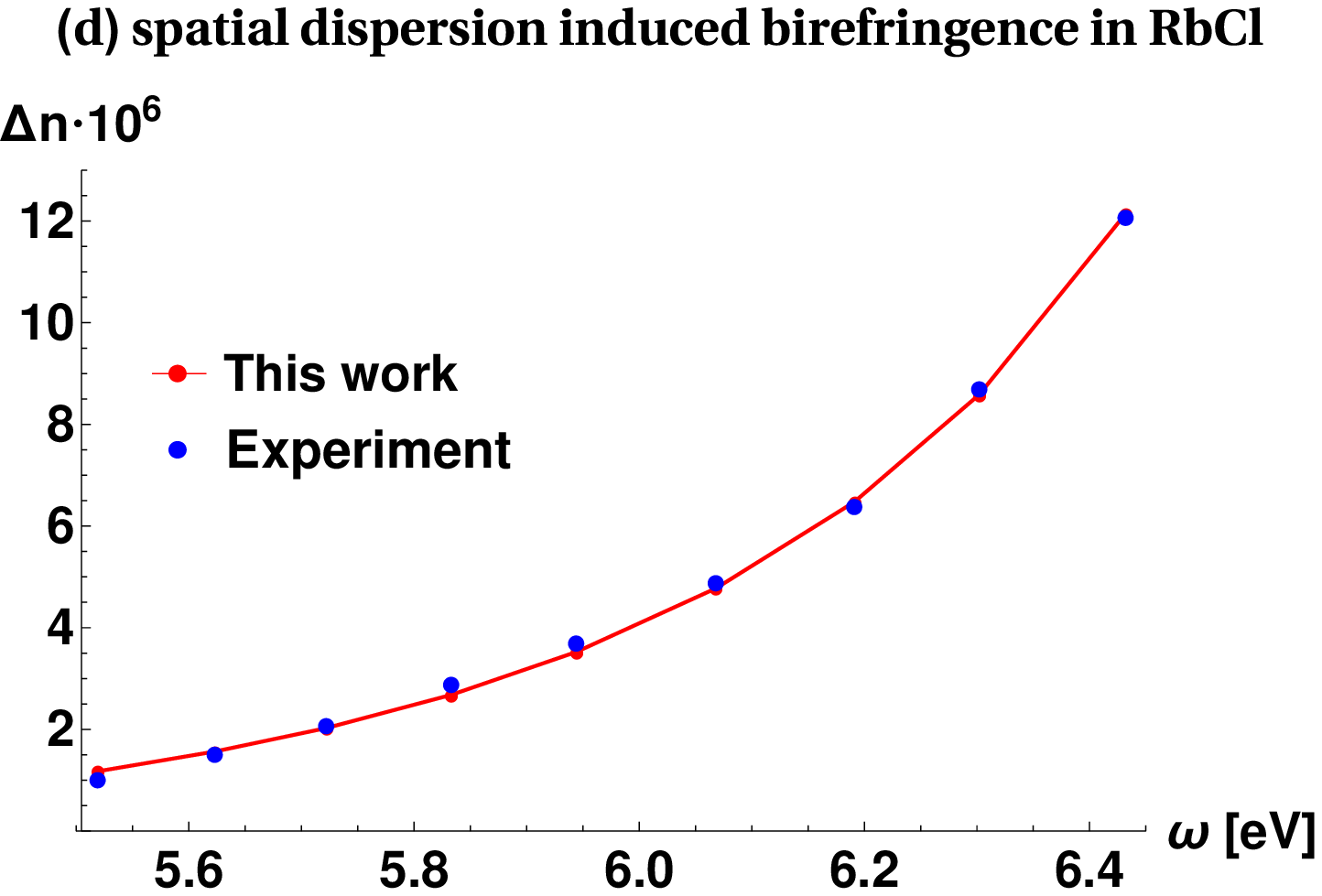} %
\end{minipage}\caption{\label{Sellmeier} Plot of index of refraction $n\left(\omega\right)$
vs. free space wavelength $\lambda=\frac{2\pi c}{\omega}$ for (a)
$CsI$ and (b) $RbCl$. Displayed are values (red dots) calculated
from the macroscopic dielectric tensor (\ref{eq: macroscopic dielectric function arbitrary M})
for $\mathbf{q}=\mathbf{0}$, solely with the lattice symmetry and
the microscopic electronic (\ref{eq: isotropic Lorentz polarizability diagonal})
and ionic polarizabilities (\ref{eq: isotropic Lorentz polarizability off diagonal})
as input, the respective parameters $\alpha_{0}^{\left(j\right)}$,
$\omega_{0}^{\left(j\right)}$,$\alpha_{0}^{\left(j,j'\right)}$ and
$\omega_{0}^{\left(j,j'\right)}$ as listed in Table \ref{fit}. The
relative error compared to a fit of experimental data for $n\left(\omega\right)$
with the multi-parameter Sellmeier formula \citep{Li1976} (blue line)
is less than 1\%. The spatial dispersion induced birefringence $\Delta\text{n}$
as calculated from the $\mathbf{q}$-dependence of the macroscopic
dielectric function (\ref{eq: macroscopic dielectric function arbitrary M})
is displayed (red) in (c) for CsI and (d) for RbCl. To compare with
experimental data \citep{Zaldo1986} (blue dots) the orientation of
the $\mathbf{q}-$vector was chosen along the diagonal of the x-y
plane. The estimated error in reading from the plots of the experimental
data in \citep{Zaldo1986} is about $\pm0.2\cdot10^{-6}$. }
\end{figure*}

\begin{figure}
\begin{minipage}[t]{1\columnwidth}%
\includegraphics[scale=0.7]{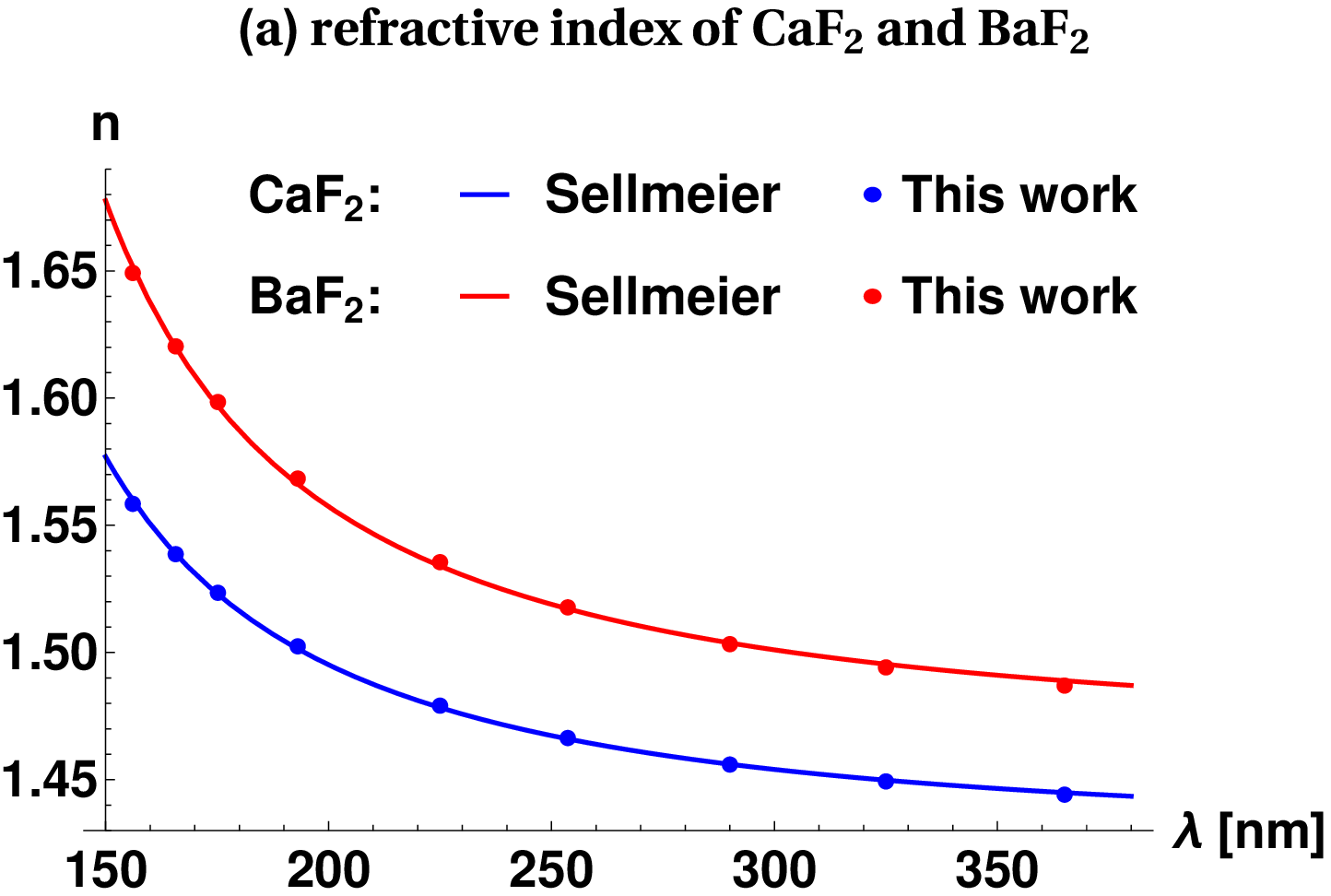}%
\end{minipage}

\vspace{1cm}

\begin{minipage}[t]{1\columnwidth}%
\includegraphics[scale=0.7]{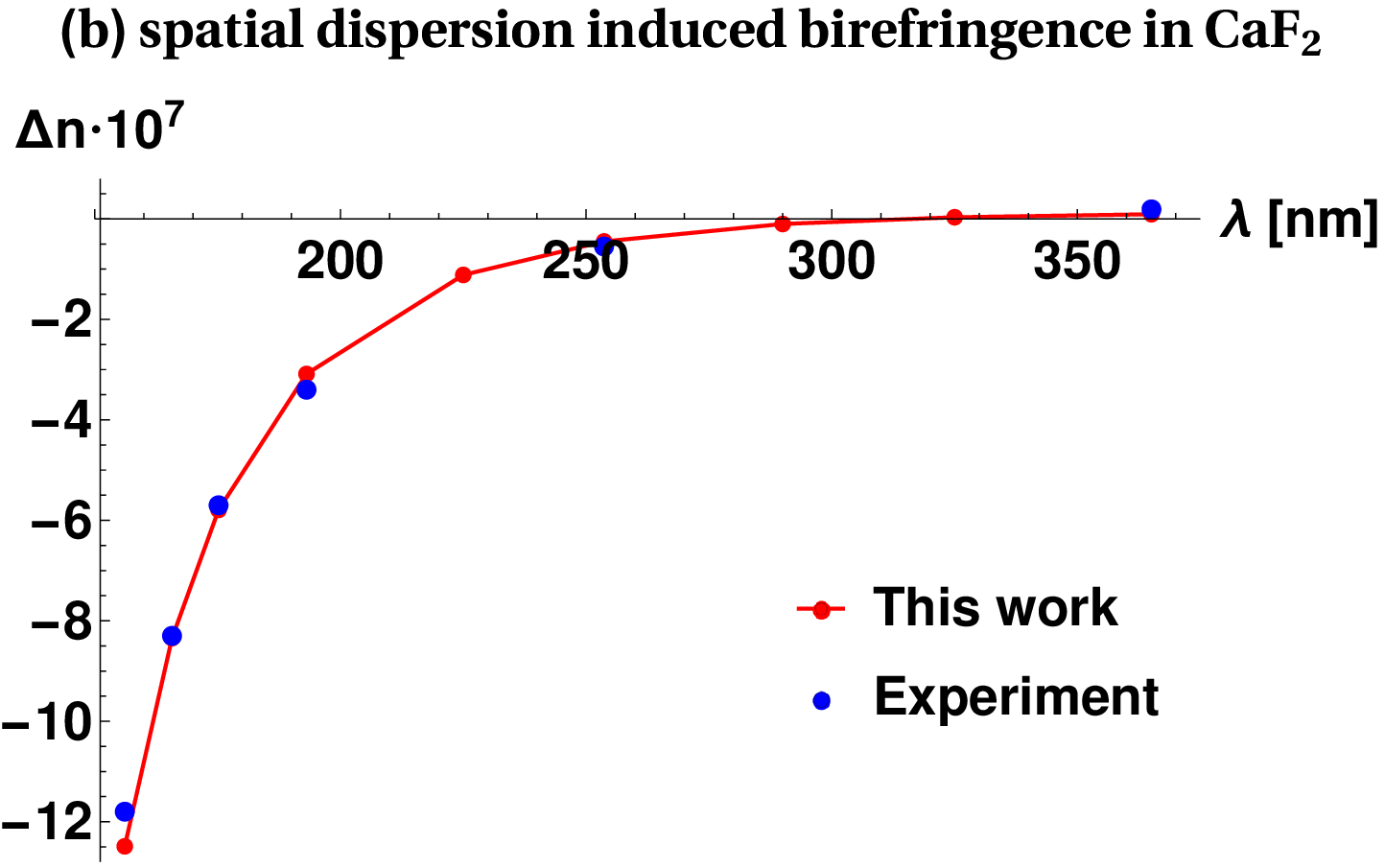}%
\end{minipage}

\vspace{1cm}

\begin{minipage}[t]{1\columnwidth}%
\includegraphics[scale=0.7]{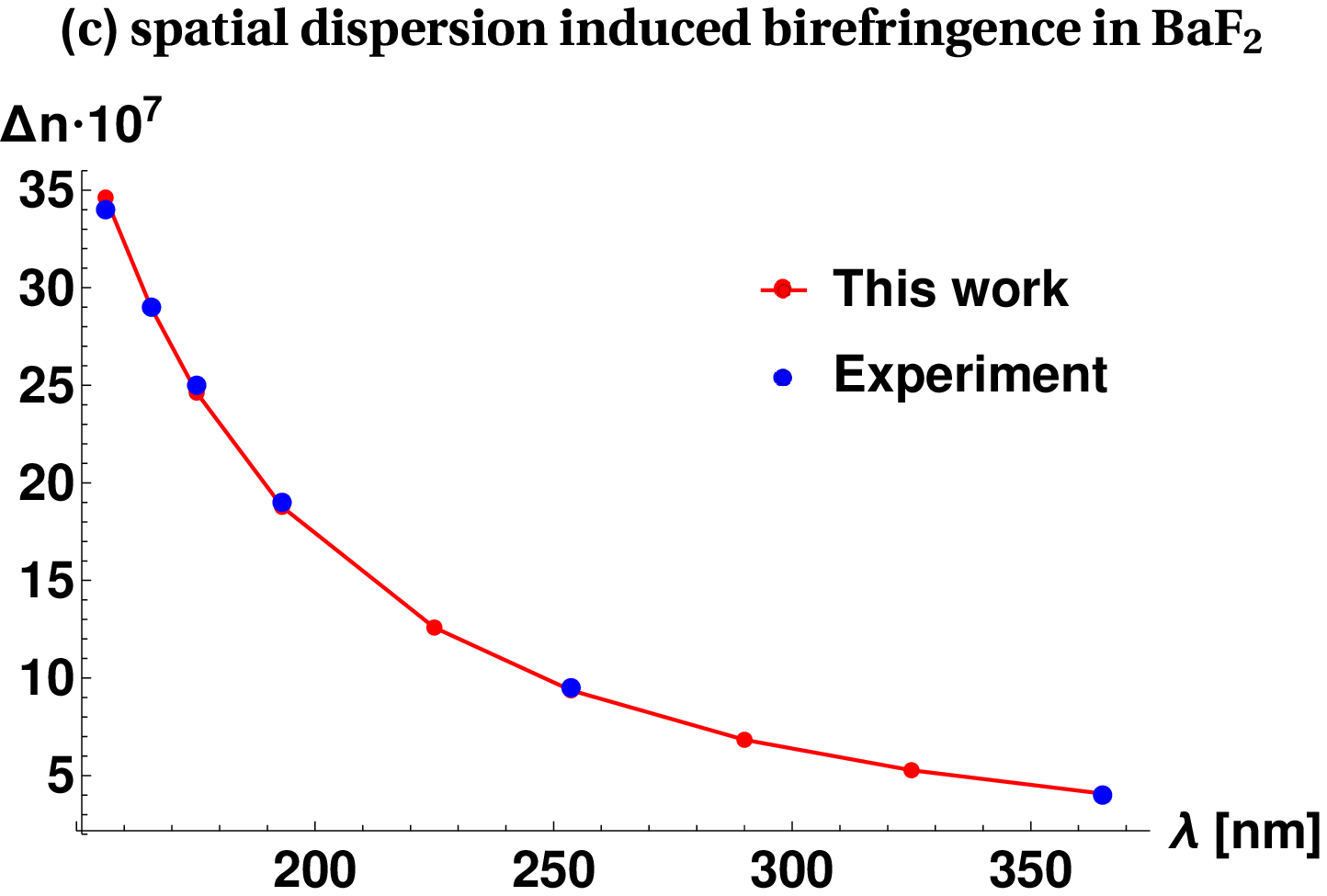}%
\end{minipage}

\caption{\label{fig:CaF2_and_BaF2}(a) Plot of index of refraction $n\left(\omega\right)$
vs. free space wavelength $\lambda=\frac{2\pi c}{\omega}$ for $CaF_{2}$
and $BaF_{2}$. Displayed are values (dots) calculated from the macroscopic
dielectric function (\ref{eq: macroscopic dielectric function arbitrary M})
for $\mathbf{q}=\mathbf{0}$, solely with the lattice symmetry and
the model for electronic polarizabilities (\ref{eq: isotropic Lorentz polarizability diagonal})
as input, the respective values of the parameters $\alpha_{0}^{\left(j\right)}$
and $\omega_{0}^{\left(j\right)}$ as in Table \ref{fit}. The relative
error compared to a fit of experimental data for $n\left(\omega\right)$
with the multi-parameter Sellmeier formula \citep{Li1980} (solid
lines) is less than 1\%. The spatial dispersion induced birefringence
$\Delta\text{n}$ as calculated from the $\mathbf{q}$-dependence
of the macroscopic dielectric function (\ref{eq: macroscopic dielectric function arbitrary M})
is displayed (red) in (b) for $CaF_{2}$ and (c) for $BaF_{2}$. To
compare with experimental data \citep{Burnett2001} (blue dots) the
orientation of the $\mathbf{q}-$vector was chosen along the diagonal
of the x-y plane. }
\end{figure}

Conceiving the polarizability of atoms or ions not as given input
from microscopic theory, but as a fit function of the type (\ref{eq: model for microscopic polarizability}),
the diagonal parts depending on two parameters $\omega_{0}^{\left(j\right)}$
and $\alpha_{0}^{\left(j\right)}$ for each atom or ion species $j\in\left\{ 1,2,...M\right\} $
and also the off diagonal parts depending on two parameters $\omega_{0}^{\left(j,j'\right)}$
and $\alpha_{0}^{\left(j,j'\right)}$ for each pair $\left(j,j'\right)$
of ions $j,j'\in\left\{ 1,2,...M\right\} $ in the unit cell $C_{\Lambda}$,
we find the well known Sellmeier fit \citep{Brueckner2011} of the
frequency dependence of the index of refraction $n\left(\omega\right)$
is nicely reproduced from the eigenvalues of the dielectric tensor
$\varepsilon_{ab}^{\left(T\right)}\left(\mathbf{q}=\mathbf{0},\omega\right)$
for a variety of ionic compounds, see exemplarily our results for
$CsI$ and $RbCl$ in Fig.\ref{Sellmeier} and $CaF_{2}$ and $BaF_{2}$
in Fig.\ref{fig:CaF2_and_BaF2}. The relative error of our calculations
compared to a fit of experimental data for $n\left(\omega\right)$
with the Sellmeier formula for the mentioned crystals is less than
1\%. 

Let us emphasize our approach warrants notably fewer fit parameters
compared to a Sellmeier fit. For example, to reproduce the experimentally
observed chromatic dispersion of $CsI$ over a wide frequency intervall,
ranging from the ultraviolet to far-infrared, a satisfactory fit to
the experimental data within our approach needs only two functions
of the type (\ref{eq: isotropic Lorentz polarizability diagonal})
to model the \emph{induced electronic polarization} of individual
ions and a third fit function of the type (\ref{eq: isotropic Lorentz polarizability off diagonal})
to model the \emph{ionic polarization} effect. So our fit relies only
on $6$ parameters modelling the microscopic polarizabilities of atoms
(ions, molecules, ion pairs) compared, for example, to the $17$ parameters
required by the Sellmeier fit \citep{Li1976}, describing chromatic
dispersion of the refractive index of $CsI$. 

Furthermore, for (ultraviolet) light propagating along the diagonal
of the x-y plane, i.e. $\mathbf{q}=\frac{\left|\mathbf{q}\right|}{\sqrt{2}}\left(\mathbf{e}^{\left(x\right)}+\mathbf{e}^{\left(y\right)}\right)$,
the dielectric tensor $\varepsilon_{ab}^{\left(T\right)}\left(\mathbf{q},\omega\right)$
reveals two transversal modes capable to propagate with slightly different
speeds inside the crystals mentioned above, thus causing an intrinsic
birefringence $\Delta n\left(\omega\right)$ induced by spatial dispersion.
Our calculation of $\Delta n\left(\omega\right)$ for the afore mentioned
ionic crystals and a comparison with experimental data can be found
in Fig.\ref{Sellmeier} and Fig.\ref{fig:CaF2_and_BaF2}. The applied
fit parameters entering the calculations of $n\left(\omega\right)$
and $\Delta n\left(\omega\right)$ are listed in Table \ref{fit}.

Having thus determined the model polarizabilities (\ref{eq: isotropic Lorentz polarizability diagonal})
and (\ref{eq: isotropic Lorentz polarizability off diagonal}) for
each (different) atom species and each ion pair, the dependence of
the dielectric function on wave vector $\mathbf{q}$ is in our approach
already fixed by the crystalline structure of the material under consideration,
i.e. the rotary power $\gamma_{abc}\left(\omega\right)$ and the dispersion
induced anisotropy $\alpha_{abcd}\left(\omega\right)$ are already
implicitely encoded in the $\mathbf{q}-$dependence of the transversal
dielectric tensor $\varepsilon_{ab}^{\left(T\right)}\left(\mathbf{q},\omega\right)$.
To what large extend our calculations agree with published experimental
data over a wide range of optical frequencies for a series of quite
different crystalline materials we summarize in Table \ref{results},
and in particular in Fig. \ref{Sellmeier} and Fig.\ref{fig:CaF2_and_BaF2}. 

While the refractive indices are deduced from the square root of the
(real) eigenvalues of the transversal dielectric function for $\mathbf{q}=\mathbf{0}$,
the rotary power is determined from the imaginary part of the off-diagonal
elements of $\varepsilon_{ab}^{\left(T\right)}\left(\mathbf{q},\omega\right)$
for wave propagation along the crystals' (optical) z-axis, i.e. $\mathbf{q}=\left|\mathbf{q}\right|\mathbf{e}^{\left(z\right)}$.
The examples of crystal structures listed in Table \ref{results}
cover the cubic crystal system as well as all uniaxial crystal systems,
where the number M of ions comprising the unit cell $C_{\Lambda}$
varies between $M=4$ (for e.g. hexagonal BeO) and $M=66$ (for e.g.
cubic $Bi_{12}SiO_{20}$ and $Bi_{12}TiO_{20}$). It should be pointed
out, that in contrast to the results shown in Fig.\ref{Sellmeier}
and \ref{fig:CaF2_and_BaF2}, our calculations for the refractive
index as well as for the rotary power, both presented in Table \ref{results},
solely rest on published data of (anisotropic) electronic polarizabilities
being reported in the particular cited references. 

As a side remark let us point out, that the described principal effects
of non locality, the optical activity $\gamma_{abc}\left(\omega\right)$
and/or the dispersion induced anisotropy $\alpha_{abcd}\left(\omega\right)$,
at first sight being small effects compared to $n_{a}^{2}-1$ with
$n_{a}^{2}$ representing the eigenvalues of the tensor $\varepsilon_{ab}^{\left(T\right)}\left(\mathbf{q},\omega\right)$
in ordinary crystalline materials, could well be comparable to $n_{a}^{2}-1$
in artificial periodic structures choosing appropriately taylored
super lattices, see \citep{Gorlach2016}.

\begin{table*}
\caption{\label{results} Calculation of the index of refraction $n^{(\text{calc})}$
and rotary power $\rho^{(\text{calc})}$ for wave propagation along
the (optical) z-axis for various crystals and comparison with experimental
results. The applied data for crystal structures and electronic polarizabilities
entering our calculations, as well as the experimental data for the
refractive index $n^{(\text{exp})}$ and rotary power $\rho^{(\text{exp})}$
are taken from the publications cited in the column ``references''.
In case of anisotropic polarizabilities, see e.g. the uniaxial crystals
TiO$_{2}$ and CaCO$_{3}$, $\left[\alpha{}^{\parallel}\right]'$
and $\left[\alpha{}^{\perp}\right]'$ denote the polarizabilities
parallel and perpendicular to the optical z-axis, respectively. }

\begin{ruledtabular}
\begin{tabular}{ccccccccc}
crystal  & space group  & $\lambda$ (nm)  & $\alpha'=\frac{\alpha}{4\pi\varepsilon_{0}}$ ({\AA }$^{3}$)  & $n^{(\text{exp})}$  & $n^{(\text{calc})}$  & $\rho^{(\text{exp})}$ ($\frac{\text{degree}}{\text{mm}}$)  & $\rho^{(\text{calc})}$ ($\frac{\text{degree}}{\text{mm}}$) & references\tabularnewline
\hline 
$\alpha$-SiO$_{2}$  & P3$_{1}$21  & 508  & $\alpha'_{\text{Si}}=0.207$  & $n_{\text{o}}=1.548$  & $n_{\text{o}}=1.543$  & -29.73  & -29.25  & \citep{Gualtieri2000,Devarajan1986,Radhakrishnan1951,Lowry}\tabularnewline
 &  &  & $\alpha'_{\text{O}}=1.213$  & $n_{\text{e}}=1.558$  & $n_{\text{e}}=1.550$  &  &  & \tabularnewline
$\beta$-SiO$_{2}$  & P6$_{2}$22  & 517  & $\alpha'_{\text{Si}}=0.185$  & $n_{\text{o}}=1.536$  & $n_{\text{o}}=1.534$  & +33.6  & +29.84  & \citep{Wyckoff,Devarajan1986,1962,Lowry}\tabularnewline
 &  &  & $\alpha'_{\text{O}}=1.250$  & $n_{\text{e}}=1.544$  & $n_{\text{e}}=1.539$  &  &  & \tabularnewline
TiO$_{2}$  & P4$_{2}$/mnm  & 589  & $\alpha'_{\text{Ti}}=0.1862$  & $n_{\text{o}}=2.613$  & $n_{\text{o}}=2.600$  & /  & /  & \citep{Restori1987,Parker1961,DeVore1951}\tabularnewline
 &  &  & $\left[\alpha{}_{O}^{\parallel}\right]'=2.6006$  & $n_{\text{e}}=2.909$  & $n_{\text{e}}=2.921$  &  &  & \tabularnewline
 &  &  & $\left[\alpha{}_{O}^{\perp}\right]'=2.2863$  &  &  &  &  & \tabularnewline
BeO  & P6$_{3}$mc  & 633  & $\alpha'_{\text{Be}}=0.007$  & $n_{\text{o}}=1.717$  & $n_{\text{o}}=1.713$  & /  & /  & \citep{Vidal-Valat1987,Dimitrov2002,Weber}\tabularnewline
 &  &  & $\alpha'_{\text{O}}=1.290$  & $n_{\text{e}}=1.732$  & $n_{\text{e}}=1.717$  &  &  & \tabularnewline
NaClO$_{3}$  & P2$_{1}$3  & 633  & $\alpha'_{\text{Na}}=0.290$  & 1.514  & 1.526  & +2.44  & +3.76  & \citep{Abrahams1977a,Devarajan1986,Chandrasekhar1967,Abrahams1977}\tabularnewline
 &  &  & $\alpha'_{\text{Cl}}=0.010$  &  &  &  &  & \tabularnewline
 &  &  & $\alpha'_{\text{O}}=1.600$  &  &  &  &  & \tabularnewline
SrTiO$_{3}$  & Pm$\bar{3}$m  & 589  & $\alpha'_{\text{Sr}}=1.0666$  & 2.410  & 2.409  & /  & /  & \citep{Abramov1995,Chaib2004,Weber}\tabularnewline
 &  &  & $\alpha'_{\text{Ti}}=0.1859$  &  &  &  &  & \tabularnewline
 &  &  & $\alpha'_{\text{O}}=2.3940$  &  &  &  &  & \tabularnewline
Bi$_{12}$TiO$_{20}$  & I23  & 633  & $\alpha'_{\text{Bi}}=0.0625$  & 2.562  & 2.553  & -5.9  & -6.12  & \citep{Swindells1988,Weber,Feldman1970}\tabularnewline
 &  &  & $\alpha'_{\text{Ti}}=0.272$  &  &  &  &  & \tabularnewline
 &  &  & $\alpha'_{\text{O}}=3.725$  &  &  &  &  & \tabularnewline
Bi$_{12}$SiO$_{20}$  & I23  & 650  & $\alpha'_{\text{Bi}}=0.150$  & 2.52  & 2.50  & -20.5  & -19.35  & \citep{Abrahams1979a,Devarajan1986,Weber,Abrahams1979}\tabularnewline
 &  &  & $\alpha'_{\text{Si}}=0.001$  &  &  &  &  & \tabularnewline
 &  &  & $\alpha'_{\text{O}}=3.540$  &  &  &  &  & \tabularnewline
$\alpha$-AlPO$_{4}$  & P3$_{1}$21  & 633  & $\alpha'_{\text{Al}}=0.050$  & $n_{\text{o}}=1.524$  & $n_{\text{o}}=1.541$  & +14.6  & +11.23  & \citep{Thong1979,Devarajan1986,Weber,Malgrange2014}\tabularnewline
 &  &  & $\alpha'_{\text{P}}=0.050$  & $n_{\text{e}}=1.533$  & $n_{\text{e}}=1.545$  &  &  & \tabularnewline
 &  &  & $\alpha'_{\text{O}}=1.370$  &  &  &  &  & \tabularnewline
BaTiO$_{3}$  & P4mm  & 589  & $\alpha'_{\text{Ba}}=1.9460$  & $n_{\text{o}}=2.426$  & $n_{\text{o}}=2.400$  & /  & /  & \citep{Xiao2008,Chaib2004,Wemple1968}\tabularnewline
 &  &  & $\alpha'_{\text{Ti}}=0.1859$  & $n_{\text{e}}=2.380$  & $n_{\text{e}}=2.380$  &  &  & \tabularnewline
 &  &  & $\alpha'_{\text{O}}=2.3940$  &  &  &  &  & \tabularnewline
CaCO$_{3}$  & R$\bar{3}$cH  & 589  & $\alpha'_{\text{Ca}}=0.792$  & $n_{\text{o}}=1.658$  & $n_{\text{o}}=1.626$  & /  & / & \citep{Maslen1993,Lawless1964,Ghosh1999}\tabularnewline
 &  &  & $\alpha'_{\text{C}}=0.000$  & $n_{\text{e}}=1.486$  & $n_{\text{e}}=1.513$  &  &  & \tabularnewline
 &  &  & $\left[\alpha{}_{\text{O}}^{\left(||\right)}\right]'=1.384$  &  &  &  &  & \tabularnewline
 &  &  & $\left[\alpha{}_{\text{O}}^{\left(\perp\right)}\right]'=1.328$  &  &  &  &  & \tabularnewline
\end{tabular}\end{ruledtabular}

\end{table*}

\begin{table}
\caption{\label{fit} Estimated fit parameters applied to the electronic and
ionic Lorentz oscillator models (\ref{eq: isotropic Lorentz polarizability diagonal})
and (\ref{eq: isotropic Lorentz polarizability off diagonal}), respectively,
regarding our calculations of the refractive index $n\left(\omega\right)$
as well as the spatial-dispersion-induced birefringence $\Delta n\left(\omega\right)$,
for ionic crystals $BaF_{2}$, $CaF_{2}$, $CsI$ and $RbCl$. }

\begin{ruledtabular}
\begin{tabular}{ccc}
ion/binding  & $\frac{\alpha_{0}}{4\pi\varepsilon_{0}}\;\left[{\lyxmathsym{\AA}}^{3}\right]$ & $\omega_{0}\;\left[\text{eV}\right]$\tabularnewline
\hline 
$\text{Ba}^{2+}$ & 1.577 & 16.353\tabularnewline
$\text{Ca}^{2+}$ & 0.759 & 27.484\tabularnewline
$\text{Cl}^{-}$  & 4.500 & 12.959\tabularnewline
$\text{Cs}^{+}$  & 2.884  & 33.220\tabularnewline
$\text{F}^{-}$ (in $BaF_{2}$) & 1.165 & 15.789\tabularnewline
$\text{F}^{-}$ (in $CaF_{2}$) & 0.866 & 15.860\tabularnewline
$\text{I}^{-}$  & 6.241  & 8.253\tabularnewline
$\text{Rb}^{+}$  & 0.285 & 7.359\tabularnewline
\hline 
$\text{Cs}^{+}$--- $\text{I}^{-}$ & 1.519 & 0.012\tabularnewline
$\text{Rb}^{+}$--- $\text{Cl}^{-}$  & 2.214 & 0.019\tabularnewline
\end{tabular}\end{ruledtabular}

\end{table}

\subsection*{Static Limit of the Dielectric Function for Monoatomic Bravais Lattices}

Assuming for simplicity $M=1$, then there is no loss of generality
setting $\boldsymbol{\eta}^{\left(1\right)}=\mathbf{0}$. Identifying
then, see (\ref{eq:block  matrix alpha}), 
\begin{eqnarray}
\alpha_{a''a'}\left(\omega\right) & \equiv & \alpha_{a'',a'}^{\left(1,1\right)}\left(\mathbf{k},\omega\right)\\
I_{aa'} & = & \delta_{a,a'},\nonumber 
\end{eqnarray}
one readily infers from (\ref{eq:  kernel K(G=00003D0,q,om)}) the
explicit representation 
\begin{equation}
\left[\bar{K}_{\Lambda}(\mathbf{q},\omega)\right]_{a',a''}=\left(\frac{\alpha\left(\omega\right)}{\varepsilon_{0}}\circ\left[I-\zeta_{\varLambda}^{\left(0\right)}(\mathbf{q},\omega)\circ\frac{\alpha\left(\omega\right)}{\varepsilon_{0}}\right]^{-1}\right)_{a',a''}.\label{eq: kernel K for M=00003D1}
\end{equation}
Elementary matrix algebra leads then to the result
\begin{eqnarray}
 &  & \varepsilon_{\Lambda}\left(\mathbf{q},\omega\right)-I\nonumber \\
 & = & \frac{1}{\left|C_{\Lambda}\right|}\frac{\alpha\left(\omega\right)}{\varepsilon_{0}}\circ\left[I-\zeta_{\varLambda}^{\left(0\right)}(\mathbf{q},\omega)\circ\frac{\alpha\left(\omega\right)}{\varepsilon_{0}}\right]^{-1}\circ\left[I+\frac{1}{\left|C_{\Lambda}\right|}\bar{\mathcal{G}}(\mathbf{q},\omega)\circ\frac{\alpha\left(\omega\right)}{\varepsilon_{0}}\circ\left[I-\zeta_{\varLambda}^{\left(0\right)}(\mathbf{q},\omega)\circ\frac{\alpha\left(\omega\right)}{\varepsilon_{0}}\right]^{-1}\right]^{-1}\nonumber \\
 & = & \frac{1}{\left|C_{\Lambda}\right|}\frac{\alpha\left(\omega\right)}{\varepsilon_{0}}\circ\left[I-\left(\zeta_{\varLambda}^{\left(0\right)}(\mathbf{q},\omega)-\frac{1}{\left|C_{\Lambda}\right|}\bar{\mathcal{G}}(\mathbf{q},\omega)\right)\circ\frac{\alpha\left(\omega\right)}{\varepsilon_{0}}\right]^{-1}.\label{eq: dielectric function M=00003D1}
\end{eqnarray}
In the supplemental material \citep{Supplementary} it is shown that
\begin{eqnarray}
 &  & \left(\zeta_{\varLambda}^{\left(0\right)}(\mathbf{q},\omega)-\frac{1}{\left|C_{\Lambda}\right|}\bar{\mathcal{G}}(\mathbf{q},\omega)\right)_{a,a'}\label{eq: sum over reciprocal lattice vectors for dielectric function}\\
 & = & \lim_{\left|\mathbf{s}\right|\rightarrow0^{+}}\left(\frac{1}{\left|C_{\Lambda}\right|}\sum_{\mathbf{G}\in\Lambda^{-1}\setminus\left\{ \mathbf{0}\right\} }e^{i\mathbf{\mathbf{G}}\cdot\mathbf{s}}\frac{\frac{\omega^{2}}{c^{2}}\delta_{a,a'}-\left(\mathbf{G}+\mathbf{q}\right)_{a}\left(\mathbf{G}+\mathbf{q}\right)_{a'}}{\left|\mathbf{\mathbf{\mathbf{G}+\mathbf{q}}}\right|^{2}-\frac{\omega^{2}}{c^{2}}-i0^{+}}-\int\frac{d^{3}k}{\left(2\pi\right)^{3}}e^{i\mathbf{k}\cdot\mathbf{s}}\frac{\frac{\omega^{2}}{c^{2}}\delta_{a,a'}-k_{a}k_{a'}}{\left|\mathbf{k}\right|^{2}-\frac{\omega^{2}}{c^{2}}-i0^{+}}\right).\nonumber 
\end{eqnarray}
The lattice sum (\ref{eq: sum over reciprocal lattice vectors for dielectric function})
is conveniently evaluated along the lines indicated by Ewald \citep{Ewald1916,Ewald1938},
splitting the sum into two absolutely converging sums, one converging
rapidly in the Fourier domain and the other converging rapidly in
the spatial domain. For details and a discussion of our (modified)
splitting method, see supplemental material \citep{Supplementary}.
For simple cubic lattices the expression (\ref{eq: sum over reciprocal lattice vectors for dielectric function})
can also be evaluated employing Jacobi theta functions \citep{Draine1993,Borwein2013}.

In the static limit, first $\left|\mathbf{q}\right|\rightarrow0$
and then $\omega\rightarrow0$, we write 
\begin{equation}
\lim_{\omega\rightarrow0}\lim_{\left|\mathbf{q}\right|\rightarrow0}\left[\varepsilon_{\Lambda}\left(\mathbf{q},\omega\right)-I\right]_{a,a'}=\left[\varepsilon_{\Lambda}-I\right]_{a,a'}.
\end{equation}
A comprehensive analysis of the general properties of electromagnetic
response functions, quite apart from a particular model description
of a material and only based on general principles such as causality
and thermodynamic stability, has been given by Kirzhnitz \citep{Kirzhnitz2012},
who derived for the isotropic case $\left[\varepsilon_{\Lambda}\right]_{a,a'}=\varepsilon I_{a,a'}$
as a lower bound of allowed \emph{static} values of the dielectric
function: 
\begin{eqnarray}
\varepsilon & \geq & 1\label{eq: stabilty criterion Kirzhnitz}
\end{eqnarray}

Introducing the (dimensionless) Lorentz factors 
\begin{equation}
\mathcal{L}_{a,a'}=\left|C_{\Lambda}\right|\lim_{\omega\rightarrow0}\lim_{\left|\mathbf{q}\right|\rightarrow0}\left[\zeta_{\varLambda}^{\left(0\right)}(\mathbf{q},\omega)-\frac{1}{\left|C_{\Lambda}\right|}\bar{\mathcal{G}}(\mathbf{q},\omega)\right]_{a,a'},\label{eq:Lorentz factor}
\end{equation}
solely dependent on the lattice geometry, and identifying $\frac{1}{\left|C_{\Lambda}\right|}=\frac{\left|\Lambda_{P}\right|}{\left|\Omega_{P}\right|}=\nu_{P}$
with the \emph{density} of polarizable atoms in the probe volume,
there follows from (\ref{eq: dielectric function M=00003D1}) for
mono-atomic Bravais lattices $(M=1)$ the following exact formula
for the static dielectric tensor: 
\begin{equation}
\left[\varepsilon_{\Lambda}\right]_{a,a'}=\left[\frac{I+(I-\mathcal{L})\frac{\alpha}{\varepsilon_{0}}\nu_{P}}{I-\mathcal{L}\frac{\alpha}{\varepsilon_{0}}\nu_{P}}\right]_{a,a'}\label{eq:static dielectric tensor}
\end{equation}
 While the polarizability $\alpha$ refers to individual atomic (ionic,
molecular) properties, the dielectric tensor $\varepsilon_{\Lambda}$
also depends via the $3\times3$ matrix of Lorentz factors $\mathcal{L}$
on how the atoms are assembled to build the crystal $\Lambda$. Note
that the particle density $\nu_{P}$ of a dielectric crystalline probe
volume is bounded from above by the a critical value $\nu_{P}^{\left(c\right)}$,
\begin{eqnarray}
\nu_{P} & < & \nu_{P}^{\left(c\right)}\equiv\frac{1}{\mathcal{L}_{max}}\frac{\varepsilon_{0}}{\alpha}\;,
\end{eqnarray}
with $\mathcal{L}_{max}$ denoting the largest positive eigenvalue
of the matrix $\mathcal{L}$. This condition indeed implies the matrix
$\varepsilon_{\Lambda}-I$ being \emph{positive definit}, thus generalizing
the stability criterion (\ref{eq: stabilty criterion Kirzhnitz})
to the anisotropic case.

It should be noted that the trace of the matrix $\mathcal{L}$ as
defined in (\ref{eq:Lorentz factor}) is normalized to unity: 
\begin{equation}
\textrm{tr}\left(\mathcal{L}\right)=1\label{eq: trace identity Lorentz factors}
\end{equation}
For a proof see supplemental material \citep{Supplementary}. Numerical
values for the Lorentz factors are readily calculated for all $14$
monoatomic Bravais lattices $\Lambda$ along the lines indicated in
the supplemental material \citep{Supplementary}. 

\begin{figure}
\includegraphics[scale=0.8]{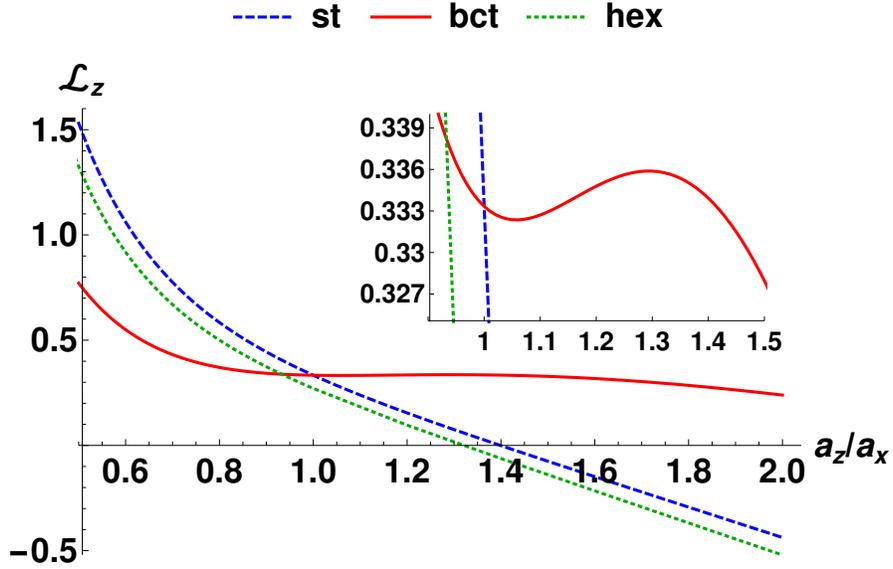}

\caption{\label{fig:Lorentz_factors} Plot of Lorentz factor $\mathcal{L}_{z}$
for simple tetragonal (blue), body-centered tetragonal (red) and hexagonal
(green) lattice symmetries vs. the ratio $\frac{a_{z}}{a_{x}}$ of
lattice constants. The inset shows a zoom into the region where $\mathcal{L}_{z}$
assumes a value around $\frac{1}{3}$, characteristic for isotropic
systems. }
\end{figure}

In the special case of a lattice with tetragonal or hexagonal symmetry
the matrix $\mathcal{L}$ becomes diagonal, $\mathcal{L}_{a,a'}=\delta_{a,a'}\mathcal{L}_{a}$.
In this case the trace identity (\ref{eq: trace identity Lorentz factors})
leads to $\mathcal{L}_{x}=\mathcal{L}_{y}=\frac{1}{2}\left(1-\mathcal{L}_{z}\right)$,
with $\mathcal{L}_{z}$ being a universal function, solely dependent
on the ratio $\frac{a_{z}}{a_{x}}$ of the lattice constants, $a_{z}$
parallel and $a_{x}$ perpendicular to the crystalline $z-$axis.
In Fig. \ref{fig:Lorentz_factors} that function $\mathcal{L}_{z}$
is plotted vs. the ratio of lattice constants  $\frac{a_{z}}{a_{x}}$
for simple tetragonal (st), body-centered tetragonal (bct) and also
hexagonal (hex) lattice symmetry. Our results for the Lorentz-factors
obtained with Eq. (\ref{eq:Lorentz factor}) coincide with previous
works, for example \citep{Colpa1971a,Colpa1971,Vanzo2014}. 

\subsubsection*{Clausius-Mossotti Relation}

For cubic symmetry there holds $\mathcal{L}_{a,a'}=\frac{1}{3}\delta_{a,a'}$
and the static dielectric tensor (\ref{eq:static dielectric tensor})
reduces to the well known Clausius-Mossotti relation for\emph{ isotropic}
systems, see supplemental material \citep{Supplementary}: 
\begin{eqnarray}
\left[\varepsilon_{\Lambda}\right]_{a,a'} & = & \delta_{a,a'}\frac{1+\frac{2}{3}\frac{\alpha}{\varepsilon_{0}}\nu_{P}}{1-\frac{1}{3}\frac{\alpha}{\varepsilon_{0}}\nu_{P}}\label{eq: Clausius-Mossotti}
\end{eqnarray}
Incidentally, the relation (\ref{eq: Clausius-Mossotti}) applies
for a wide class of (isotropic, non polar) materials, including dielectric
liquids and gases. On a final note: our derivation of (\ref{eq: Clausius-Mossotti})
completely avoids the usual trick of introducing the Lorentz sphere,
where the medium outside of this sphere is considered as a continuum.
For an in-depth explanation of that trick see for example \citep{Ashcroft1981}.

\section{Conclusions}

The field-integral equation approach presented in this article differs
from traditional presentations of crystal optics, for example \citep{Authier2012,Born1999,Fluegge2013}.
Based on the Helmholtz-Hodge theorem the source term in the microscopic
Maxwell equations representing the current density has been decomposed
into longitudinal and transversal parts, thus establishing the formulation
of equivalent (inhomogenous) field-integral equations with a kernel
modelling the induced microscopic polarization $\mathbf{\tilde{P}}\left(\mathbf{r},\omega\right)$
inside a dielectric crystal as a convolution integral of the dielectric
susceptibility $\chi_{aa'}\left(\mathbf{r},\mathbf{r}',\omega\right)$
(\ref{eq: phenomenological dielectric kernel}) and the Fourier amplitude
$\mathbf{\tilde{E}}\left(\mathbf{r},\omega\right)$ of the microscopic
local electric field, see (\ref{eq:polarization vs local field}).
Exploiting the lattice periodicity of the dielectric susceptibility
tensor $\chi_{aa'}\left(\mathbf{r}+\mathbf{R},\mathbf{r}'+\mathbf{R},\omega\right)=\chi_{aa'}\left(\mathbf{r},\mathbf{r}',\omega\right)$
it is then natural to expand the field amplitude $\mathbf{\tilde{E}}\left(\mathbf{r},\omega\right)$
into a complete and orthonormal basis of eigenfunctions of the operator
$T_{\mathbf{R}}$ generating translations by a lattice vector $\mathbf{R}\in\Lambda$,
see (\ref{eq: shift operator}). But instead of expanding the solution
to the field-integral equations in the well known basis of plane waves
$e^{i\left(\mathbf{q}+\mathbf{G}\right)\mathbf{r}}$, constructed
from eigenfunctions of the \emph{momentum} operator, thus requiring
to handle for each wave vector $\mathbf{q}$ in the Brillouin zone
$C_{\Lambda^{-1}}$ of the lattice $\Lambda$ then (infinite dimensional)
matrices labelled by reciprocal lattice vectors $\mathbf{G},\mathbf{G}'\in\Lambda^{-1}$,
we use non-standard Bloch functions representing a complete and orthonormal
system of eigenfunctions of the translation operator $T_{\mathbf{R}}$
constructed from eigenfunctions of the \emph{position} operator, see
(\ref{eq: basis w(r,s,k)}). Our choice of basis functions indeed
enables to sidesteps the inversion (and truncation) of large matrices
with regard to the reciprocal lattice vectors $\mathbf{G},\mathbf{G}'\in\Lambda^{-1}$,
thus easing significantly the numerical effort. 

Considering the \emph{homogenous} field-integral equations the electromagnetic
modes and the photonic band structure $\omega_{n}\left(\mathbf{q}\right)$
of a dielectric crystal have been identified solving a \emph{small
}sized $3M\times3M$ matrix eigenvalue problem, with $M$ denoting
the number of polarizable atoms (ions, molecules) in the elementary
cell $C_{\Lambda}$ of the lattice. A radiation damping term $\propto\omega^{3}$,
that originates \emph{not} from the damping terms in the atom-individual
polarizabilities (\ref{eq: atom polarizability}), but from the retarded
Helmholtz propagator evaluating the lattice sum without the self field
term, see (\ref{eq: Im Zeta_0}), has been taken into account in all
our calculations. Exemplarily we then presented the photonic band
structure of diamond ($M=2$). An overview of our results for the
photonic bandstructure of (primitive) sc-, fcc- and bcc lattices we
present in the supplemental material \citep{Supplementary}. In all
cases we find quantitative agreement with previously published work
on photonic band structures calculated with other methods.

Considering the \emph{inhomogenous} field-integral equations, choosing
a (circular) frequency $\omega$ and a wavevector $\mathbf{q}$ obeying
to $\omega=\omega_{n}\left(\boldsymbol{q}\right)$, the microscopic
local electric field amplitude $\mathbf{\tilde{E}}\left(\mathbf{r},\omega\right)$
in the presence of a slowly varying time harmonic external field with
amplitude $\mathbf{\tilde{E}}_{ext}\left(\mathbf{r},\omega\right)$
is found to display in general a radiative transversal part (T) and
also a longitudinal part (L), the size of the longitudinal amplitude
$\mathbf{\tilde{E}}^{\left(L\right)}\left(\mathbf{r},\omega\right)$,
see (\ref{eq: longitudinal microscopic field}), compared to the size
of the transversal amplitude $\mathbf{\tilde{E}}^{\left(T\right)}\left(\mathbf{r},\omega\right)$,
see (\ref{eq:transversal microscopic field}), being strongly depend
on the density of polarizable atoms (ions, molecules) in the crystal,
see Fig.\ref{fig:field_contributions_and_index}. In a sufficiently
dense packed dielectric crystal we find the longitudinal amplitude
$\mathbf{\tilde{E}}^{\left(L\right)}\left(\mathbf{r},\omega\right)$
being substantially larger compared to the transversal amplitude $\mathbf{\tilde{E}}^{\left(T\right)}\left(\mathbf{r},\omega\right)$,
but in a dilute (artificial) superlattice with polarizable subunits
positioned at the lattice sites the longitudinal field amplitude becomes
small compared to the transversal one, see Fig.\ref{fig:field_contributions_and_index}.
While the microscopic local electric field amplitude $\mathbf{\tilde{E}}\left(\mathbf{r},\omega\right)$
displays inside a densely packed lattice rapid spatial variations
with large amplitude traversing a distance set by the microscopic
inter-particle distance $a$, it turns out that the transversal (radiative)
part $\mathbf{\tilde{E}}^{\left(T\right)}\left(\mathbf{r},\omega\right)$
essentially coincides with the slowly varying \emph{macroscopic} electric
field amplitude $\mathcal{\boldsymbol{\tilde{E}}}\left(\mathbf{r},\omega\right)$,
that field $\mathbf{\tilde{\mathcal{\boldsymbol{E}}}}\left(\mathbf{r},\omega\right)$
being conceived as the low pass filtered microscopic local field amplitude,
see (\ref{eq: low pass filtered microscopic electric field}). If
the external field $\mathbf{\tilde{E}}_{ext}\left(\mathbf{r},\omega\right)$
was purely transversal and slowly varying, then the amplitude of the
residue $\delta\mathbf{\tilde{E}}^{\left(T\right)}\left(\mathbf{r},\omega\right)=\mathbf{\tilde{E}}^{\left(T\right)}\left(\mathbf{r},\omega\right)-\mathbf{\tilde{\mathcal{\boldsymbol{E}}}}^{\left(T\right)}\left(\mathbf{r},\omega\right)$
is tiny, see Fig.\ref{fig:local_vs_macro_fields}.

Conceiving correspondingly the \emph{macroscopic} Polarization $\tilde{\mathcal{\boldsymbol{P}}}\left(\mathbf{r},\omega\right)$
as the low pass filtered \emph{microscopic} polarization $\mathbf{\tilde{P}}\left(\mathbf{r},\omega\right)$,
see (\ref{eq: low pass filtered microscopic polarization}), then
the dielectric $3\times3$ tensor $\varepsilon_{\Lambda}\left(\mathbf{q},\omega\right)$
of macroscopic electrodynamics emerges from the requirement $\bar{\mathcal{\boldsymbol{P}}}\left(\mathbf{q},\omega\right)=\varepsilon_{0}\left[\varepsilon_{\Lambda}\left(\mathbf{q},\omega\right)-I\right]\mathcal{\bar{E}}\left(\mathbf{q},\omega\right)$,
with $\bar{\mathcal{\boldsymbol{P}}}\left(\mathbf{q},\omega\right)$
and $\mathcal{\bar{E}}\left(\mathbf{q},\omega\right)$ denoting the
spatial Fourier transformation of those amplitudes. The derived exact
formula for the dielectric tensor $\varepsilon_{\Lambda}\left(\mathbf{q},\omega\right)$,
see (\ref{eq: macroscopic dielectric function arbitrary M}), then
solely depends on the microscopic polarizabilities $\alpha_{aa'}\left(\boldsymbol{\eta}^{\left(j\right)},\boldsymbol{\eta}^{\left(j'\right)},\omega\right)$
of atoms (molecules, ions) together with the crystalline symmetry
$\Lambda$ as input into the theory.

Regarding the propagation of light signals inside a dielectric crystal
then the question was clarified, which parts of the dielectric tensor
$\varepsilon_{\Lambda}\left(\mathbf{q},\omega\right)$ describe the
renormalization of the speed of light and possibly birefringence,
chromatic dispersion, rotary power and spatial-dispersion induced
birefringence, and which parts govern electric-field screening. Accordingly
we derived directly from the field-integral equations, see (\ref{eq: low pass filtered field integral equation}),
a set of (coupled) differential equations for the longitudinal and
transversal components determining the macroscopic electric field
$\mathcal{\boldsymbol{\tilde{E}}}\left(\mathbf{r},\omega\right)$
directly, without any prior knowledge of the microscopic local electric
field $\mathbf{\tilde{E}}\left(\mathbf{r},\omega\right)$. If the
external field was purely transversal then an effective wave equation
for the transversal (radiative) macroscopic field emerged, thus identifying
the pieces of $\varepsilon_{\Lambda}\left(\mathbf{q},\omega\right)$
comprising the transversal dielectric tensor $\varepsilon_{aa'}^{\left(T\right)}\left(\mathbf{q},\omega\right)$
, see (\ref{eq: transversal dielectric tensor}). Conversely, if the
external field was purely longitudinal, then an effective Poisson
type equation for a scalar potential $\tilde{\phi}\left(\mathbf{r},\omega\right)$
turned up, so that the longitudinal macroscopic field is represented
by $\mathbf{\tilde{\mathcal{\boldsymbol{E}}}}^{\left(L\right)}\left(\mathbf{r},\omega\right)=-\boldsymbol{\nabla}\tilde{\phi}\left(\mathbf{r},\omega\right)$,
thus identifying the pieces of the dielectric tensor comprising the
longitudinal dielectric tensor $\varepsilon_{aa'}^{\left(L\right)}\left(\mathbf{q},\omega\right)$
being liable for electric-field screening (like in electrostatics),
see (\ref{eq:longitudinal dielectric tensor}). In the static limit,
$\mathbf{q}\rightarrow\mathbf{0}$ and $\omega\rightarrow0$, an exact
expression for the dielectric tensor $\varepsilon_{\Lambda}$ is derived
that applies for all $14$ mono-atomic Bravais lattices, see (\ref{eq:static dielectric tensor}).
Our result for $\varepsilon_{\Lambda}$ in particular conforms with
general (thermodynamic) stability criteria \citep{Kirzhnitz2012},
and for cubic symmetry the well known Clausius-Mossotti relation is
recovered.

The Taylor expansion of the transversal dielectric tensor, $\varepsilon_{ab}^{\left(T\right)}\left(\mathbf{q},\omega\right)=\varepsilon_{ab}^{\left(T\right)}\left(\omega\right)+i\sum_{c}\gamma_{abc}\left(\omega\right)q_{c}+\sum_{c,d}\alpha_{abcd}\left(\omega\right)q_{c}q_{d}+..$
around $\mathbf{q}=\mathbf{0}$, provides then insight into various
optical phenomena connected to retardation and non locality of the
dielectric response, in full agreement with the phenomenological reasoning
of Agranovich and Ginzburg \citep{Agranovich1984}: the eigenvalues
of the tensor $\varepsilon_{ab}^{\left(T\right)}\left(\omega\right)$
describing chromatic dispersion of the index of refraction and birefringence,
the first order term $\gamma_{abc}\left(\omega\right)$ specifying
rotary power (natural optical activity) in crystals lacking inversion
symmetry, the second order term $\alpha_{abcd}\left(\omega\right)$
shaping the (weak) effects of a spatial-dispersion induced birefringence,
nowadays a critical parameter for the design of lenses made from $CaF_{2}$
and $BaF_{2}$ for optical lithograpy systems in the ultraviolet.

Considering various dielectric crystalline materials comprising atoms
(molecules, ions) with known polarizabilities from the literature,
in all cases the calculated indices of refractions, the rotary power
and the dispersion induced birefringence have been shown to coincide
well with the experimental data displayed in table \ref{results},
thus illustrating the utility of the theory. For ionic crystals, exemplarily
for $CsI$ and $RbCl$, a satisfactory agreement between theory and
the measured chromatic dispersion of the index of refraction is manifest
over a wide frequency interval, ranging from ultraviolet to far infra-red,
accomplishing this with an appreciably smaller number of adjustable
parameters directly linked to microscopic atomindividual polarizabilities
compared to the well known Sellmeier fit. Further we computed for
the cubic fluorites $CaF_{2}$ and $BaF_{2}$ the frequency dependence
of the intrinsic birefingence, as represented by three independent
components of the tensor $\alpha_{abcd}\left(\omega\right)$, and
found good agreement with published data \citep{Burnett2001,Burnett2002}.

Even though all our calculations are based on a phenomenological model
of microscopic polarizabilities, see (\ref{eq: isotropic Lorentz polarizability diagonal})
and (\ref{eq: isotropic Lorentz polarizability off diagonal}), due
to the conformity with the fundamental frequency dependence of atom-individual
polarizabilities as predicted by quantum mechanics, see (\ref{eq: atom polarizability}),
our results for the dielectric tensor $\varepsilon_{\Lambda}\left(\mathbf{q},\omega\right)$
may well claim general validity in the range of optical frequencies
and below. While the frequency dependence of $\varepsilon_{\Lambda}\left(\mathbf{q},\omega\right)$
is deeply anchored in the \emph{retarded} response of polarizable
atoms (ions, molecules) comprising a dielectric crystal, the dependence
of $\varepsilon_{\Lambda}\left(\mathbf{q},\omega\right)$ on the wave
vector $\mathbf{q}$ of the propagating light describes the effects
attributed to the \emph{non-locality} of that response, for instance
optical activity and intrinsic birefringence, all theses effects being
primarily dependent on details of the crystalline symmetry. Finally
it should be noted, that our theory of the macroscopic electric field
doesn't make use of the notion of a displacement field $\mathbf{D}$,
and therefore we avoided to mention it.

\subsubsection*{Outlook}

We are confident the presented field-integral equation approach can
be extended to calculations of the microscopic local electric field
near to the surface of a dielectric crystal and also to thin films.
In addition we consider it possible to extend our approach to disordered
systems, for instance crystals subject to substitutional site disorder,
thus enabling a theory of the dielectric tensor for disordered systems
within the frame of the coherent potential approximation (CPA). 
\begin{acknowledgments}
The corresponding author thanks Oleg Dolgov for insightfull remarks
and for bringing the references \citep{L.V.Keldysh2012,Maksimov2012,Kirzhnitz2012}
to his attention, all of this a good while ago. He also thanks Mario
Liu for sharing with him over the years his thoughts on macroscopic
electrodynamics. Last not least we thank Oliver Eibl, Klaus-Peter
Federsel, József Fortágh, Reinhold Kleiner and Claus Zimmermann for
useful discussions.
\end{acknowledgments}

\appendix
\bibliographystyle{unsrt}
\bibliography{References}

\end{document}